\documentclass[%
 reprint,
 amsmath,amssymb,
 aps,
rmp,
floatfix,
]{revtex4-1}
\pdfoutput=1

\usepackage{graphicx}% Include figure files
\usepackage{dcolumn}% Align table columns on decimal point
\usepackage{bm}% bold math
\usepackage{csquotes}
\MakeOuterQuote{"}
\usepackage{tabularx}
\usepackage{float}

\usepackage{hyperref}% add hypertext capabilities
\usepackage{xcolor}
\hypersetup{
    colorlinks,
    linkcolor={black},
    citecolor={blue!50!black},
    urlcolor={blue!80!black}
}
\usepackage[mathlines]{lineno}% Enable numbering of text and display math
\newcommand{\kpcf}{kagom\'{e}}
\newcommand{\pgpcf}{HC-PBG}
\newcommand{\arpcf}{HC-PCF}
\newcommand{\hc}{HC-PCF}
\newcommand{\bhc}{broadband-guiding HC-PCF}
\usepackage{txfonts}
\newcommand{\um}{$\muup$m}
\usepackage{color}

\begin{document}

\title
{Hybrid Photonic-Crystal Fiber}% Force line breaks with \\

\author{Christos Markos}

\affiliation
{
 DTU Fotonik, Department of Photonics Engineering, Technical University of Denmark, DK-2800, Kgs. Lyngby, Denmark
}%

\author{John C. Travers}
 
\affiliation
{%
 Institute of Photonics and Quantum Sciences, School of Engineering and Physical Sciences, Heriot-Watt University, Edinburgh, EH14 4AS, United Kingdom
}%

\author{Amir Abdolvand}

\affiliation
{%
 School of Electrical \& Electronic Engineering, College of Engineering, Nanyang Technological University, 50 Nanyang Avenue 639798, Singapore \\
 \& \\
 Max Planck Institute for the Science of Light, Staudt-Strasse 2, Erlangen D-91058, Germany
}%

\author{Benjamin J. Eggleton}

\affiliation{
 Centre for Ultrahigh-bandwidth Devices for Optical Systems (CUDOS), Institute for Photonic Optical Sciences, School of Physics, University of Sydney, New South Wales 2006, Australia
}%

\author{Ole Bang}
\affiliation{%
  DTU Fotonik, Department of Photonics Engineering, Technical University of Denmark, DK-2800, Kgs. Lyngby, Denmark \\
}%

\date{\today}% It is always \today, today,

\begin{abstract}
This article offers an extensive survey of results obtained using hybrid photonic crystal fibers (PCFs) which constitute one of the most active research fields in contemporary fiber optics. The ability to integrate novel and functional materials in solid- and hollow-core PCFs through various post-processing methods has enabled new directions towards understanding fundamental linear and nonlinear phenomena as well as novel application aspects, within the fields of optoelectronics, material and laser science, remote sensing and spectroscopy. Here the recent progress in the field of hybrid PCFs is reviewed from scientific and technological perspectives, focusing on how different fluids, solids and gases can significantly extend the functionality of PCFs. In the first part of this review we discuss the most important efforts by research groups around the globe to develop tunable linear and nonlinear fiber-optic devices using PCFs infiltrated with various liquids, glasses, semiconductors and metals. The second part is concentrated on the most recent and state-of-the-art advances in the field of gas-filled hollow-core PCFs. Extreme ultrafast gas-based nonlinear optics towards light generation in the extreme wavelength regions of vacuum ultraviolet (VUV), pulse propagation and compression dynamics in both atomic and molecular gases, and novel soliton - plasma interactions are reviewed. A discussion of future prospects and directions is also included.

\end{abstract}

\maketitle

\tableofcontents

\section{\label{1}Introduction} 

\subsection{\label{History}Historical Background}
Practical data transmission using optical fibers with reduced attenuation was proposed for the first time by \textcite{Kao1966a}. This achievement marked an era where optical fibers could be a realistic communication medium. Since the first experimental realization of low loss optical fibers in 1970 \cite{Kapron1970}, world-wide communications---among other fields---experienced an extraordinary growth.

In conventional optical fibers, the silica core of the fiber is usually doped with another material in order to increase the refractive index and consequently satisfy Snell's law of total internal reflection (TIR) \cite{Kao1966a}. A small core diameter of $\sim 5$ $\mu$m is also necessary in order to maintain only the zero order (fundamental) propagation mode, in the visible spectral region. However, doping of the core can lead to high attenuation levels, while the small core size leads to nonlinear interactions that are strongly pronounced over length scales of a few kilometers---these are useful for nonlinear optics, but usually a significant disadvantage for communications. New guiding mechanisms were thus essential in order to minimize or even eliminate these effects. One such early attempt was the idea of using an air-glass microstructure for optical fiber guidance; proposed and demonstrated by \textcite{standley1974experimental} and \textcite{Kaiser1974}. Although unsuccessful, this work would be reinvented with great success a few decades later.

In 1987, \textcite{John1987} and \textcite{Yablonovitch1987} predicted the photonic bandgap effect, an extension of the concept of band structure in semiconductors (a foundation on which modern electronics relies on) to photonics. The electronic band structure is the outcome of the interaction between electrons with the periodic modulation of the potential created by a crystalline lattice. Solving the Schr\"{o}dinger wave equation for a periodic potential, it can be shown that there are fixed "forbidden energy regions", which separate the energy bands. Photonic crystals are considered as the extension of the latter in which the forbidden energy states are now replaced by optical bandgaps, i.e. forbidden frequency ranges for any polarization states, formed via the periodic modulation of the material's effective refractive index. The main aim of a structure with a full 3-D photonic bandgap would be to block the propagation of photons whose frequency (energy) falls within the photonic bandgap frequency range, irrespective of their polarization states. This has attracted a great deal of (scientific) interest, allowing the fabrication of a whole new generation of photonic devices and the emergence of a new research field in optics and photonics \cite{Joannopoulos1997}. The first experimental observation of the photonic bandgap effect---in the microwave transmission range---was reported in 1991 by Yablonovitch \textit{et al.} in a bulk high index material with 1-mm drilled holes \cite{Yablonovitch1991}. 

\subsubsection{Photonic crystal fiber (PCF)}
Inspired by Yablonovitch's work \cite{Yablonovitch1987}, in 1991, Philip Russell conceived the idea of a hollow-core waveguide based on a 2-D out-of-plane photonic bandgap, i.e a bandgap for angles of incidence out of the plane of periodicity \cite{Russell2001}. This reduces the constraint of using high refractive index contrast materials, required for in-plane photonic bandgaps; a key element that enabled the usage of a simple air-glass interface and made it compatible with fiber optics technology \cite{Birks1995}. Russell's idea particularly targets waveguiding in an air-core fiber in order to overcome the aforementioned limitations of conventional fibers, i.e. material attenuation, dispersion and nonlinear effects \cite{Russell2001}. A bandgap is provided through a 2-D glass-airhole pattern running along the fiber length parallel to the core. However, the experimental realization of such structures required infrastructure not available at that time. Russell and his colleagues developed their own fabrication method by stacking silica capillaries of  1 mm diameter in a specific hexagonal pattern. In 1996 they demonstrated the first photonic crystal fiber \cite{Birks1996}. This first attempt incorporated a solid-core rather than a hollow-core in its center, which was surrounded by an array of micrometer-scaled holes arranged in a triangular lattice with an overall hexagonal shape. Consequently, the fiber did not guide based on the photonic bandgap effect, but rather a mechanism akin to total internal reflection. But because the main motivation at that time was to fabricate a fiber that confined light through the photonic bandgap effect, the term photonic crystal fiber (PCF) was introduced, and remains in use for the whole class of microstructured fibers, including those without any kind of `crystal' structure. This is the term we also use in this review article. `Holey' or microstructured fibers are other terms for PCF. The guidance principle of solid-core PCF is similar to conventional step-index fibers with an added unique feature. If properly designed, the fiber can support only the zero-order (fundamental) mode for all wavelengths \cite{Birks1997}. This endlessly single mode guidance provided a great degree of freedom to engineer the properties of the fiber in ways which do not exist in standard fibers \cite{Ranka2000}.

The initial idea of Russell \textit{et al.} to guide light in air was successfully achieved in 1999, demonstrating for the first time the fabrication of a hollow-core PCF (HC-PCF) capable of confining more than 99\%  of light in its core over distinct narrow transmission windows \cite{cregan_single-mode_1999}. This breakthrough was a significant step forward towards a new and exciting science in the field of optics and photonics, causing several research groups around the globe to focus their activities on this new class of optical fibers. Indeed, with rapid improvements to the design and the fabrication process, in only five years time losses as low as 1 dB/km were reported \cite{Smith2003,Roberts2005}.

\subsubsection{Broadband guiding \hc{}}
In 2002, \textcite{benabid_stimulated_2002} demonstrated the first \hc{} capable of transmitting light not in a narrow spectral region, but over a broad range of optical frequencies. The cladding structure of this fiber was somewhat different from the conventional triangular or honey comb \hc{} structures reported previously, consisting of thin (nanometer scale) silica webs arranged in a \kpcf{} lattice. While it was originally anticipated that the guidance might be due to high-order photonic bandgaps \cite{Couny2006}, using numerical simulation it was soon realized that the cladding structure does not support any photonic bandgaps, but rather regions of low cladding density of photonic states \cite{Pearce2007}. Moreover, in these regions core and cladding modes with a similar frequency and propagation constant co-exist \cite{couny_generation_2007}. The guidance mechanism in these and other \bhc{} (to be discussed below) has been the subject of significant study and debate, which we review in Section~\ref{broadband}.

\begin{figure*}
\centering
\includegraphics[width=\linewidth]{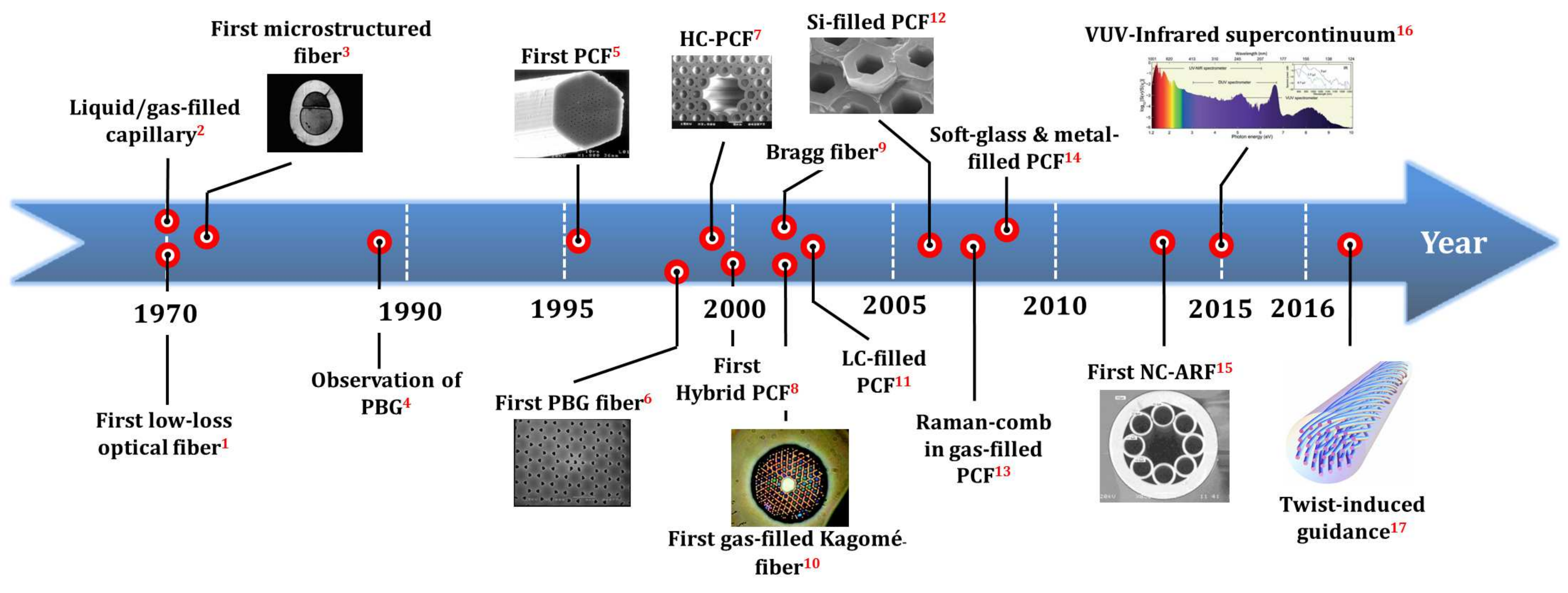}
\caption{Chronological overview of the most important breakthrough in the fields of PCF and hybrid PCF\footnote{\textsuperscript{1}\textcite{Kao1966a},\textsuperscript{2}\textcite{ippen_low-power_1970},\textsuperscript{3}\textcite{Kaiser1974},\textsuperscript{4}\textcite{John1987} and \textcite{Yablonovitch1987},\textsuperscript{5}\textcite{Birks1996},\textsuperscript{6}\textcite{knight1998},\textsuperscript{7}\textcite{cregan_single-mode_1999},\textsuperscript{8}\textcite{Westbrook2000},\textsuperscript{9}\textcite{Temelkuran2002},\textsuperscript{10}\textcite{benabid_stimulated_2002},\textsuperscript{11}\textcite{Larsen2003},\textsuperscript{12}\textcite{Sazio2006},\textsuperscript{13}\textcite{couny_generation_2007},\textsuperscript{14}\textcite{Schmidt2008,Schmidt2009}, \textsuperscript{15}\textcite{Pryamikov2011}, \textsuperscript{16}\textcite{belli_vacuum-ultraviolet_2015} and \textsuperscript{17}\textcite{beravat2016twist}.}.}
\label{fig:chrono_overview}
\end{figure*}

Several observations have been decisive in the recent progress in the fabrication of low-loss \bhc{s}. First of all it was realized that the exact geometry of the cladding is not the crucial factor in displaying broadband guidance in \hc{s}. Fibers with square or even honey comb cladding structures are also capable of broadband guidance \cite{Beaudou:08, Couny:08}. Second, in contrast to strict \pgpcf{,} the loss in broadband-guiding \hc{s} is independent of the number of the cladding layers surrounding the core, so in principle even a single cladding layer should also demonstrate similarly broadband guidance \cite{Pearce2007, Argyros:07}. Third, the geometry of the first layer surrounding the core is crucial in the performance of the fiber \cite{Argyros:07, Wang2011}. Wang \textit{et al.} showed that by optimizing the core shape of a \kpcf{} \hc{} and creating a negative (inward) curvature in the core surround it is possible to considerably enhance the performance of the fiber. \textcite{Wang2011} experimentally demonstrated that such a core shape has lower loss than the \kpcf{} \hc{} with conventional core geometry.

A breakthrough came about when in 2011 Prymamikov \textit{et al.} presented a truly single cladding layer \hc{} in which the core was formed by a single row of silica capillaries \cite{Pryamikov2011}. This type of fibre has become known as negative-curvature \hc{} (NC HC-PCF). The fiber described by \textcite{Pryamikov2011} transmitted light over a wide range of frequencies in the mid-IR spectral region with a relatively low loss. The low loss regions were intercepted by high-loss bands corresponding to the resonances of the cladding determined by the thickness of the capillary walls. This property is generally true for all the broadband-guiding \hc{s.} Nowadays, NC \hc{} has attracted significant attention from the scientific community as an alternative to \kpcf{} \hc{} with significantly reduced fabrication complexity, and often improved performance.

A note on the terminology is necessary when considering different types of broadband-guiding \hc{s} mentioned above, such as \kpcf{,} negative curvature \hc{,} or single-ring \hc{.}  In these fibers there is neither a unique cladding design nor a generally accepted guidance mechanism for the broadband guidance. These fibers may consist of a honey comb or a \kpcf-style lattice. Alternatively, in the case of the negative-curvature \hc{s} the cladding usually consists of a reduced, or even single layer, of annular tubes (see \cite{yu_negative_2016} and the references therein), although more complex geometries such as nested tubes also exist \cite{Belardi2014, Poletti2014, Habib:15}. All these structures are capable of showing a degree of broadband guidance when operating in the so-called large-pitch regime, where a characteristic length scale in the cladding of the fibre (e.g. pitch, or the diameter of the annular tubes in negative-curvature \hc{s}) is larger than the optical wavelength. In this paper, we use the generic term \bhc{} for all of these types of \hc{,} with the exception of hollow-core photonic bandgap fibers (\pgpcf{}). This is irrespective of the possible guidance mechanism and even though the cladding may not consist of a truly crystalline, i.e. periodic, structure. It has been common in the literature to also refer to some of these fibers as inhibited-coupling, or anti-resonant guiding, \hc{,} based on the perceived guidance mechanism. We avoid that in this review due to the still somewhat open nature of the guidance mechanism debate (see Section~\ref{broadband}).

Losses as low as 7.7 dB/km have been reported in \bhc{} in the near-infrared (NIR) region \cite{Debord:16, Debord:17}. Perhaps more surprising, is that such \bhc{} made with silica glass can even guide with low attenuation in regions where silica itself has very high material loss, such as the mid-infrared (MIR) \cite{Pryamikov2011,Yu2012,yu_spectral_2013, yu_negative_2016}, and even vacuum ultraviolet (VUV) \cite{belli_vacuum-ultraviolet_2015, ermolov_supercontinuum_2015}.

Recently it has also been found that various types of negative-curvature and single-ring \hc{} offer increased higher-order mode suppression \cite{Yu:16b,Uebel:16b,Newkirk:16,Hayes:17}. This is achieved by optimizing the phase-matching between the higher-order core modes and the modes of the cladding structure so as to achieve a higher differential modal loss. \textcite{Yu:16b} obtained a low attenuation of 0.025~dB/m at 1064~nm for the fundamental mode, but a 200 times higher attenuation for the next lowest loss higher-order mode. Such a differential was obtained across the visible and near-infrared region while maintaining record attenuation levels at each wavelength, proving that single-mode performance can be achieved while maintaining low loss. Twisting the fiber during fabrication can also be used to improve single-mode performance \cite{Edavalath:17}. The optimization of bending loss in simplified \bhc{s} has also been recently studied \cite{Gao:16,Gao:17}.

Figure~\ref{fig:chrono_overview} summarizes in a timescale illustration the major breakthroughs in the PCF field starting from the first report on low-loss optical fiber (\cite{Kao1966a}) until the recent demonstration of VUV-Infrared supercontinuum generation based on a gas-filled HC-PCF \cite{belli_vacuum-ultraviolet_2015} and twist-induced guidance in a coreless PCF \cite{beravat2016twist}.

\subsubsection{Hybrid PCF}
The unique ability to tailor and control light based on the geometrical characteristics of PCF, opened the door to the exploitation of new fundamental optical phenomena and applications \cite{Ranka2000}. However, one of the most important features of these fibers is perhaps their ability to act as a \textit {substrate} capable of hosting novel functional optical materials inside their  air-holes. Therefore, many researchers used both solid and \hc{s} as low-loss platforms with high aspect-ratio and long interaction lengths. By merging the fields of materials science and optics, a new branch of fiber optics research known as \textit{hybrid PCFs} is created.

The first hybrid PCF was reported by \textcite{Westbrook2000}. They demonstrated the ability to post-process and tune the cladding-mode resonances by simply varying the external temperature in a solid-core fiber with polymer-infused holes. The first hybrid gas-filled hollow-core PCF was reported by Benabid \textit{et al.} in 2002 demonstrating, for the first time, enhanced stimulated Raman scattering generated by ultra-low pulse energies \cite{benabid_stimulated_2002}. The latter attracted significant attention from the scientific community moving nonlinear optics in gases to remarkable and previously inaccessible parameter regimes of high intensity and extended interaction length \cite{russell_hollow-core_2014}. In a similar manner, atomic vapors were also used as active materials for the investigation of quantum phenomena inside \hc{s}  \cite{Ghosh2006,Venkataraman2011,Epple2014, Sprague2014, Kaczmarek2015}. Finally, the term \textit {multi-material} has been also used mainly to describe fiber structures with more than two integrated materials \cite {Abouraddy2007}.

In this review article we consider only silica PCFs as the host platform, with additional materials (solid, liquid or gas) incorporated to form the \textit{hybrid PCFs}.

%%%%%%%%%%%%%%%%%%%%%%%%%%%%%%%%%%%%%%%%%%%%%%%%%%%%%%%%%%%%%%%%%%%%%%%%%%%%%%%%%%%%%%%%%%%%%%%%%%%%%%%%%%%%%%%%%%%%%%%%%

\subsection{\label{B} Guiding mechanisms in PCFs}

\subsubsection{\label{Solid} Solid-core PCF - Modified total internal reflection}

\begin{figure}
\centering
\includegraphics[width=0.98\linewidth]{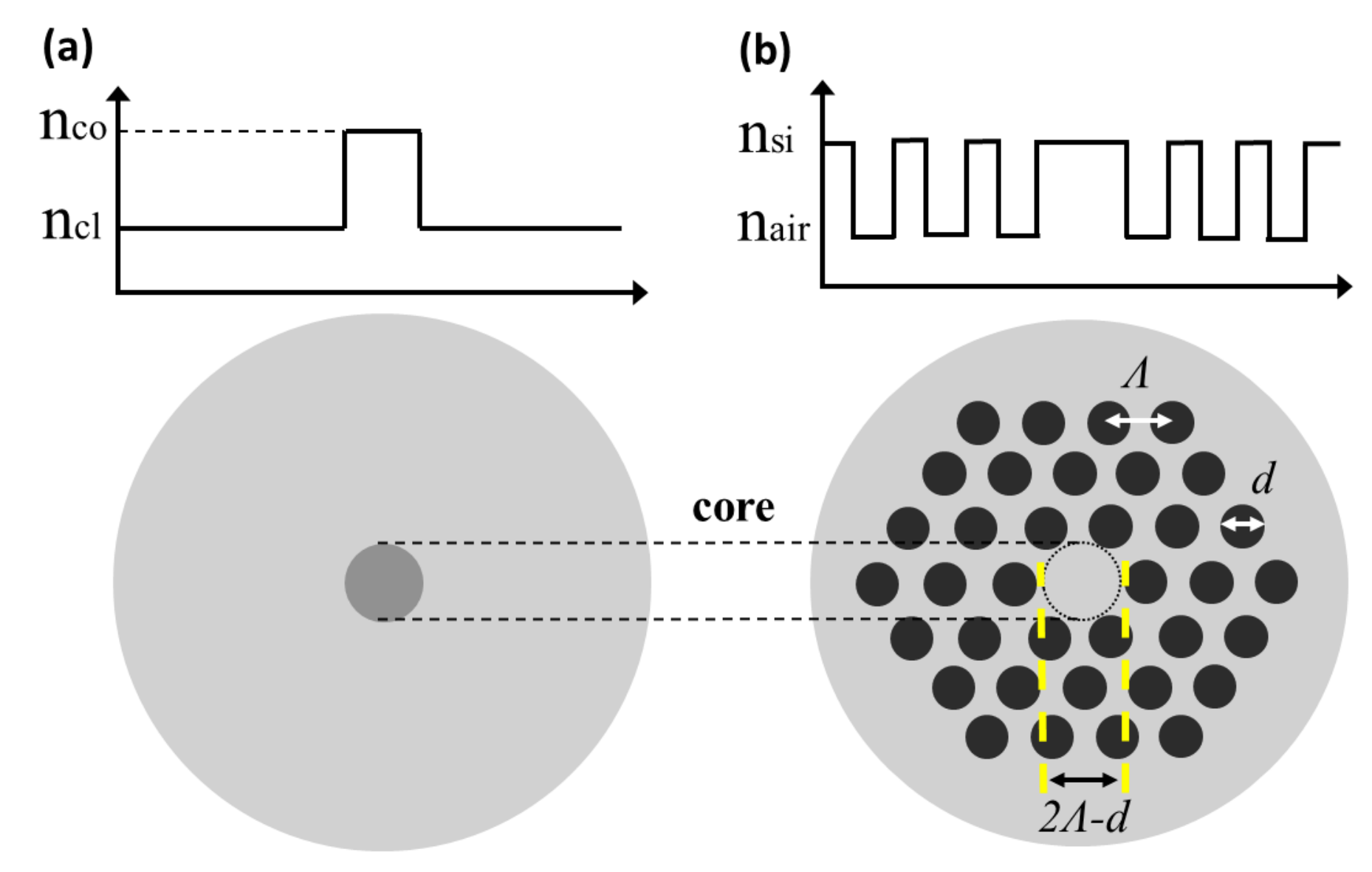}
\caption{Cross section and refractive index profile of (a) a conventional step-index optical fiber and (b) solid-core PCF. The grey and black region corresponds to silica material and air, respectively. The photonic crystal cladding is an ideal triangular lattice with parameters pitch $\Lambda$, i.e. the distance between air holes, and inner hole diameter $d$ which is related to the air-filling fraction. The core diameter is defined as $\sim$ $2\Lambda$.}
\label{fig:refra_profile}
\end{figure}

In standard optical fiber, light propagation purely relies on total internal reflection at the core-cladding interface. The maximum cladding mode index is limited by the cladding refractive index, which must be lower than the core index. To shape the refractive index profile in standard optical fibers various dopants such as GeO\textsubscript{2} or Al\textsubscript{2}O\textsubscript{3} are used in the preform fabrication stage. In solid-core PCFs, however, this is done by stacking capillaries around a solid rod in the preform fabrication process. The resulting air holes in the fiber's cladding, which run along its entire length, reduce the average refractive index of the cladding to be less than that of the fiber's core. Figure \ref{fig:refra_profile}(a) shows the refractive index profile of a standard optical fiber with $n\textsubscript{cl}$ indicating the refractive index of the silica \big (low-index material\big) and $n\textsubscript{cor}$ the refractive index of the doped core \big (high-index material\big). Figure \ref{fig:refra_profile} (b) shows the corresponding refractive index profile of a solid-core PCF. The photonic crystal cladding is an ideal triangular lattice with a lattice constant, or pitch, $\Lambda$, i.e. the inter-hole spacing, and an inner hole diameter $d$ which is related to the air-filling fraction $f=\frac{\pi}{2\sqrt{3}}\left (\frac{d}{\Lambda} \right )^2$. The guidance properties of the PCF are purely determined by $\Lambda$, and $d$.

\begin{figure}
\centering
\includegraphics[width=0.98\linewidth]{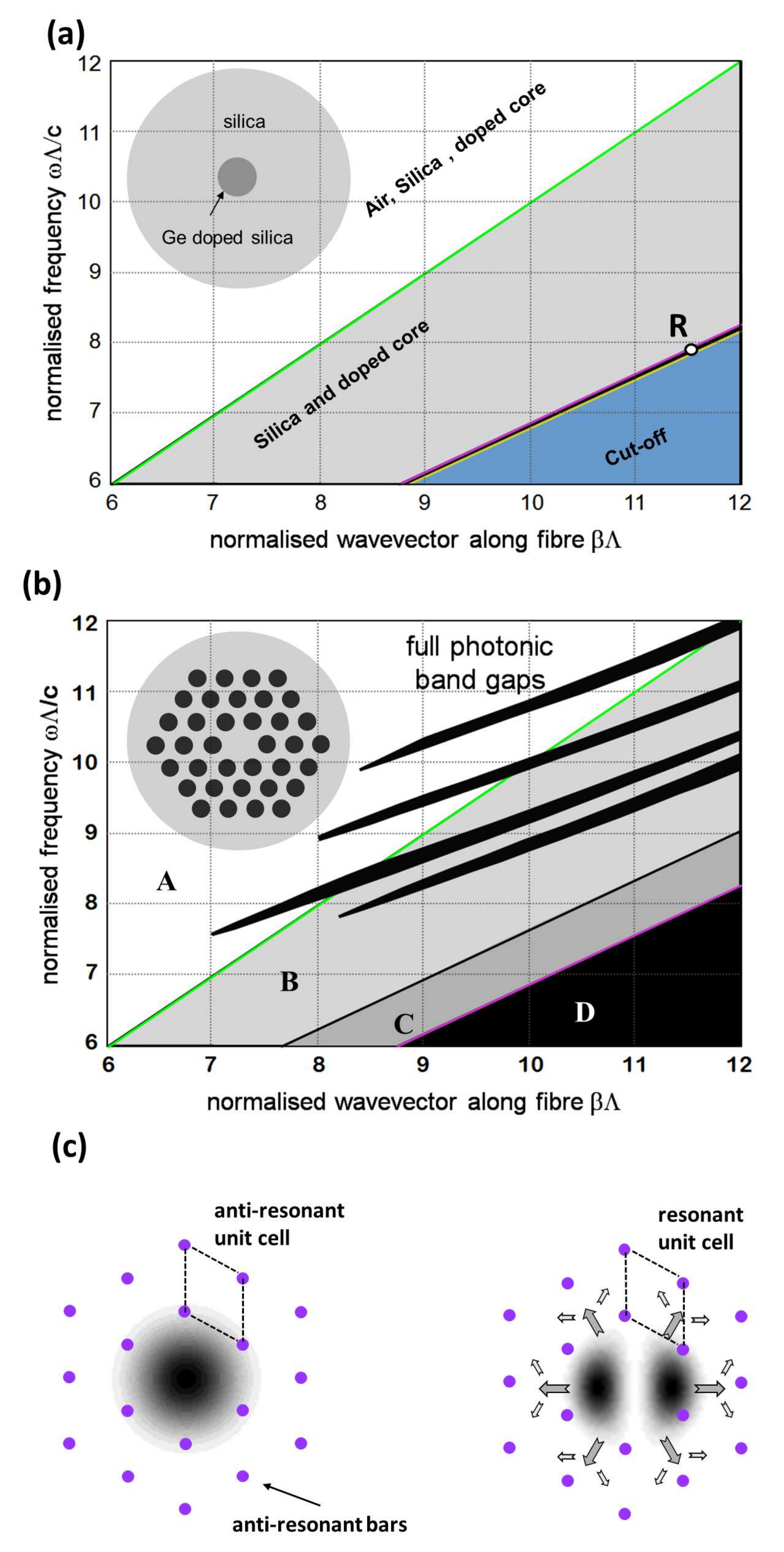}
\caption{(a) Propagation diagram of a conventional step-index fiber with a Ge-doped core and pure silica cladding. (b) Propagation diagram for a triangular hole lattice in silica glass with air-filling fraction 45\%. The green line in both plots corresponds to the air-line. (c) Schematic representation of how the air-holes act as bars keeping the fundamental mode confined (left) while the higher order modes are leaking out (right). Adapted from \textcite{Russell2003}}
\label{fig:step_guiding}
\end{figure}

An efficient way to represent the guidance mechanism in optical fibers is in the form of propagation diagram whose axes are the dimensionless quantities $\beta\Lambda$, where $\beta$ is the axial wave-vector, i.e. the component of the wave-vector along the propagation axis of the fiber, and $ \omega\Lambda/c$, where $\omega$ is the angular frequency and $c$ is the speed of light in vacuum \cite{Russell2003}. This diagram indicates the ranges of frequencies and axial wave vectors $\beta$, where light is evanescent, i.e. unable to propagate. At a fixed optical frequency (wavelength), the maximum possible value of $\beta$ is set by $\kappa n=\omega n/c$ where $n$ is the refractive index of the material under consideration \cite{Russell2003}. Solid lines show these values for each material. For instance the solid green line in Fig. \ref{fig:step_guiding} (a) is the light line (line separation between white and grey region) given by $\kappa=\omega/c$. For an axial wave vector in the  $\beta < \kappa n$, light is able to propagate in a medium of refractive index $n$ while for $\beta > \kappa n$ it is evanescent. In standard optical fibers, guided modes appear at points like \textit{R} lying in the dark red region (bold line) in Fig. \ref{fig:step_guiding} (a), i.e. where light is free to propagate in the high index core but evanescent in the cladding. 

In the case of PCF, the propagation diagram or "finger plot" is slightly different, given the fact that the cladding is now formed by a triangular photonic crystal lattice. Figure \ref{fig:step_guiding} (b) shows the band diagram for a PCF with air-filling fraction 45\% \cite{Russell2003}. The black area D in the bottom right corner of Fig. 2 (b) is the region where the light is evanescent. In area A, the light is free to propagate in every region of the fiber while in area B, light can propagate only in the photonic crystal and the silica glass. Propagation modes occur in area C where the light is confined in the silica material but is evanescent in the photonic crystal cladding due to the reduced average refractive index of the cladding. We note that the cladding's triangular lattice may also support 2-D photonic bandgaps. The slant, finger-shaped black regions show these bandgaps, that is, where for a given pitch and frequency no propagating mode with an axial wave-vector $\beta$ lying in the black region can be excited in the lattice. It should be emphasized that the photonic bandgaps extend into the region where $\beta<\kappa$, supporting the evidence that air guidance is possible without light propagating to the photonic crystal cladding \cite{Russell2003}. If the core of the fiber, whether hollow or not, supports a mode with a $(\beta,\omega)$ falling in a cladding bandgap, the mode will be confined in the core and does not leak out into the cladding as it cannot find any corresponding photonic state to couple to.

One of the most important properties of solid-core PCFs is their ability to be single mode for all $\beta<\kappa n$ if properly designed \cite{Birks1997, Knight:98}. This property can be understood by considering the cladding air holes as "anti-resonant bars" of the fundamental mode \cite{Russell2003}. The Gaussian-like single-lobe intensity distribution of the fundamental mode is "trapped" in the hexagonal array of the air holes of a fiber with a core diameter $\sim$ 2$\Lambda$. The intensity distribution of the higher order modes constitute smaller lobes which due to their intensity distribution are able to "escape" through the silica gaps between the air-holes \cite{Russell2003}. A schematic representation of the intensity distribution of the zero-order (fundamental) and the first order mode is shown in Fig.\ref{fig:step_guiding} (c). It should be noted that if the relative hole size $d/\Lambda >0.4$ then the core-cladding index contrast increases. In that case the higher-order modes are able to remain confined in the core and thus the fiber will support more than one mode.\footnote{A detailed theoretical study on the conditions for endlessly single mode operation can be found in \textcite{Bjarklev2003, Russell2006, Poli2007}.}

%%%%%%%%%%%%%%%%%%%%%%%%%%%%%%%%%%%%%%%%%%%%%%%%%%%%%%%%%%%%%%%%%%%%%%%%%%%%%%%%%%%%%%%%%%%%%%%%%%%%%%%%%%%%%%%%%%%%%%%%%

\subsubsection{\label{bandgap} \hc{} - the photonic bandgap mechanism}

\begin{figure}
\centering
\includegraphics[width=0.98\linewidth]{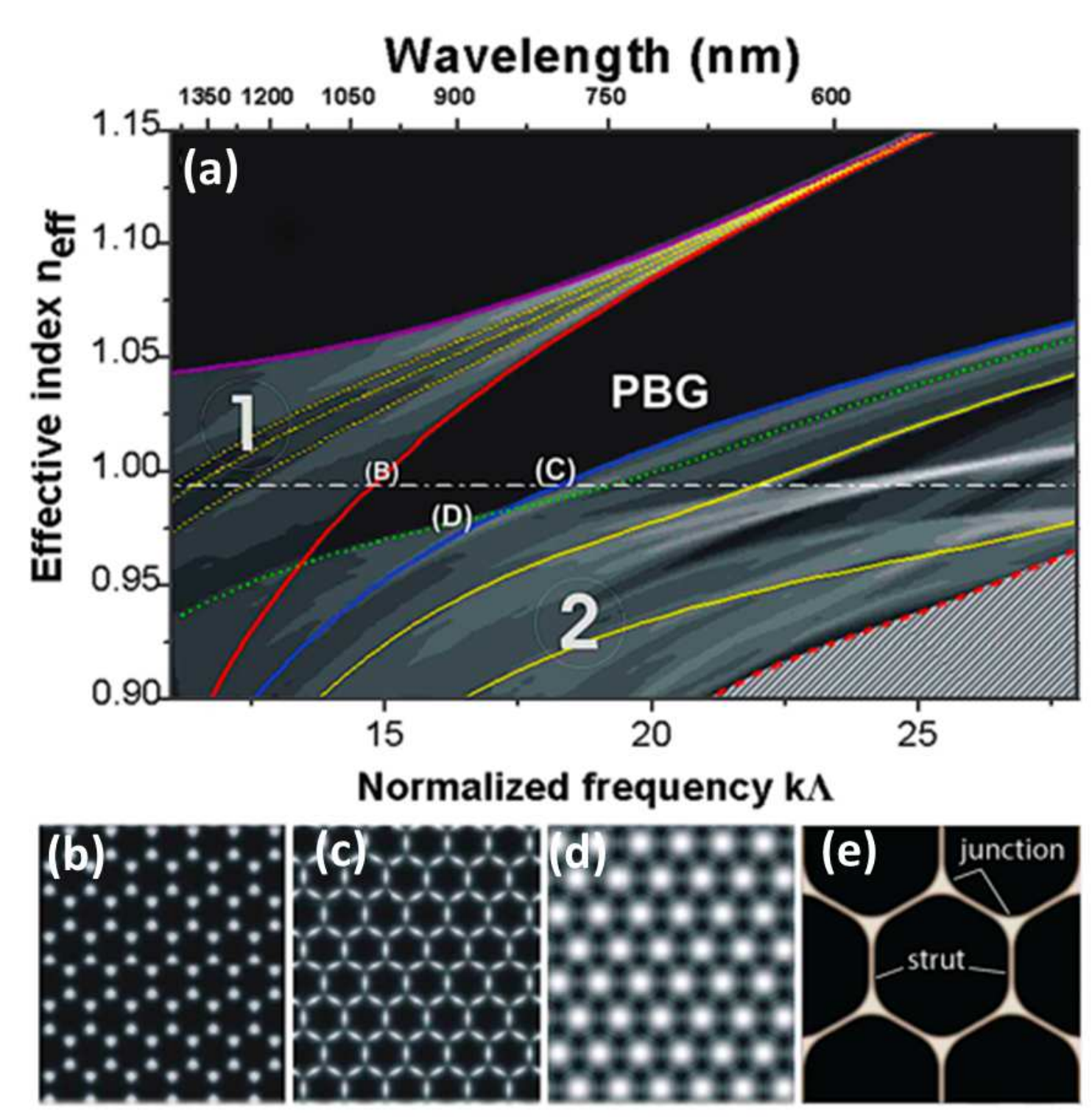}
\caption{(a) Numerically calculated density of states (DOS) plot for a triangular HC-PCF cladding lattice. Black represents zero DOS and white maximum DOS. The slanted coloured lines show several cladding modes trajectories. Modes trajectories for the cladding modes on the edges of the photonic bandgap are represented in red for the interstitial apexes mode (labeled by the letter b), in blue for the strut mode (labeled by the letter c), and in green for the hexagonal air holes mode (labeled by the letter d). The dash-dotted white line represents $n\textsubscript{eff}\sim0.995$. Panels (b), (c), and (d) show the near-filed pattern of the modes calculated for $n\textsubscript{eff}\sim0.995$ in (b) and (c) for the apex and strut modes, respectively, and for $n\textsubscript{eff}\sim0.973$ in (d) for the hexagonal air holes mode. Adapted from \textcite{Couny:07}. (e) Unit cell of the cladding structure of a honey comb triangular lattice of air holes indicating the struts and apexes.}
\label{fig:Bandgap_mechanism}
\end{figure}

When the hollow-core of an \hc{} is either evacuated or filled with air or other gas, its core index $n\textsubscript{cor}\sim 1$ is lower than the effective cladding index, and therefore guidance via modified TIR, as in solid-core PCFs, is not possible \footnote{Note that for photonic bandgap guidance the core does not necessarily need to be hollow. A solid-core PCF with cladding holes infiltrated with high-index material may also show  guidance based on a photonic bandgap.}. Confinement of light in the core of a \hc{} can instead be mediated by a photonic bandgap (black regions in Fig. \ref{fig:step_guiding} (b)), created by the surrounding two-dimensional photonic crystal cladding \cite{cregan_single-mode_1999,Russell2001,Knight2003a,Russell2003}. The guidance mechanism in such a hollow-core photonic bandgap fiber (\pgpcf{}) can be better understood by considering the unit cell of the photonic crystal cladding and the air-core as fixed localized resonators \cite{Birks1995}. The unit cell is defined as the smallest section of the entire cladding structure possessing the overall symmetry of the 2-D crystal geometry from which the entire crystal can be generated by its translation through all the primitive vectors in the 2-D Bravais lattice, see Fig. \ref{fig:Bandgap_mechanism} (e) for the unit cell of a honey comb lattice. For certain wavelengths, light is coupled to the unit cell resonators and the guided light is leaked out to the entire structure. Light remains confined in the core if the guided modes do not overlap or have common resonances with the crystal cladding. One way to determine the cladding resonances and consequently define the operating wavelengths at which light is confined in the air-core, is by calculating the Density of States (DOS) of the unit cell. The wall thickness of the glass and pitch of the unit cell exclusively govern the locations and widths of the photonic bandgaps \cite{Pottage2003}. Figure \ref{fig:Bandgap_mechanism} (a) shows as an example the calculated DOS for the honey comb lattice shown in Fig. \ref{fig:Bandgap_mechanism} (e) \cite{Couny:07}. The black region below the light line (the dotted line at $n\textsubscript{eff}=1$ in Fig. \ref{fig:Bandgap_mechanism} (a)), for effective refractive indices smaller than $n\textsubscript{eff}<n\textsubscript{cor}$, is where the core guided modes can exist. Outside the photonic bandgap, in regions 1 and 2 in Fig. \ref{fig:Bandgap_mechanism} (a) the cladding can support propagating modes. The slanted coloured lines in these regions show the dispersion of some of the cladding modes. From this plot, one can see that the upper and lower edges of the out-of-plane photonic bandgap are formed by three cladding modes or resonances indicated by the solid lines in red (b) and blue (d), and the green dotted line (c). Figures \ref{fig:Bandgap_mechanism} (b-d) show the near-field pattern of these cladding modes calculated at a representative normalized vacuum wavenumber $k_0\Lambda$. The plot indicates that a few unit cell resonator modes define the edges of the bandgap. The upper edge of the bandgap for small $\kappa \Lambda$ is determined by the resonances of the interstitial apex modes (Fig. \ref{fig:Bandgap_mechanism} (b)), while the lower edge for larger $\kappa\Lambda$ is determined by the resonance of the struts (Fig.\ref{fig:Bandgap_mechanism}(c)) and the resonances of the hexagonal array of air-holes (Fig. \ref{fig:Bandgap_mechanism}(d)). Detailed descriptions have been reported in \cite{Birks1995,Bjarklev2003,benabid_hollow-core_2006,Couny:07,Poli2007,benabid_linear_2011}.

In the  category of bandgap fibers we should also include the \textit{Bragg fiber}. It was first reported by \textcite{Yeh1978} while later it was further investigated theoretically by \textcite{Fink1999} and experimentally demonstrated by \textcite{Temelkuran2002}. The mechanism is different than the conventional honey comb cladding fibers. The hollow core in this case is surrounded by multiple alternating sub-micrometre-thick layers of high-refractive-index contrast materials such as  high-index glass and low-index polymer. These layers act as "perfect reflectors" confining the mode in the core of the fiber. The fundamental and high-order transmission windows are determined by the layer dimensions and can be scaled from visible to infrared wavelengths \cite{Temelkuran2002}.

%%%%%%%%%%%%%%%%%%%%%%%%%%%%%%%%%%%%%%%%%%%%%%%%%%%%%%%%%%%%%%%%%%%%%%%%%%%%%%%%%%%%%%%%%%%%%%%%%%%%%%%%%%%%%%%%%%%%%%%%%

\subsubsection{\label{broadband} Broadband-guiding \hc{s} - Inhibited coupling and anti-resonance}
As noted in the historical introduction (see Section \ref{History}), the guidance mechanism of \bhc{} is not settled. Currently there are two main proposed mechanisms: inhibited coupling (IC) and antiresonant guiding. Although a recent review by Yu and Knight describes in detail several other models that can be also used to describe the guidance in NC \hc{s} \cite{yu_negative_2016}.

\paragraph{Inhibited coupling (IC).}
\begin{figure}
\centering
\includegraphics[width=0.98\linewidth]{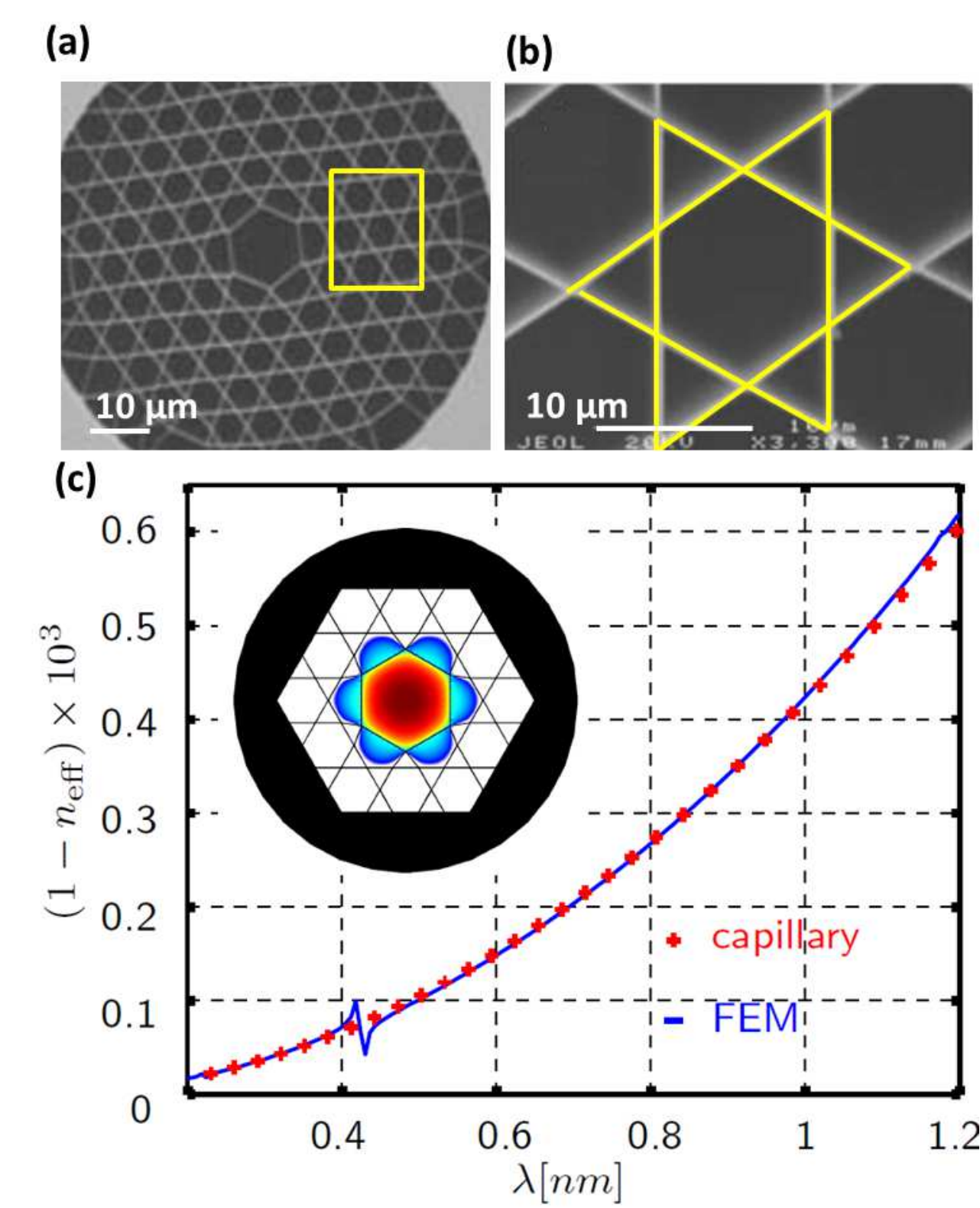}
\caption{(a) Scanning electron micrograph of a \kpcf-style HC-PCF. (b) Star-of-David pattern (light grey). The highlighted region represents the unit cell. (c) Calculated effective indices for one ring Kagom\'{e} structure in comparison with a capillary model. The inset shows the HE11 mode at $800$~nm with the structure superimposed.Adapted from \textcite{Nold2011, Benabid2009}.}
\label{fig:Kagome_capillary}
\end{figure}
In a \pgpcf{} with a honeycomb lattice cladding, with suitable parameters, e.g. pitch and air-filling fraction, light introduced into the core is unable to escape from it thanks to the bandgap that prohibits the propagation of light in the cladding region. The different cladding design introduced by \textcite{benabid_stimulated_2002}, called the \kpcf-lattice, consists of thin (nanometer scale) silica webs arranged as shown in Figs. \ref{fig:Kagome_capillary} (a-b). If properly designed, a \kpcf{} \hc{} is capable of guiding light over a significantly broader frequency window than \pgpcf{.} 
It is known that the cladding of \kpcf{} \hc{} does not have a full photonic bandgap, but rather a low DOS \cite{benabid_hollow-core_2006,couny_generation_2007,Argyros2008, Argyros:07}, and that the core and the cladding resonances do not spatially overlap for most frequencies. This is despite the co-existence of the core and the cladding modes with a similar frequency and propagation constant \cite{couny_generation_2007}. These characteristics fully differentiate a \kpcf{} \hc{} from a \pgpcf{.} A wide number of different theoretical approaches to explain the underlying guidance mechanism of \kpcf{} \hc{} have been reported \cite{benabid_hollow-core_2006, Couny2006, couny_generation_2007,Pearce2007,Fevrier2010,benabid_linear_2011, Debord:17}. \textcite{couny_generation_2007} first proposed that the high degree of transverse field mismatch between the core and the cladding modes practically inhibits the coupling between them and this is known as inhibited coupling (IC).  In the IC picture the guided core modes only weakly interact with the cladding modes due to the rapid phase modulation of the latter. Analogy has been drawn to solid-state physics by considering IC as the photonics analogue of Von Neumann-Wigner bound states (i.e. core modes) of the Schr\"{o}dinger equation within a continuum (i.e. cladding modes) \cite{VonNeumann1993,couny_generation_2007}.

\paragraph{Antiresonance guiding.}
\begin{figure}
\centering
\includegraphics[width=0.98\linewidth]{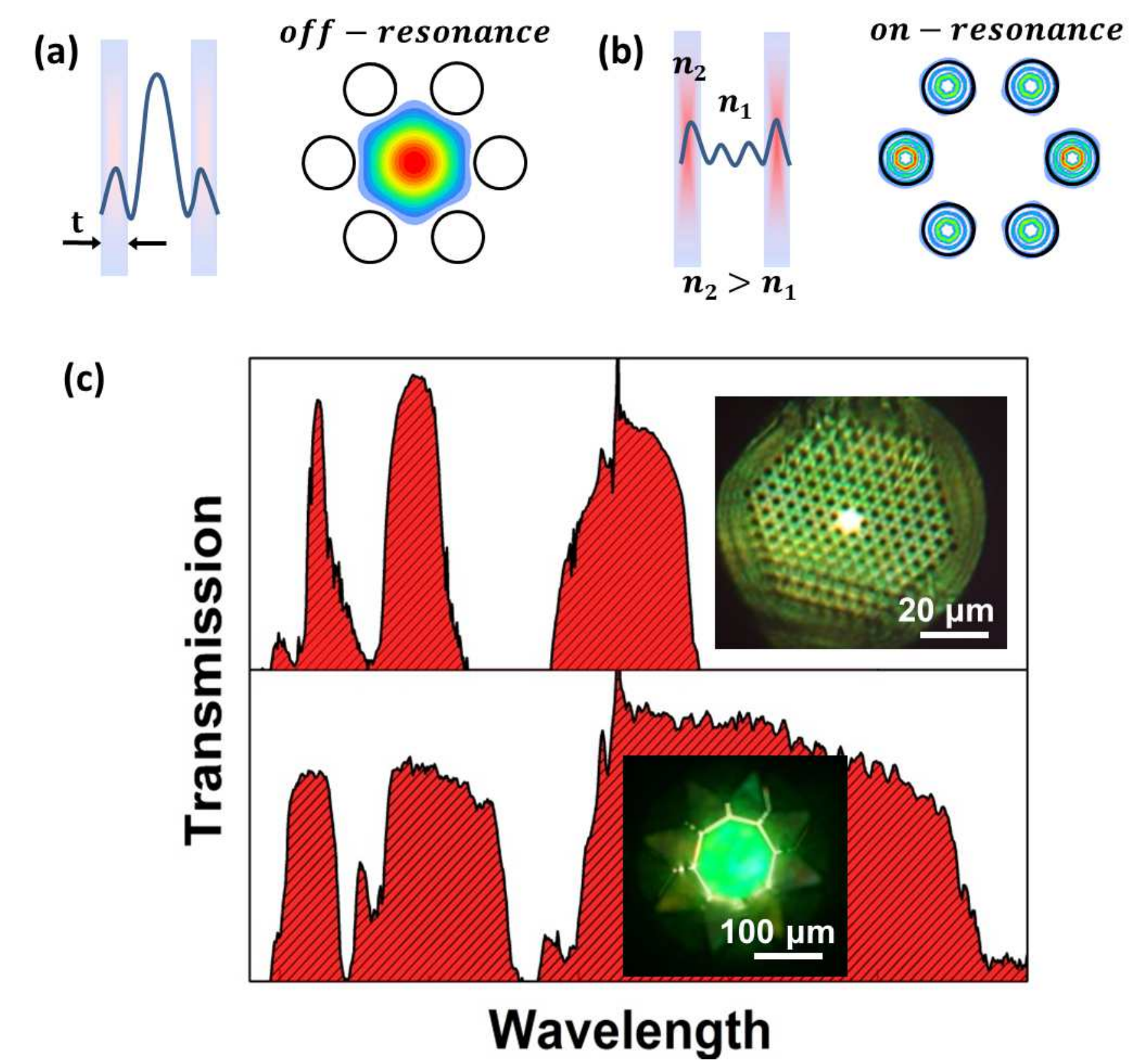}
\caption{(a),(b) Principle of ARROW guidance when light is in antiresonance or off-resonance state light remains confined in the core(top). At resonance state (bottom), light couples to the high-index cylinders, and thus escapes from the core introducing a dip in the transmission spectra. Example of the transmission spectrum of a high-index infiltrated solid-core PCF (top) and antiresonant hollow-core fiber (bottom). Inset images: White-light near-field profiles of the propagating light in the core of the fibers.}
\label{fig:antiresonant}
\end{figure}
The broadband guidance windows in a \kpcf{} \hc{}, and also the more general class of \bhc{s} (such as negative curvature, single-ring \hc{}) are intercepted by narrower high-loss frequency bands. It has been shown that the position of these loss windows correspond to the resonances of the thin cladding glass struts and can be found using the following equation: 
\begin{equation}
\lambda_r=\frac{2t}{m} \sqrt {n^2-1}
\end{equation}
where $n$ is the refractive index of the glass, $t$ the thickness of the surrounding struts, and $m$ is an integer which corresponds to the number of the resonance \cite{Litchinitser2002, Pearce2007}. Eq. 1 refers also to the antiresonant reflection optical waveguide (ARROW) model~\cite{Duguay1986}. This has led researchers to consider the ARROW model as the the underlying guidance mechanism in broadband-guiding \hc{s}, and this is perhaps the most widely accepted model. The ARROW model was first reported back in 1986 describing how the light is confined in a planar silicon waveguide having as cladding a number of low and high index layers \cite{Duguay1986}. In 2002, Litchister \textit{et al.} explained that the same principle can be used to describe the light transmission not only in planar multi-layer structures but also in solid-core PCF with high index inclusions as well as photonic bandgap fibers. It approximates the cladding of the fiber as an array of high and low refractive index regions \cite{Litchinitser2002}. Each higher refractive index layer can be considered as a Fabry--P\'{e}rot resonator as shown in Fig. \ref{fig:antiresonant} (a) and (b). In the on-resonance state, the F-P cavity is transparent, allowing the light to escape from the structure, while the reflectivity can be very high in the off-resonance state thereby strongly confining the light in the core of the fiber. Figure \ref{fig:antiresonant} (c) shows the ARROW transmission spectrum in a solid-core PCF infiltrated with high index material as well as the transmission spectrum of an HC-PCF fiber. In both cases there are discrete resonances for which the ARROW model could be used to predict the exact location of the band edge. It should be mentioned that ARROW-based guidance in solid-core PCFs with high-index inclusions was extensively investigated prior the appearance of \hc{s} \cite{Litchinitser2002, Litchinitser2004, Kuhlmey2009}.

We can describe the guidance in many hybrid PCFs by incorporating the refractive index of the host and inclusion material into Eq. 1,
\begin{equation}
\lambda_r'=\frac{2n_1d}{m} \sqrt{\left(\frac{n_2}{n_1}\right)^2-1},
\end{equation}
where $n_1$ is the refractive index of the low-index host material of the solid-core PCF, $n_2$ the refractive index of the high-index inclusions. However, this model is not valid for long wavelengths satisfying $\lambda/t>2\sqrt{n_2^2-n_1^2}$ \cite{Litchinitser2004}.

\paragraph{Dispersion.}
Practically, the wide transmission windows in \bhc{s} when compared to \pgpcf{s}, along with propagation losses approaching that of \pgpcf{}---currently at 7.7 dB/km \cite{Roberts2005,Wang2011,Debord:17}---and low chromatic dispersion, have made them ideal vessels for high-field ultrafast nonlinear optics \cite{travers_ultrafast_2011,russell_hollow-core_2014}. The dispersion in these fibers can be approximated to a good degree by Marcatili and Schmeltzer's model  \cite{Marcatili1964}. Figure \ref{fig:Kagome_capillary} (c) shows the calculated effective indices (solid blue line) for the simplified \kpcf{} \hc{} shown as inset. The red plus signs show the dispersion of a hollow dielectric capillary with the same geometrical characteristics as the \kpcf{} fiber calculated using Marcatili and Schmeltzer's model. See Section~\ref{sec:gas_disp} for further details.

%%%%%%%%%%%%%%%%%%%%%%%%%%%%%%%%%%%%%%%%%%%%%%%%%%%%%
\subsubsection{\label{twist} Twist-induced guidance}
\begin{figure}
\centering
\includegraphics[width=0.98\linewidth]{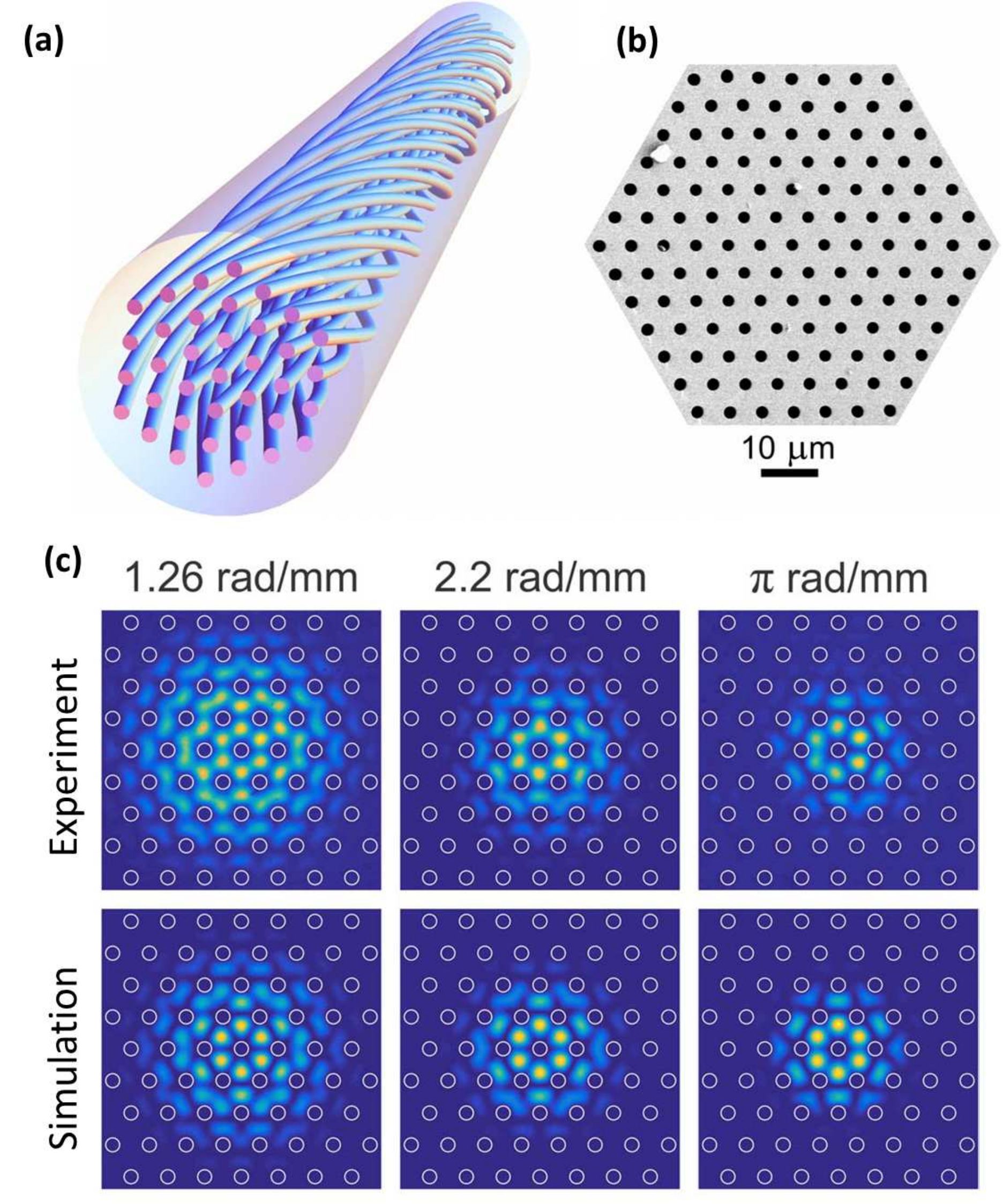}
\caption{(a) Illustration of a twisted coreless PCF (b) Scanning electron micrograph of the coreless microstructured PCF. (c) Experimental and simulated Poynting vector distributions at $818$~nm at different twist rates. Adapted from \textcite{beravat2016twist}}.
\label{fig:twist}
\end{figure}

\textcite{beravat2016twist} recently reported  a new guidance mechanism revealed in a coreless PCF that has been twisted around its z-axis. Interestingly, the authors demonstrated for the first time that the twisted PCF creates a "helical channel" in which the light can be robustly trapped in the center of the microstructure without the need of any discernible core structure. The authors explain the guiding properties of the twisted coreless PCF using Hamiltonian optics analysis and they show that the light follows closed oscillatory paths within its spiral structure. \textcite{beravat2016twist} showed that the confinement of these unusual guided modes can be controlled by changing the twist rate enabling thus the use of such guiding mechanism in sensing, high-power delivery and nonlinear optics. Figure \ref{fig:twist} (a) and (b) show the schematic representation of the twisted coreless PCF and the SEM image of the fabricated microstructured fiber, respectively. Figure \ref{fig:twist} (c) demonstrates the measured and calculated Poynting vector distributions at different twist rates at a fixed wavelength ($818$~nm).

\subsection{\label{summary} Summary}

In this section, we have provided a brief overview of the different existing guiding mechanisms in conventional solid-core, hollow-core and coreless PCFs. Summarizing, the light confinement can be categorized into five main mechanisms: \textit {i)} modified total internal reflection in which light guidance is similar to the conventional step-index fibers, \textit {ii)} photonic bandgap, \textit {iii)} inhibited coupling/anti-resonance guidance , and the most recent \textit {iv)} twist-induced guidance which briefly introduced in this article. From a physics point of view, the first three guidance mechanisms can also adequately describe the guidance in hybrid PCFs that we will review later in this article. 
%%%%%%%%%%%%%%%%%%%%%%%%%%%%%%%%%%%%%%%%%%%%%%%%%%%%%%%%%%%%%%%%%%%%%%%%%%%%%%%%%%%%%%%%%%%%%%%%%%%%%%%%%%%%%%%%%%%%%%%%%

\section{\label{2}Hybrid solid-core PCF}

In this section, we focus and highlight the main scientific achievements with solid-core PCFs infiltrated with liquid crystals (LCs), metals and other high-index materials. It should be emphasized that we focus on post-processed silica PCFs and we present how the use of different high-index functional materials inside the PCF allows a step towards tunable and sensing devices, plasmonics as well as enhanced nonlinear interactions \footnote{For a recent partial review on hybrid fibers based on solid materials see \textcite{Schmidt2016}.}. In this review, reports based on silicate or oxide glass-filled PCFs have not been considered.  

\subsection{\label{Linear} Tunable linear devices }

The possibility to functionalize silica PCFs with active fluids has paved the way for the development of unique all-fiber devices. The work from Westbrook \textit{et al.} \cite{Westbrook2000} was followed up by \textcite{Bise2002} a couple of years later demonstrating for the first time the possibility of transforming a high-index core PCF into a solid-core photonic bandgap fiber (PBG) by infiltrating a high-index liquid into the air-holes \cite{Bise2002}. By varying the external temperature, the refractive index of the infused polymer was changing giving the possibility to control the location of the bandgaps. This report constituted the kick-off for many other research groups around the world to investigate various different fluids with enhanced thermo-electrical properties as a route to the development of sensors, switches, Gaussian and tunable bandpass filters, fiber probes for biological applications, etc.\footnote{See e.g.\cite{Kuhlmey2009,mach2002tunable,kerbage2002tunable,domachuk2004transverse, Markos2010a,Markos2012b,markos2013broadband,markos2011guiding,ma7085735,vaiano2016lab} and early reviews \cite{kerbage2004manipulating,eggleton2005laboratory}.} In this section we review and focus on the main materials that have been used by the scientific community to develop the aforementioned silica fiber devices. 

\subsubsection{\label{LC}Liquid crystals-filled PCFs}
Liquid crystals have properties between those of conventional liquids and those of solid crystals. For instance, a LC shows fluidity like a liquid, but it also demonstrates optical anisotropy like a crystal. Research on LCs has been carried out for many years as they find applications in computer screens, mobile phones as well as tunable focusing elements, beam steerers, polarization control materials, etc. \cite{Andrienko2006} The molecules of LCs are directionally oriented and disordered in position. The structure of a LC involves rigid $\pi$-electron systems bearing flexible long alkyl chains. Many LC molecules are calamitic shaped with a group for polarization, but planar molecules are also known. Different external perturbations such as temperature, voltage, strain can directly affect the LC phase by modifying the length of alkyl chain. A practical liquid crystal has a mesophase at room temperature. In addition to an application for a liquid crystal display, LC materials are expected to be organic semiconductors \cite{Iino2015}. A semiconductor having a LC phase, the so-called LC semiconductor, spontaneously undergoes a molecular orientation and self-assembly. There are mainly nematic, smectic, cholesteric, and discotic (disc-like) phases \cite{Andrienko2006}. The shape of LC is rod-like while its size is a few nanometers of a LC molecule is in the order of a few nanometers. In 2003, \textcite{Larsen2003}, demonstrated for the first time how the combination of LCs with solid-core PCFs could lead to unique optical devices. This work was followed by several other different groups who thoroughly investigated similar effects taking advantage of the unique thermo-electro-optical properties of LCs \cite{Du2004,Zografopoulos2006a, Brzdakiewicz2006, Zografopoulos2006}.

\paragraph{\label{Thermal}Thermal tunability} \hfill \break

Temperature has a crucial effect on the LCs as it directly affects the ordinary ($n_0$) and the extraordinary ($n_i$) refractive index of the material. This effect can be used to control the optical properties of the hybrid LC-filled PCF and can be exploited for the development of thermally controlled tunable devices. The birefringence $\Delta n$ (or refractive index contrast) of the LCs is defined as the difference between the $n_0$ and $n_i$ and is represented by the average refractive index $\langle n \rangle$ which is defined as \cite{Yang2014}:

\begin{equation}
\langle n \rangle=\frac{n_i+2n_0}{3} 
\end{equation}

Redefining the $n_0$ and $n_i$ we have:

\begin{equation}
n_0=\langle n\rangle - \frac{1}{3}\Delta n 
\end{equation}
\begin{equation}
n_i=\langle n\rangle + \frac{2}{3}\Delta n 
\end{equation}

For temperatures which are not too close to the critical or clearing point (defined as the temperature at which LCs become fully isotropic), the Haller \cite{Haller1975} approximation can be used to describe the temperature dependence $\Delta n$ as: 

\begin{equation}
\Delta n(T)\approx (\Delta n)_0 \Big(1-{\frac{T}{T_C}}\Big)^\beta
\end{equation}

$(\Delta n)_0$ is defined as the birefringence of the LC in the crystalline state ($T=0$ K), $T_C$ is the temperature at the clearing  point, and $\beta$ is the material constant (varying based on the LC). The final temperature dependence is linear and with respect to wavelength ($\lambda$) and can be described as \cite{Li2004a}:
\begin{equation}
\langle n\rangle = A-BT 
\end{equation}

Combining Eqs. 6 and 7 with the Eqs. 4 and 5, we can derive a four-parameters model that describes to a good approximation, the temperature dependence of ($n_0$) and ($n_i$) \cite{Li2004a}: 

\begin{equation}
n_0(T,\lambda)\approx A-BT- \frac{(\Delta n)_0}{3}\Big(1-{\frac{T}{T_C}}\Big)^\beta
\end{equation}
\begin{equation}
n_i(T,\lambda)\approx A+BT- \frac{2(\Delta n)_0}{3}\Big(1-{\frac{T}{T_C}}\Big)^\beta
\end{equation}

The parameters \textit{A} and \textit{B} can be extracted via a 2\textsuperscript {nd} order fitting of the measured refractive index of the LC. Equations 8 and 9 predict the temperature dependence of the $n_0$ , $n_i$ which are important parameters for the performance of the LC-filled PCF and they are directly related to the molecular properties of the LC used \cite{Li2004a}. 

Infiltration of LCs in a solid-core PCFs is a relatively simple procedure as only a few centimeters-long infiltrated length is required to thermally control the output light from the fiber. The simplest way of filling the PCF with LC is by immersing the end-facet in a reservoir filled with the desired LC while the other end of the fiber is open to atmospheric pressure. Due to the surface affinity of the LC with the fiber’s material, a few minutes is required for the infiltration of a few centimeters of the fiber. However, it should be mentioned that the infiltration highly depends on the diameter of the PCF's cladding holes as well as the viscosity of the LC \cite{nielsen_selective_2005}. To expedite the filling procedure a pressure system can be used to further assist the LC infiltration or the temperature can be increased which decreases eventually the viscosity of the LC. Several reports in the literature have extensively investigated the filling times and conditions for fluid-infiltration of silica PCFs with varying cladding hole sizes \cite{nielsen_selective_2005, Kuhlmey2009}.

\begin{figure}
\centering
\includegraphics[width=0.98\linewidth]{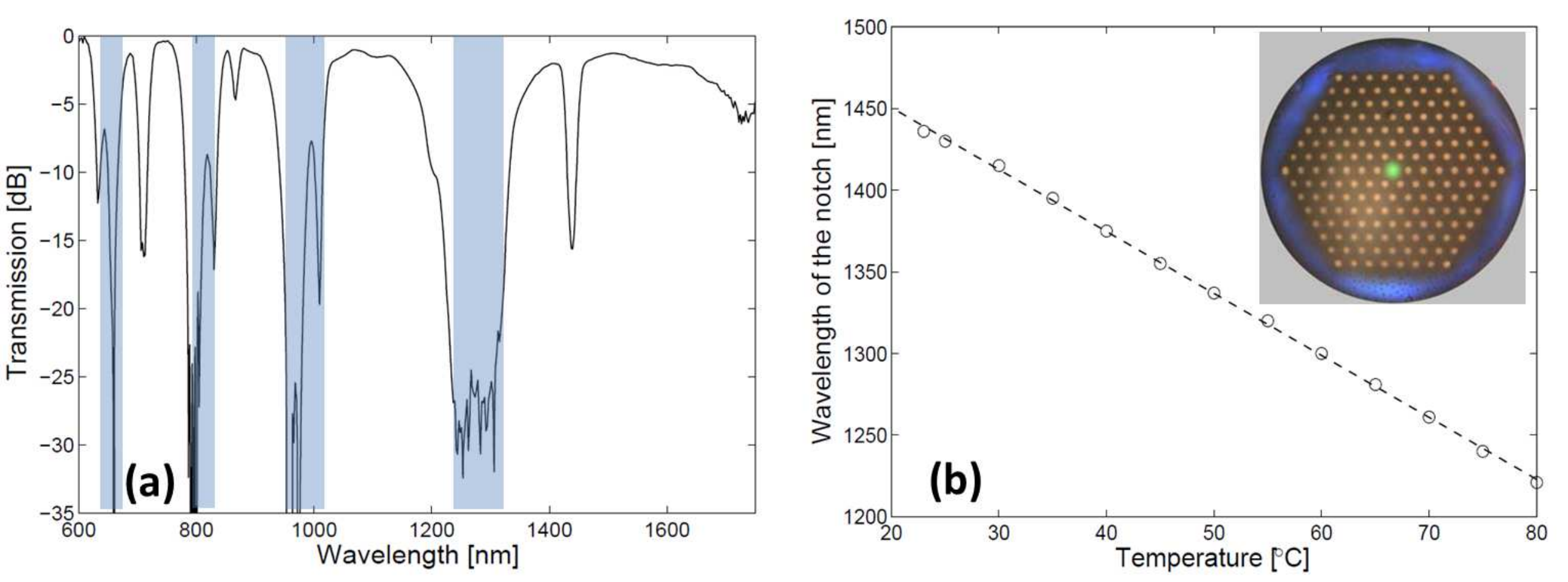}
\caption{(a) Transmission spectrum of solid-core PCF filled with LCs (highlighted areas indicate the on-resonance state in which the light is coupled to the high index rods). (b) Thermal tuning of the fiber filled with LCs from 20$^{\circ}$C up to 80$^{\circ}$C. The plot shows the spectral position of the notch in the long $1300-1600$~nm wavelength edge. Inset: near-field profile of the LC-PCF. Adapted from \textcite{Scolari2009}.}
\label{fig:LC_tuning}
\end{figure}

The most crucial factor which defines the effective tunability of the photonic bands in such hybrid structure is the thermo-optic coefficient of the infused LCs ($\partial n/\partial T$) \cite{Wei2009}. For silica as host material, the thermo-optic coefficient is very low ($\sim 10^{-6}$ K${}^{-1}$) \cite{Malitson1965} when compared to other materials. If the material of the host PCF is sensitive to thermal variations, such as polymer PCFs, then a more complex methodology is required to discriminate the effects arising from the infused and the host material. LCs exhibit one of the highest thermo-optic coefficients up to $\partial n/\partial T \approx 10^{-2}$ K${}^{-1}$ \cite{Li2004b}, due to their high birefringence ($\partial n \approx 0.8$) among other liquids\cite{Li2004b}. Figure \ref{fig:LC_tuning} (a) shows an example of the revealed high extinction-ratio bandgaps in a commercially available PCF (LMA-13 by NKT Photonics) filled with LCs (MDA-00-3969).

The presence of the high extinction bandgaps as shown in Fig. \ref{fig:LC_tuning}(a) suggests that the fiber can be used as an all-fiber spectral filter having narrow and wide bandwidth windows. Figure \ref{fig:LC_tuning} (b) shows the linear response of the long edge of the bandgap shifting towards shorter wavelengths as the temperature increases. The sensitivity of the fiber was found to be -3.6 nm/ $^{\circ}$C in the infrared. The linearity of the tunability response is attributed to the fact that the refractive index of the LC decreases linearly with respect to the temperature. 

In 2006, \textcite{Alkeskjold2006} reported the development of a hybrid LC-filled PCF with significantly enhanced thermal tunability. This was achieved by using a solid-core PCF with large core diameter of 25 $\mathrm{\mu m}$ and effective modal area of 440 $\mathrm{\mu m}^2$ filled with a specially designed and synthesized nematic LC which has high birefringence and low clearing temperature \cite{Alkeskjold2006, Li2004b}. The thermal sensitivity of the spectral position of the bandgap was measured to 27 nm/$^{\circ}$C, which is 4.6 times higher than what the commercially available LCs offer and to the best of our knowledge it holds the record temperature sensitivity in LC-filled PCFs until today \cite {Alkeskjold2006}. Several different configurations are demonstrated over the past 10 years employing LCs for thermally-tuned PCFs. For example, \textcite{Saitoh2008} proposed a hybrid waveguide structure using a multi-core PCF filled with nematic LCs. The design involves two identical cores separated by a third one which acts as a LC resonator. Based on the design parameters associated with the LC core, phase matching at a single wavelength can be achieved, enabling thermo-tunable narrow-band resonant directional coupling between the input and the output cores \cite{Saitoh2008}.

\paragraph{\label{electric}Electric tunability} \hfill \break

The most important property of LCs is perhaps the possibility of changing the orientation of the LC molecules electrically. This provides a much wider dynamic tunability range of the refractive index than using optical or thermal means of tuning \cite {Alkeskjold2007}. The response of the LC molecules to an externally applied field is due to the different values of dielectric constants along the long molecular axis ($\varepsilon_{||}$) and any axis perpendicular to it ($\varepsilon_{\perp}$). The dielectric anisotropy $\Delta\varepsilon$, defined as $\varepsilon_{||}-\varepsilon_{\perp}$ is related to the dipole moment $\mu$ and distribution $\theta$ {defined as the angle between the molecular axis and the director} with respect to the long molecular axis as well as the applied temperature and frequency \cite{Scolari2009}. The most common LCs are the nematic with positive $\Delta\varepsilon$ which are mainly used in Liquid Crystal Display (LCD) \cite{Demus1995}. Similarly, LCs with negative dielectric anisotropy ($\Delta\varepsilon$) are used in glass cells with a vertical alignment configuration. In order to exhibit a negative $\Delta\varepsilon$, molecules must have their dipole $\mu$ perpendicular to the principal molecular axis \cite{Scolari2009}. A combination or mixture of two different LCs with opposite dielectric anisotropy leads to a new class of LCs known as dual-frequency LCs. The dual-frequency LCs exhibit a positive $\Delta\varepsilon$ at low frequencies and negative at high frequencies as they are made of both positive and negative $\Delta\varepsilon$ compounds. A major advantage in using these LCs is that, despite their high rotational viscosity, they exhibit a fast response time to an applied electric field. The threshold voltage, $V_{th}$ at which the LC starts to respond to the external applied voltage is directly related to the dielectric anisotropy and can be described as \cite{Lagerwall2003}: 

\begin{equation}
V_{th}=\pi \sqrt{\frac{K_{11}}{\Delta\varepsilon}}
\end{equation}
where $K_{11}$ is the splay elastic constant (describing the LC molecules as pure splay deformation). Therefore, as the dielectric anisotropy, $\Delta\varepsilon$ increases, the LC responds to lower applied voltages. It should be emphasized that other factors such as the rise and decay times, elastic constants, Frederick's transition of the LCs also play a role in the final performance of the LC-based devices \cite{Lagerwall2003, Onuki2004}.

A PCF with a cladding with six-fold symmetry is not polarization maintaining. By applying an external electric field with a preferential direction in space on a LC-filled solid-core PCF, it is possible to break the six-fold symmetry of the fiber, introducing polarization dependence of the transmission. The ability to control the polarization state in a hexagonal PCF is not possible using thermal and optical methods. In 2004, Du \textit{et al.} reported for the first time the development of tunable light switch using a photonic crystal fiber filled with nematic LC \cite{Du2004}. The electrical tunability switching in the hybrid PCF was over $30\ \mathrm{dB}$ attenuation at $60\ \mathrm{V_{rms}}$ operating at $633$~nm \cite{Du2004}. 

\begin{figure}
\centering
\includegraphics[width=0.98\linewidth]{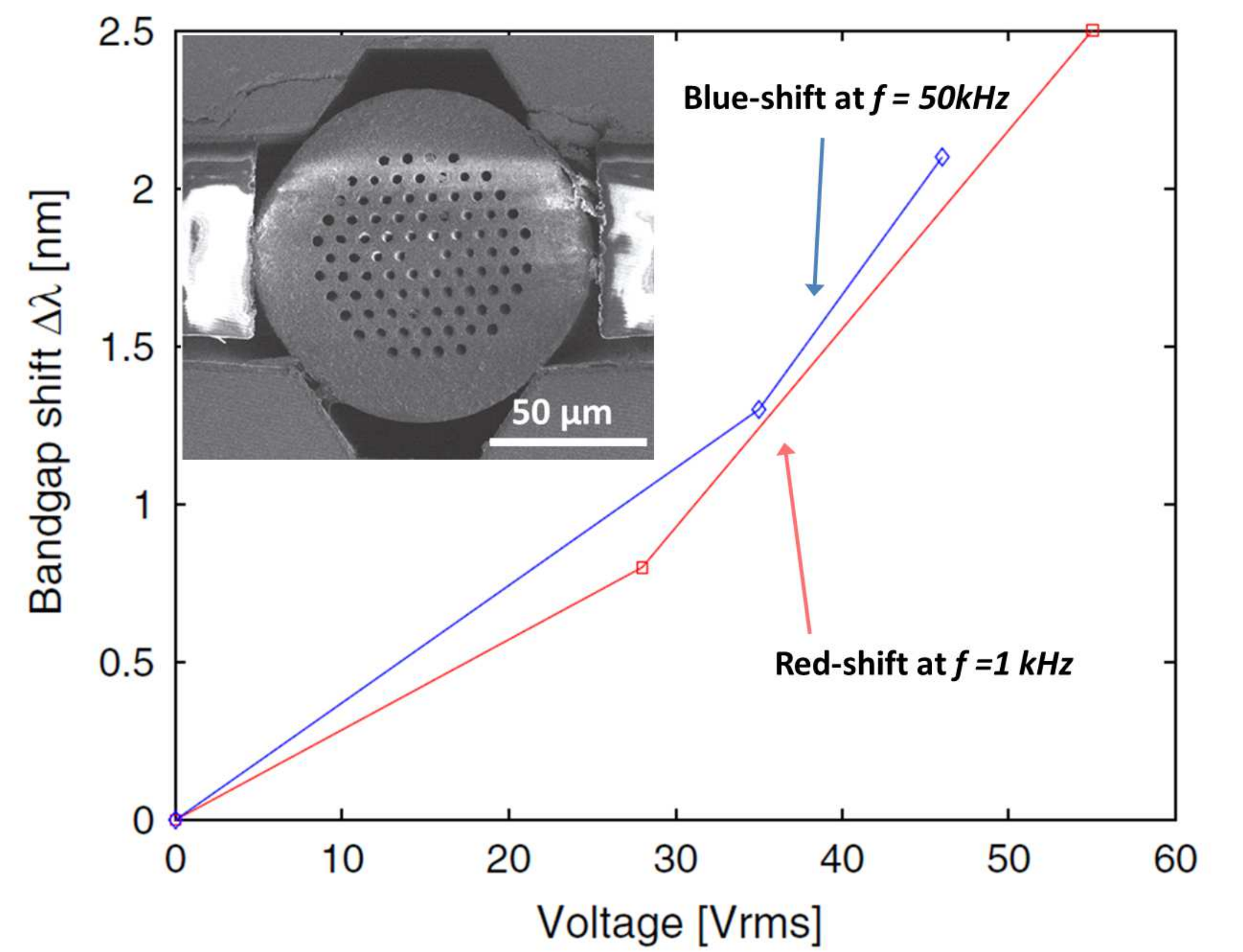}
\caption{Blue- and red-edge bandgap tunability with respect the applied voltage. Inset: Integrated LC-filled PCF sandwiched between the two electrodes used to drive the device at $1600$~nm wavelength edge. Adapted from \textcite{Alkeskjold2007} and inset image adapted from \textcite{Li2016}.}
\label{fig:shift_LC}
\end{figure}

A year later, \textcite{Haakestad2005} investigated how the electrically controlled  polarization-dependent loss of nematic LC-filled PCF is related to the response time of LC. They found that the responsivity of their tunable fiber device was lying in the millisecond range above a certain $V_{th}$. \textcite{Scolari2005} reported for the first time that dual-frequency LC-filled PCF allows continuous tunable spectral positioning of the photonic bandgaps towards both \textit{blue}- and \textit{red}-edge depending on the frequency of the applied voltage. The proposed fiber device was a significant step forward in the development of a stable fiber-based birefringence controller or a low-speed electro-optical modulator. Figure \ref{fig:shift_LC} shows as an example, the tunability of the transmission window in a dual-frequency LC-filled PCF at around $1550$~nm.

In 2009, \textcite{Wei2009a} moved a step forward and reported a fully compact electrically controlled broadband LC-filled PCF polarizer. By controlling the electrodes attached to the LC-filled PCF independently, the direction of the electrical field is rotatable and effectively rotates the optical axis of the polarizer in steps of $45^{\circ}$. The fiber device was tested under different driving voltages, $V_{\mathrm{rms}}$, in the C-band ($1300-1600\ \mathrm{nm}$ wavelength range) and the polarization extinction ratio was found to be as high as $21.3\ \mathrm{dB}$ \cite{Wei2009a}. The main drawback of the proposed fiber device was the high insertion loss (up to $5.1\ \mathrm{dB}$) limiting its use in real all-fiber communication systems where low-loss fiber components are required. However, the same research group interestingly demonstrated for the first time, that the proposed fiber could be effectively used as optically fed microwave true-time delay \cite{Wei2009b}. They showed that a modulated microwave signal which was coupled into the hybrid LC-filled PCF and converted back to electrical microwave signal could be significantly phase-shifted with a corresponding power change.

\begin{figure}
\centering
\includegraphics[width=0.98\linewidth]{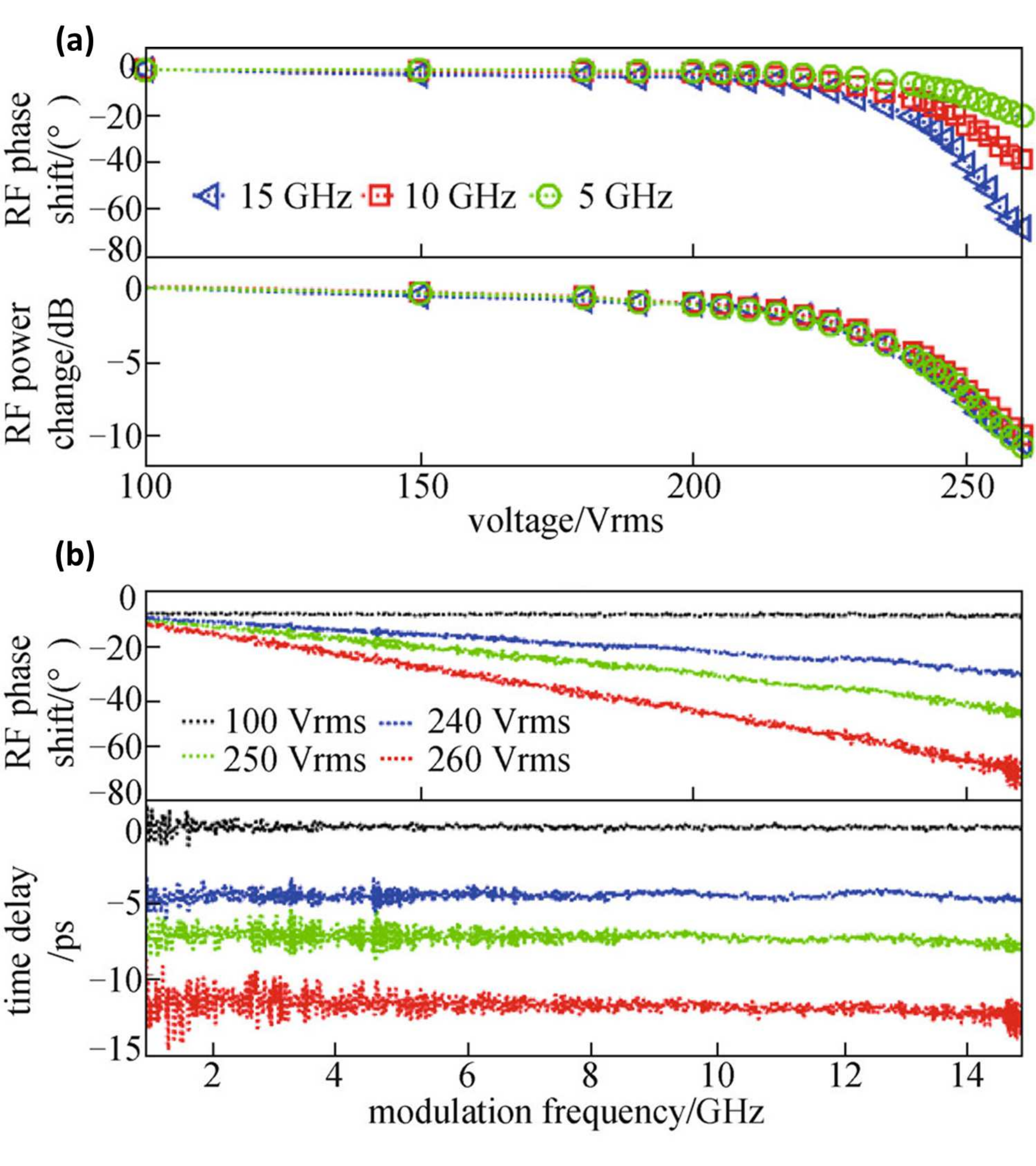}
\caption{(a) RF phase shift and relative power change versus the driving voltage for different modulation frequencies; (b) RF phase shift and corresponding time delay versus the modulation frequency for different driving voltages (top flat line corresponds to 100 $V_{\mathrm{rms}}$ in both panels) . Adapted from \textcite{Wei2009b}.}
\label{fig:RF_LC}
\end{figure}

Figure \ref{fig:RF_LC} (a) shows how the RF signal and the change in power can be dramatically affected with respect to the driving voltage for different modulation frequencies. The results demonstrate a continuously tunable RF phase shifter by simply changing the driving voltage with a maximum RF phase shift of around $70^{\circ}$C at a modulation frequency of $15\ \mathrm{GHz}$. Similarly, Fig. \ref{fig:RF_LC} (b) shows how the phase shift and the corresponding time delay change with respect to the modulation frequency for different driving voltages. At 260 $V_{\mathrm{rms}}$, an averaged $12.9\ \mathrm{ps}$ time delay over the whole measurement bandwidth is achieved, which indicates a broadband microwave true-time delay. The reported results from Wei \textit{et al.} show that although the compact LC-filled PCF might not be suitable for long communication systems but it can be easily integrated into microwave photonic systems and act as tunable true-time delay at different microwave or millimeter-wave frequency bands \cite{Wei2009b}. 

\begin{figure}
\centering
\includegraphics[width=0.98\linewidth]{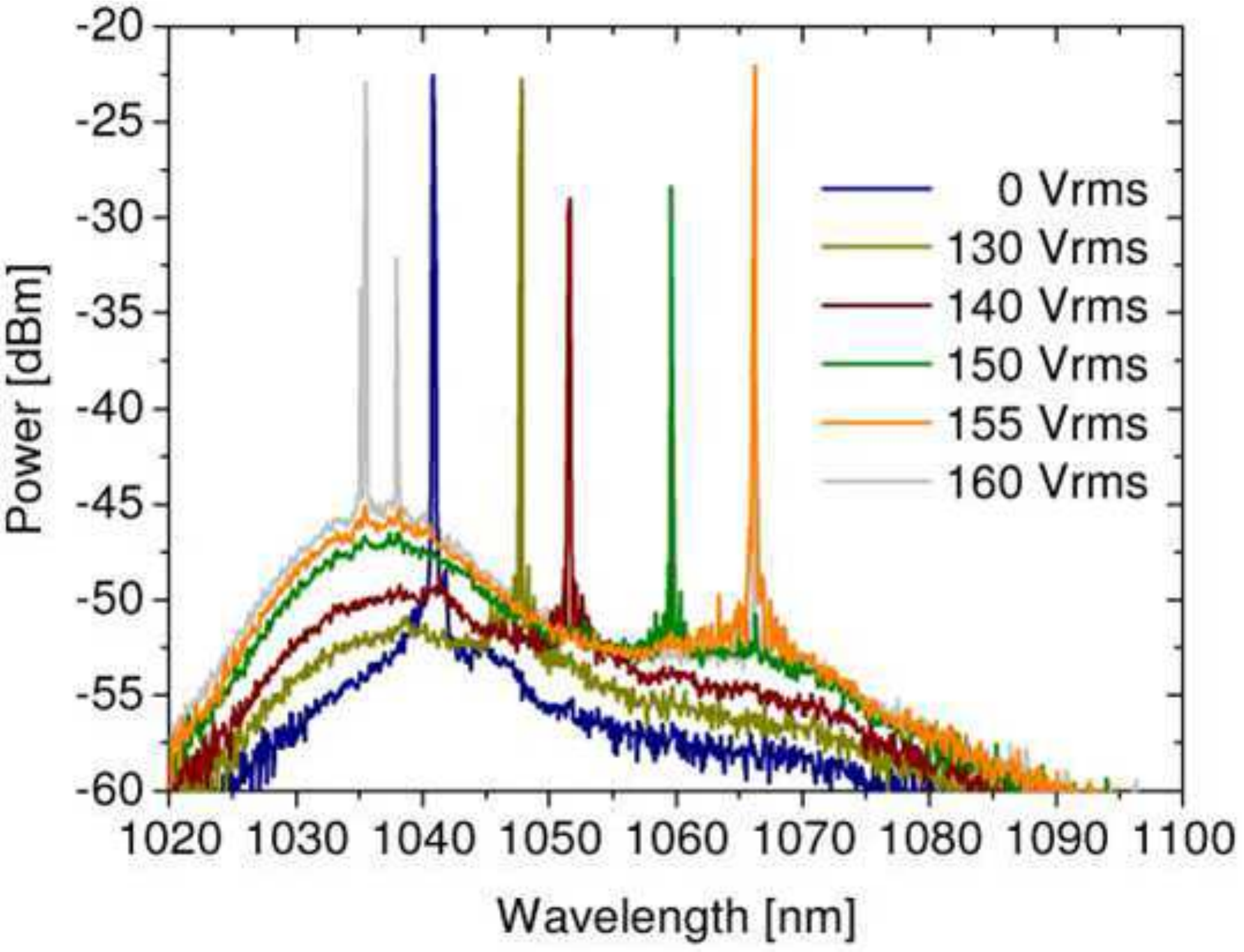}
\caption{The laser wavelength is shifted towards longer wavelengths as $V_{\mathrm{rms}}$ increases. At 160 $V_{\mathrm{rms}}$ parasitic lasing sets in at shorter wavelengths ($\sim1035$~nm). A total shift of $25\ \mathrm{nm}$ is observed. Adapted from \textcite{Olausson2010}.}
\label{fig:tunable_laser}
\end{figure}

In 2010 A. Shirakawa \textit{et al.} reported that combining the wavelength filtering effect due to PBG confinement with an ytterbium-doped core can be used for the development of solid-core PBG fiber lasers and amplifiers \cite{Shirakawa2009}. The bandgap efficiently suppresses the amplified spontaneous emission (ASE) at the conventional ytterbium gain wavelengths around $1030\ \mathrm{nm}$ and enables both short and long wavelength operation. \textcite{Olausson2010} reported that merging the PBG fiber laser concept with the tunability offered by LCs could be an efficient way to tune the laser line. They measured a dynamic tuning range of $25\ \mathrm{nm}$ ($1040\ \mathrm{nm}$ to $1065\ \mathrm{nm}$) by direct modulation of the driving voltage in a LC-filled PCF as shown in Fig. \ref{fig:tunable_laser}. The proposed approach could be further improved by reducing the cavity losses or by reducing the number of components and interfaces in the cavity between LC-PCF and Yb-doped PCF \cite{Olausson2010}. However, this is the first report showing the potential of LCs to be used as active tunable media for low-cost, compact and easy-to-use tunable seed sources for fiber amplifiers. 

Various and more advanced fiber components were also demonstrated based on LC-PCFs. For example, devices based on mechanically induced long period gratings (LPGs) in which the resonant dip can be tuned either by pressure or applied electric field has been demonstrated by \textcite{Noordegraaf2007}. They showed that the resonance wavelength could be widely tuned up to $11$~nm/$^{\circ}$C in the temperature interval  from 55$^{\circ}$C to 59$^{\circ}$C. Later in 2009, \textcite{Scolari2009a} found that the functionality of LCs could be further expanded by introducing barium titanate (BaTiO\textsubscript{3}) nanoparticles inside the original LC matrix.

In conclusion, the use of LCs with solid-core PCFs opened a new way of developing active and passive fiber-based components. Many interesting tunable devices have been demonstrated over the past years such as variable waveplates and switchable polarizers \cite{Scolari2005,Wei2009,Lee2010, Li2016}, filters \cite{Noordegraaf2008,Du2008, Lee2010, Lee2013,Liu2014, Sun2015, Li2016}, sensors \cite{Alkeskjold2007, Wolinski2009, Mathews2011a, Wolinski2012}, tunable grating filters \cite{Noordegraaf2007,Wei2009c}, polarization devices \cite{Tartarini2007,Lesiak2007, Wei2010, Liou2011}, and tunable lasers \cite{Olausson2010}. Further optimization of the optical losses would be an essential step in order for the proposed devices to find their way to the market. Novel LC synthesis or combinations with different electro-optic characteristics could be also an alternative route to further enhance the performance of the LC-filled PCF devices.

%%%%%%%%%%%%%%%%%%%%%%%%%%%%%%%%%%%%%%%%%%%%%%%%%%%%%%%%%%%%
\subsubsection{\label{highindexliquids}Other high-index liquids}
We discussed the most important advances in the topic of LC-filled PCFs for the development of tunable devices. Here we focus our attention on other liquid and solid high-index materials (i.e. those with refractive indices higher than silica) that have been infiltrated inside the air-holes of solid-core PCFs enabling the realization of similar devices and sensors. Due to the extensive existing literature, we decided to refer only to a number of selected works that we believe have had a dominant impact on the field.\paragraph{\label{liquids}Liquids} \hfill \break 

\begin{figure}
\centering
\includegraphics[width=0.98\linewidth]{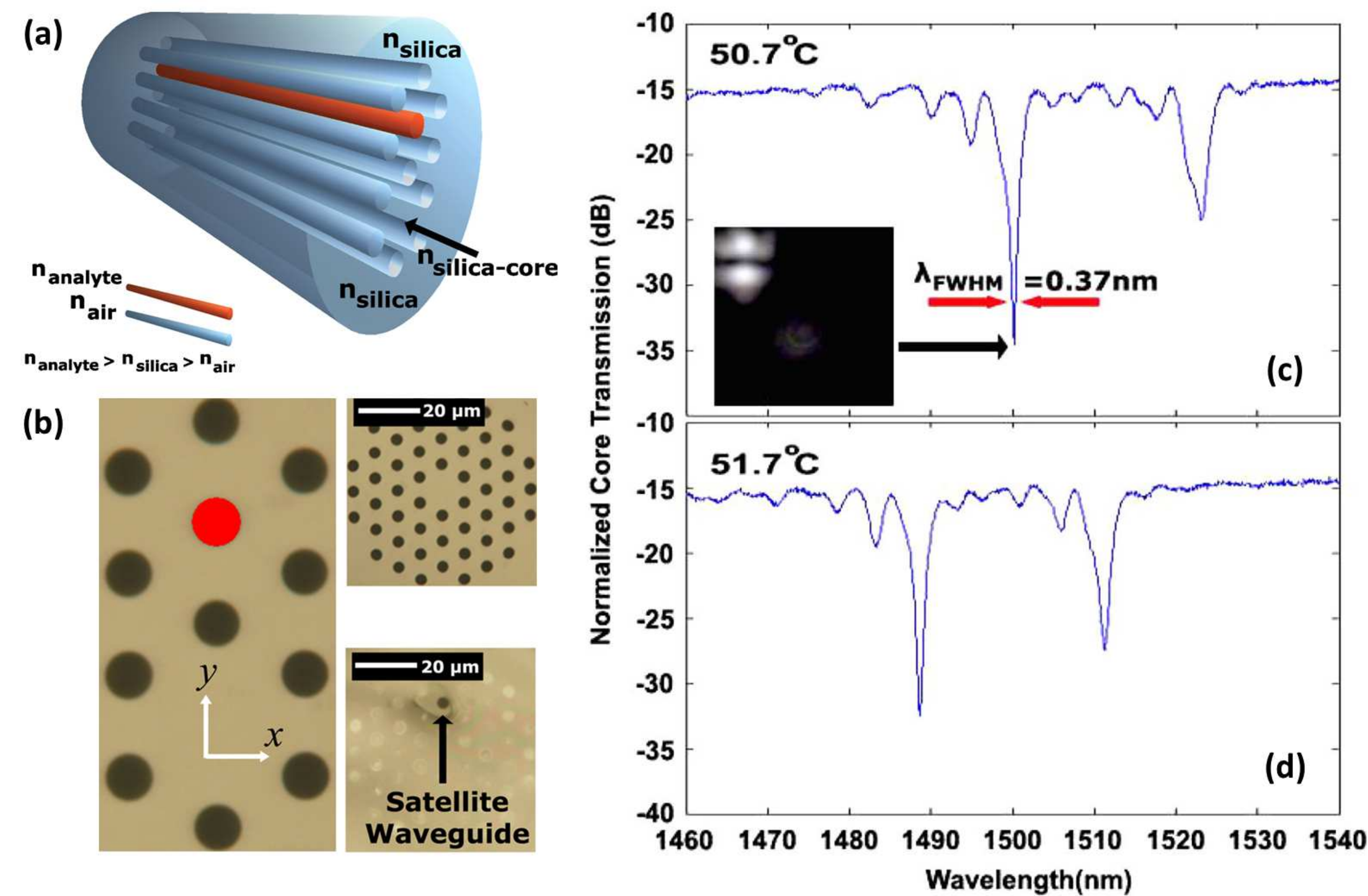}
\caption{(a) Schematic of the directional coupler. (b) Optical images of the empty and filled PCF with the high index inclusion. Transmission as a function of wavelength for (c) 50.7$^{\circ}$C and (d) 51.7$^{\circ}$C of the selectively-filled hybrid PCF based refractive-index sensor. Inset of (c): Measured near-field profile of the core and high index waveguide mode profiles. Adapted from \textcite{Wu2009}.}
\label{fig:RI_sensor}
\end{figure}

In 2005, Steinvurzel \textit{et al.} reported a highly tunable bandpass filter by applying a variable thermal gradient in a liquid-filled PCF \cite{Steinvurzel2005}. The active thermo-refractive infused material used was a commercially available high-index fluid with refractive index $\sim1.65$ and $\partial n/\partial T \approx -4.65 \times 10^{-4}$ $^{\circ}\mathrm{C}^{-1}$. The authors experimentally measured continuous tuning of the width of each band from $\sim 35\ \mathrm{THz}$ to effectively zero and achieved a high out-of-band suppression. Similar high-index fluid with refractive index $\sim1.5$ was also used in a different configuration by \textcite{Wu2009} for the development of a refractive index sensor. In that work the authors reported a novel sensor architecture integrating only one single high-index inclusion - in the form of a microfluidic analyte channel - within the solid-core PCF (see Fig. \ref{fig:RI_sensor} (a) and (b)). The selectively filled PCF formed a directional coupler, with the analyte channel being a waveguide itself. The transmission of the PCF presents dips when the fundamental mode couples to a mode of the analyte channel, with the coupling wavelength depending on the refractive index of the analyte. The main novelty of the proposed sensing device is that the microstructure allows the core mode to couple to higher-order modes of the analyte channel just above the cutoff, leading to extremely high sensitivities. Typically, a refractive index sensor will have a resonance feature (e.g., a resonance peak in the spectrum) whose resonant wavelength $\lambda_{\mathrm{r}}$ depends on the refractive index $n_{\mathrm{analyte}}$ to be measured. Therefore, the sensitivity which is one of the most important factor determining the performance of a sensor is defined as \cite{White2008}:
\begin{equation}
S = \frac {\partial\lambda_{\mathrm{r}}}{\partial n_{\mathrm{analyte}}}
\end{equation}
When higher-order modes of the high-index fluid waveguide in the work of Wu \textit{et al.} become leaky modes, their mode field expand, and rapidly they become higher order modes of the microstructured high index core. These modes have effective refractive index, $n_{\mathrm{eff}}$ below that of the host material's fundamental mode. There is thus a wavelength close to the cutoff wavelength of the high-index fluid cylinder at which both the fundamental silica core mode and the higher order mode of the high index channel are phase-matched and couple \cite{Wu2009}. This coupling introduces a strong resonance in the transmission as shown in Fig. \ref{fig:RI_sensor} (c) and (d). 

The resonant peak is dramatically blue-shifted with a small increase of the temperature (see Fig. \ref{fig:RI_sensor} (c) and (d)). The sensitivity was found to be as high as $30100$~nm/Refractive Index Unit (RIU). It should be noted however that the proposed directional coupler geometry described above can be applied only for analytes with refractive index higher than the host material \cite{Wu2009}. This is certainly a problem for several bio-applications such as the detection of biomolecules typically found in aqueous, low-index solutions. The authors reported that one possible solution to this problem could be to coat the holes \cite{Kuhlmey2009a} or directly draw a fiber using a low-index material (CYTOP) which has a refractive index of only 1.34 \cite{Wu2009}. Nevertheless, the aforementioned directional coupler holds the record refractive index sensitivity in fibers until today.  

Magnetic liquids such as ferrofluids belong also to the family of functional high-index materials as they have been used for the development of all-fiber magnetically tunable devices \cite{Blue2016}. In 2011, \textcite{Thakur2011} infiltrated the Fe\textsubscript{3}O\textsubscript{4} nano-fluid  inside a polarization-maintaining solid-core PCF for the first time and they demonstrated a magnetic field sensor. The sensing mechanism was relying upon tracking the wavelength shift of the interference fringes introduced by the two orthogonally polarization modes. The measured sensitivity was measured as high as 242 pm/mT \cite{Thakur2011}. \textcite{Zu2012} reported that by raising the refractive index of the magnetic liquid (using toluene) and modifying the guidance from index-guiding to ARROW the magnetic sensitivity can be significantly enhanced. The authors introduced the high-index fluid into the holes of a 20 mm section of a PCF. The revealed resonances were shifting by modulating the external magnetic field. The sensitivity was found to be as high as 15600 pm/mT over a range of 3 mT magnetic field strength \cite{Zu2012} and the proposed device holds the record sensitivity in PCF grating-free magnetic field sensors. However, it should be mentioned that the fiber device had an insertion loss of $\sim$ 6 dB \cite{Zu2012}. In 2013, Gao \textit{et al.} infiltrated a water-based magnetic fluid into the air holes of a $\sim$ 10.4 cm long solid-core PCF demonstrating for the first time an intensity-based magnetic sensor \cite{Gao2013}. As the magnetic field was increasing, the refractive index of the fluid was increasing introducing thus high confinement loss making the transmission drops. The reported sensor exhibited a linear response from 20 up to 60 mT with sensitivity 0.011 $\mu$W/mT \cite{Gao2013}. In 2015, Mahmood \textit{et al.} demonstrated for the first time a novel magnetic field sensing configuration based on whispering gallery modes (WGM) \cite{Mahmood2015}. The authors showed that infiltration of a magnetic fluid doped with nanoparticles inside a 1.3 cm long PCF can form a micro-resonator. They demonstrated the shft if the WGM resonances toward higher wavelengths upon applying a magnetic field. The sensitivity of the proposed sensor was found to be as high as 110 pm/mT in the magnetic field range from 0 to 38.7mT \cite{Mahmood2015}.  

\subsubsection{\label{softglasses}Chalcogenide glasses} \hfill \break

Chalcogenide glasses, containing the chalcogen elements sulfur (S), selenium (Se) and tellurium (Te), have existed for  more than 60 years, yet they have  only  recently  been  proven  as  emerging  materials  offering  new  possibilities  in photonics \cite{Seddon1995, Eggleton2010,Eggleton2011}. Their  most  important  optical properties  are perhaps  their  extremely  high nonlinear coefficients which can be even 1000 times higher than silica (depending on the glass composition), high refractive indices  $n=2-3.5$, photo-tunability properties when illuminated with light at wavelength near their band-gap edge and most importantly their ability to be transparent in the infrared region. Both linear and nonlinear optical properties of chalcogenide glasses have been extensively investigated over the past years by the glass and ceramic research community \cite{Eggleton2011}. \textcite{Schmidt2009} reported for the first time the development of a post-processed all-solid PBG fiber by pumping molten tellurite glass into a silica-air photonic crystal fiber at high pressures. The hybrid tellurite/silica PCF revealed strong resonances in its transmission due to the coupling of the core mode to the high index infused glass. However, the micrometer-diameter tellurite strands are found to contain micro-heterogeneities arising most probably from devitrification, resulting in an elevated fiber attenuation \cite{Schmidt2009}. The proposed technique from Schmidt \textit{et al.} offered for the first time an alternative for the development of hybrid geometries with glasses unsuitable for direct fiber drawing that could be potentially used for fiber-based amplifiers, filters and nonlinear devices \cite{Schmidt2009}. The later pressure-assisted infiltration method was further investigated a year later by Da \textit{et al.} where they developed a full model on what are the rheological properties and flow of highly-viscous molten of two chalcogenide glasses (As$_2$S$_3$ and 75TeO$_2$-10ZnO-15Na$_2$O) in highly constrained $\mu$m-scale geometries under high mechanical load \cite{Da2010}. 

\begin{figure}
\centering
\includegraphics[width=0.98\linewidth]{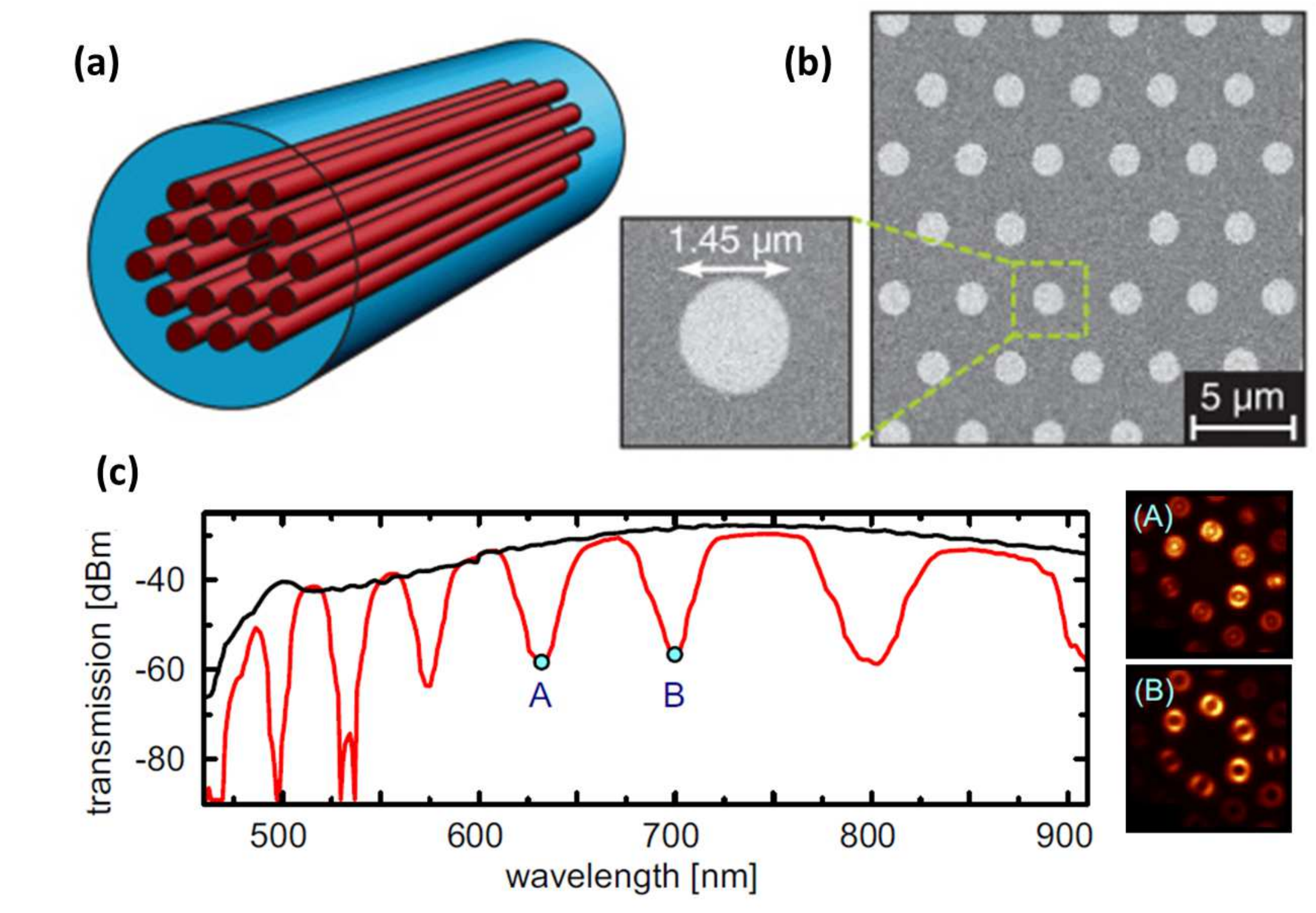}
\caption{(a) Schematic representation of the hybrid chalcogenide glass/silica PCF. (b) SEM image of the hybrid PCF. (c) Transmission spectrum of the unfilled (black line) and the hybrid PCF (red line). The point A and B correspond to the output mode patterns at $630$ and $700$ nm, respectively. Adapted from \textcite{Granzow2011}.}
\label{fig:chalco_hybrid}
\end{figure}

In 2011 Granzow \textit{et al.} reported the fabrication of a hybrid chalcogenide/silica PCF using a similar pressure-assisted melt filling method and an arsenic-free chalcogenide glass (Ga$_4$Ge$_{21}$Sb$_{10}$S$_{65}$) \cite{Granzow2011}. Figures \ref{fig:chalco_hybrid} (a) and (b) show the schematic of the hybrid chalcogenide/silica PCF and the scanning electron micrograph (SEM) of the fabricated device, respectively. 

By launching light into the hybrid PCF from a supercontinuum source, distinct resonances appeared from $450$~nm up to $\sim$900 nm as shown in Fig. \ref{fig:chalco_hybrid} (c) due to the coupling of light from the core to the high index strands. The near field output profiles of the cladding strands are also shown in Fig. \ref{fig:chalco_hybrid} (c) (points A and B) verifying the core mode coupling to the high index strands which eventually introduces the observed strong transmission resonances \cite{Granzow2011}.  

\begin{figure}
\centering
\includegraphics[width=0.98\linewidth]{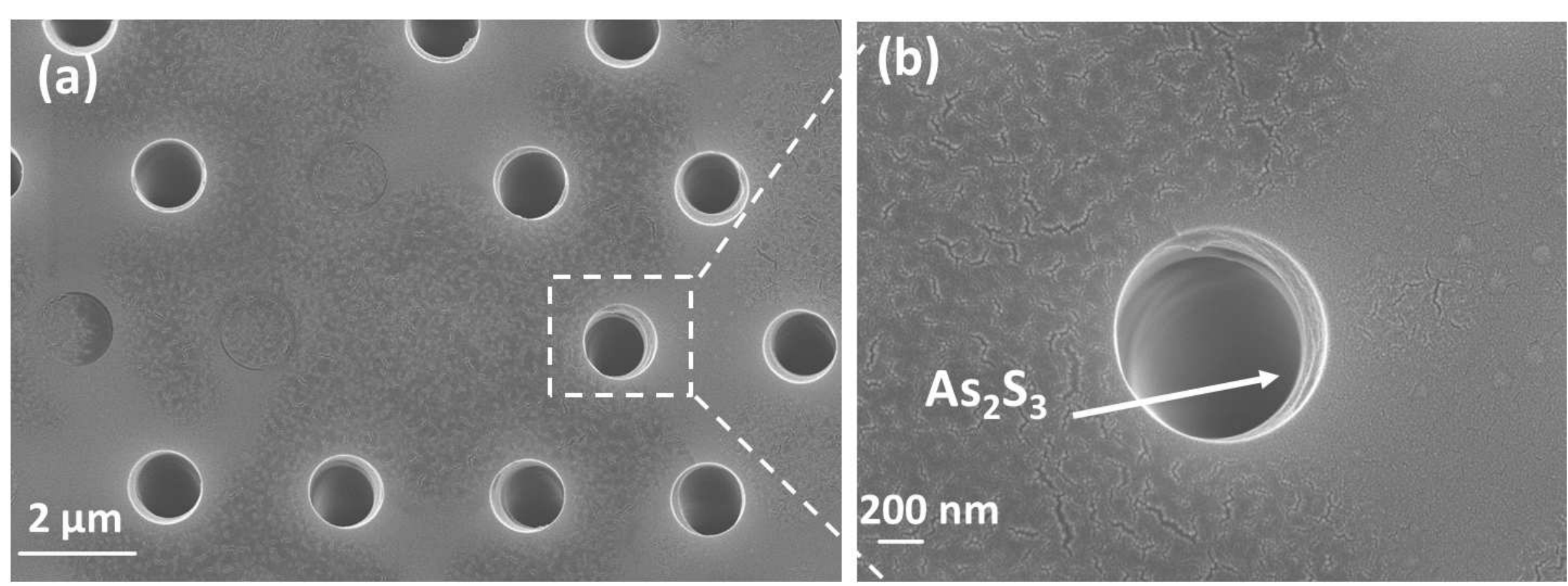}
\caption{Cross-section SEM image of the core of a solid-core PCF with deposited As$_2$S$_3$ layers. (b) Zoomed-in SEM image of a single hole. Adapted from \textcite{Markos2012}.}
\label{fig:markos_layers}
\end{figure}

The pressure-assisted approach to melt filling of the chalcogenide glass presented previously is certainly an efficient and straight forward way for the development of hybrid soft-glass/silica PCFs. It is pertinent to note however that the infiltration of the molten chalcogenide glass is achieved under inert atmosphere (Ar) at high pressure ($\sim$50 bars) and high temperature ($\sim$665$^{\circ}$C) where custom-made and specialized equipment is required.  In  2012,  it  was reported  for   the  first time by \textcite{Markos2012} a  new  facile  and  cost-effective  method  of  depositing  As$_2$S$_3$ chalcogenide   glass  films  inside  PCF  by  using  a  solution-derived approach simply requiring  ambient   conditions   and   no sophisticated   equipment. This method involves the dissolution of the initial bulk material into liquid. After the infiltration of the PCF with the liquid glass, a soft-anneal post-processing is required for the formation of solid chalcogenide glass layers inside the holes of the PCF. One   of   the   most  important advantages of this approach was the possibility to modify the guidance mechanism of the fiber from index-guiding to ARROW with only a few nanometers thick high-index glass films with the air-holes of the fiber still available for further functionalization or post-processing \cite{Markos2012, Markos2014a,Markos2016,markos2015thermally}. Figure \ref{fig:markos_layers} (a) and (b) show the SEMs of the As$_2$S$_3$ coated PCF based on the glass-solution approach. The main advantage of this method is that it can be adopted for any chalcogenide glass composition and can be extended to any solution-dissolved material \cite{Konidakis201476,Konidakis:15} or even for stacked multi-material layer deposition. However, uniform film deposition over long fiber lengths still remains a challenging task.  

\subsubsection{\label{semiconductors}Semiconductors} \hfill \break

Semiconductor-based photonics started back in late 1950s, during which the optical properties of many materials were investigated. Today, silicon (Si) is still perhaps the most widely used material in integrated photonic circuits owing to its high transparency in the C-band \cite{Jalali2006}, while germanium (Ge) is attracting a lot of interest in the mid-IR "molecular fingerprint" region \cite{Soref2010}. Fabrication of crystalline or amorphous semiconductor fibers for optoelectronic applications is a difficult task due to their thermal, chemical and mechanical mismatch with the conventional silica glass fibers. However, it has been already reported by \textcite{ballato2008silicon} the successful fabrication of step-index fibers with a silicon core using either direct fiber drawing \cite{ballato2008silicon, suhailin2016tapered} or by using the recent laser-assisted recrystallization method \cite{coucheron2016laser}.  Similarly, many efforts have been focused on the combination of semiconducting materials using silica PCFs as template \cite{peacock2016semiconductor, Tyagi2008, Peacock2014}\footnote{Direct drawing of novel optoelectronic fibers of thermo-mechanically compatible materials have been thoroughly investigated by the group of Prof. Fink at Massachusetts Institute of Technology (MIT). See review \textcite{Abouraddy2007}.}. 

\begin{figure}
\centering
\includegraphics[width=0.98\linewidth]{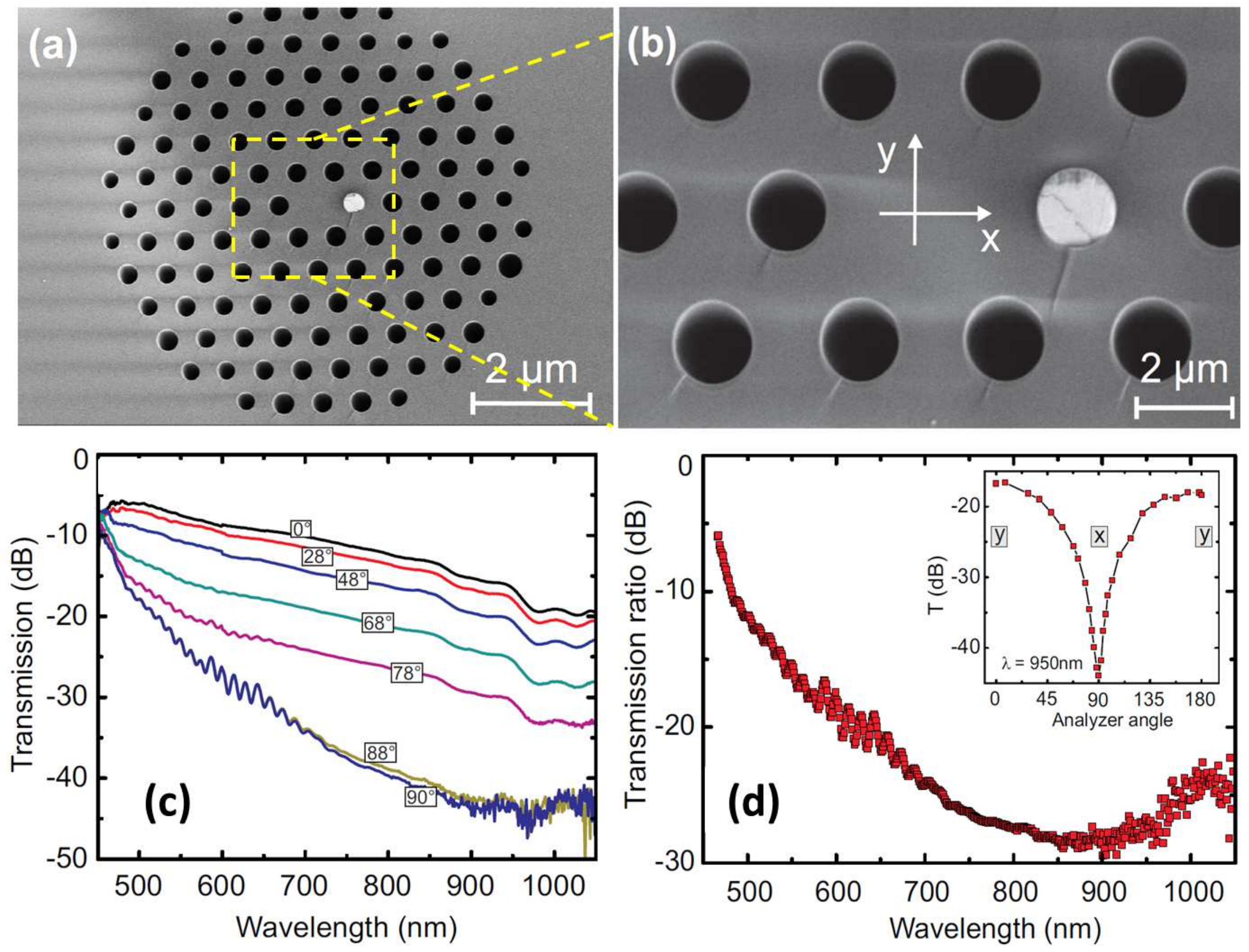}
\caption{SEM image of the core of a solid-core PCF with a single high index Ge wire. (b) Zoomed-in SEM image of a the cross section of the fiber. (c) Transmission spectra with respect the wavelength at various positions of the output polarizer. (d) Transmission profile of the ratio between the two orthogonal polarization states as a function of wavelength. Inset: Intensity response with respect the analyzer angle at fixed wavelength ($950$~nm). Adapted from \textcite{Tyagi2011}.}
\label{fig:germanium_hybrid}
\end{figure}

Germanium is considered an indirect semiconductor which means that the maximum energy of the valence band occurs at a different value of momentum to the minimum in the conduction band energy. It is one of the most widely used semiconductors for optical devices due to its low loss transparency window from 2 $\mu$m up to 16 $\mu$m \cite{Soref2010}. Using the pressure-assisted melting method it has been demonstrated that Ge can be successfully integrated inside the holes of the PCF for in-fiber devices and sensors \cite{Tyagi2008}. The main reason for choosing Ge over other semiconductors is its lower melting point in comparison to silicon and silica (400$^{\circ}$C lower than fused silica). This large thermal difference ensures stability during the filling of molten germanium into the empty holes of PCFs. In contrast for example integration of Si inside PCF can be challenging as the melting points of the two are quite close and therefore the risk of destroying the PCF microstructured air-hole pattern during infiltration is high. Using a selective-filling method, \textcite{Tyagi2008} developed a hybrid PCF with only one hole in the vicinity of the glass-core is filled with Ge. Figure \ref{fig:germanium_hybrid} (a) and (b) show the SEM image of the integrated single Ge wire inside PCF.  

As it is expected, the high index Ge strand adjacent to the core breaks the degeneracy of the fundamental mode creating thus two strong polarization states. By coupling broadband light, the transmission spectrum decreases as the polarization axis of the output changes from y- to x- polarization as shown in Fig. \ref{fig:germanium_hybrid} (c) \cite{Tyagi2011}. Figure \ref{fig:germanium_hybrid} (d) shows the spectral distribution of the ratio between the two polarization states over the visible and near-IR region. No strong oscillations up to $1000$~nm wavelength were observed and with a polarization discrimination as high as 30 dB in the visible confirming thus the ability of the current fiber device to act as polarizer \cite{Tyagi2008}. Moreover, it was shown that such a hybrid structure could act as an in-fiber thermometer by tracking the shift of the loss peaks of the transmission. The sensing mechanism is relying upon the fact that with the increasing temperature the indirect electronic band gap of Ge decreases shifting thus the absorption band to longer wavelengths. The dielectric function increases due to Kramers Kroenig's relation and thus a shift of the loss peaks is observed. The thermal sensitivity was found to be $0.18$~nm/$^{\circ}$C which is comparable or even higher than other temperature fiber sensors such as Bragg ($\sim$ 0.01 nm/$^{\circ}$C) or long period ($\sim 0.1\ \mathrm{nm/{}^{\circ}C}$) gratings sensors \cite{Tyagi2008}.

\begin{figure}
\centering
\includegraphics[width=0.98\linewidth]{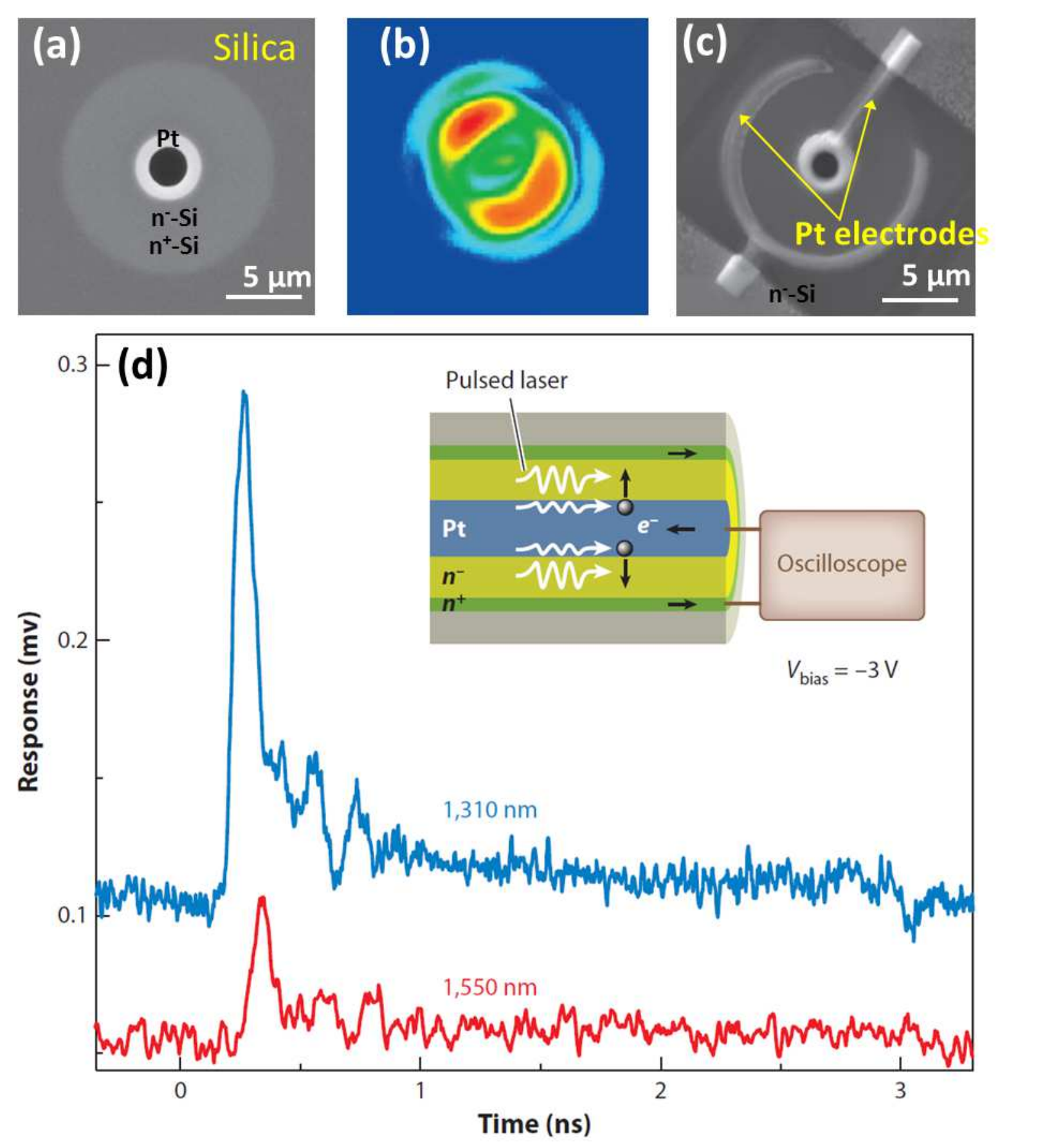}
\caption{(a) A Schottky Pt/n-Si junction formed in a capillary by multi-material deposition using HPCVD of phosphorus-doped n$^{+}$-Si, n$^{-}$-Si, and Pt layers. (b) Fundamental guiding mode in the n$^{-}$Si layer.  (c) Electrodes in the Pt/n-Si Schottky junction formed using FIB (d) Photo-detection response of the diode triggered  with $\sim$10-ps laser pulses at wavelengths of $1310$~nm (blue) and $1550$~nm (red) and measured using an oscilloscope. Inset: Internal photoemission process, carrier transport, and collection scheme. Adapted from \textcite{He2012}.}
\label{fig:multimaterial_layers}
\end{figure}

Even though the pressure-assisted filling approach for filling PCF with low-melting-temperature semiconductors constitutes a straight forward method for the development of hybrid structures \cite{Lee2008, Schmidt2008}, it also has some limitations.  For example, this technique is limited to materials with melting temperatures below the glass-transition temperature (Tg) of silica ($\sim$1300$^{\circ}$C) \cite{Peacock2014}. Furthermore, integration of precisely structured films by means of this technique as well as multi-material integration appears to be difficult and rather challenging tasks. One way to tackle these limitations is by employing the well-known high-pressure chemical vapor deposition (HPCVD) technique. Appropriately chosen chemical precursors along with a carrier gas (e.g. helium) are pumped at high pressures into a reservoir. Once the PCF is placed inside the reservoir, the precursors are running along the length of the fiber through the air-holes due to the high pressure environment. Thermal treatment of the fiber gives rise then to chemical reaction of the precursors which eventually lead to the formation of the desired layer or wire inside PCF. \textcite{Sazio2006} first reported this method towards development of polycrystalline elemental or compound semiconductors within silica PCF in a flexible and controllable way. Interestingly, Badding's group further investigated this approach for the development of optoelectronic fibers \cite{Sparks2013}. In 2012, it was shown that fast and efficient detection of light in the C-band is possible by using directly functional optoelectronic fibers avoiding the use of lossy planar junctions. Figure \ref{fig:multimaterial_layers} (a) shows the multilayer deposition of doped crystalline semiconductors inside silica capillaries and selected voids of PCFs \cite{He2012}.  The Pt/n-Si junction structures confines and guides the light in the n$^{-}$ layer because the n$^{+}$ layer has a lower refractive index and there is metallic reflection at the n-/Pt interface as it can be seen from Fig. \ref{fig:multimaterial_layers} (b). The fundamental mode at $1550$~nm is HE\textsubscript{11}. Using focused ion beam (FIB), these junctions could be contacted as shown in Fig. \ref{fig:multimaterial_layers} (c). Optical experiments were performed at both $131$~nm and $1550$~nm using 10-ps laser pulses. Figure \ref{fig:multimaterial_layers} (d) shows the photoresponse of the junction inside the fiber having a 60-ps rise time and a 100-ps fall time, at $1550$~nm under bias voltage of − 3 V \cite{He2012}. Similar response time but with bigger photoresponse observed at $1310$~nm wavelength. The authors suggested how the photo-response of the built-in device can be further enhanced by considering different metal electrodes with lower work function, smaller-bandgap semiconductors and optimization of the layer geometry \cite{Sparks2013}.

%%%%%%%%%%%%%%%%%%%%%%%%%%%%%%%%%%%%%%%%%%%%%%%%%%%%%%%%%%%%
\subsection{\label{plasmons} Hybrid metal-filled plasmonic PCF}

One of the challenges faced by modern optics nowadays is breaking the diffraction limit and achieving sub-wavelength scale optical waveguides. Perhaps the most promising platform for achieving this is surface plasmon polaritons (SPPs) where electromagnetic waves (EM) can be guided in a metal-dielectric nanostructure below the diffraction limit having still high bandwidth \cite{Maier2007}. The most prominent materials which exhibit collective oscillations of electron densities with EM are metals\footnote{Silver plasmon resonance has been recently reported using a silver metaphosphate glass-filled PCF by \textcite{konidakis2014}.}. One of the first SPPs observation was by Sommerfeld and Zenneck in 1899 and 1907, respectively \cite{Sommerfeld1899,Zenneck1907}. These studies were mainly focused on long wavelengths while Kretschmann and Ritchie reported the excitation of SPPs at visible range in 1968 using a prism coupling \cite{Kretschmann1968} and metal grating \cite{Ritchie1968}, respectively. More recently, \textcite{Ebbesen1998} reported that the field is strongly enhanced in arrays of sub-wavelength holes in metal substrates reporting the first strong light-SPPs coupling in nanostructured metal surfaces. Since then excitation of SPP via surface structuring has became more popular. For a detailed and more comprehensive overview on plasmonics and nanophotonics the books by Novotny and Hecht \cite{Novotny2006}, Maier \cite{Maier2007}, Bozhevolnyi and Raether \cite{Raether1988} are recommended. Here, we overview the most important reports on generated SPPs using a hybrid metal-filled PCFs as a route towards plasmonic structures for on-chip applications or in-fiber polarization devices \cite{Lee2008, Lee2012}.
%%%%%%%%%%%%%%%%%%%%%%%%%%%%%%%%%%%%%%%%%%%%%%%%%%%%%%%%%%%%
\subsubsection{\label{linear}Linear properties of metals}

It is well known that metals are electron-rich materials with concentrations \textit{N} of the order $N \approx 10^{28} m^{-3}$. The optical response of metals is mainly originate from the conduction electrons. With the free electron theory of metals \cite{Maier2007}, the dielectric function and the absorption of light due to the electron transition to higher energy states with respect to the conduction band can be closely approximated. The electron transport in metals is often described by the Drude model \cite{Drude1900} which for the design of plasmonic structures is perhaps the most important parameter. The plasma frequency is usually in the range of $\omega_{\mathrm{P}} \approx 10^{16}$ Hz. To derive the expressions for $\omega_{\mathrm{P}}$ and  $\epsilon$ (dielectric permittivity) of a metal we start by calculating the oscillation amplitude of a free electron under the action of an external electric field $E(t)=E_0 e^{-i \omega t}$ \cite{Sempere2010}. The differential equation governing this is given by:  
\begin{equation}
m_{e}\bigg (\frac{\partial^{2}x}{\partial t^{2}}+\gamma \frac{\partial x}{\partial t} \bigg)=-eE_{0}e^{-i\omega t} 
\end{equation}
where $\omega$ is the frequency of light, $m_{e}$ the electron mass and $E_{0}$ its amplitude. The factor $\gamma = 1/\tau$ is damping constant of the material which is related to the scattering rate of the electrons. $\tau$ is the relaxation time of the electron gas \cite{Ordal1983}. The left hand side of Eq. 12 describes the acceleration and the frictional damping of the electron while the right hand side describes the force exerted by the light. Assuming a solution of the form $x(t)=x_{0} e^{-i \omega t}$ into Eq. 12 we get:
\begin{equation}
x(t)=\frac{eE(t)}{m_{e}\big (\omega^{2}+i\gamma\omega \big)}
\end{equation}

The induced polarization is given by $P(t)=-Nex(t)$, where $N$ is the number density of free electrons. Also, $D(t)=\epsilon_{0}E(t)+P(t)$ and by replacing $P$ we end up with:
\begin{equation}
D=\epsilon\epsilon_{0}E=\epsilon_{0}E+P=\epsilon_{0}E-\frac{Ne^{2}E}{m_{e}(\omega ^{2}+i\gamma \omega)}
\end{equation}
Therefore: 
\begin{equation}
\epsilon(\omega)=1-\frac{Ne^{2}}{\epsilon_{0}m_{e}}\frac{1}{\omega^{2}+i\gamma\omega}=1-\frac{\omega_{\mathrm{P}}^{2}}{\omega^{2}+i\gamma\omega}
\end{equation}
where the \textit{plasmonic frequency} is:  
\begin{equation}
\omega_{\mathrm{P}}=\sqrt{\frac{Ne^{2}}{\epsilon_{0}m_{e}}}
\end{equation}

The Drude model equation which represents the permittivity of metals is described by Eq. 15 and can be divided into its real ($\epsilon'$) and imaginary part ($\epsilon''$) as following: 

\begin{equation}
\epsilon'(\omega)=1-\frac{\omega_{\mathrm{P}}^{2}}{\omega^{2}+\gamma^{2}}
\end{equation}
\begin{equation}
\epsilon''(\omega)=\frac{\omega_{\mathrm{P}}^{2}\gamma}{\omega^{3}+\omega\gamma^{2}}
\end{equation}

\begin{figure}
\centering
\includegraphics[width=0.98\linewidth]{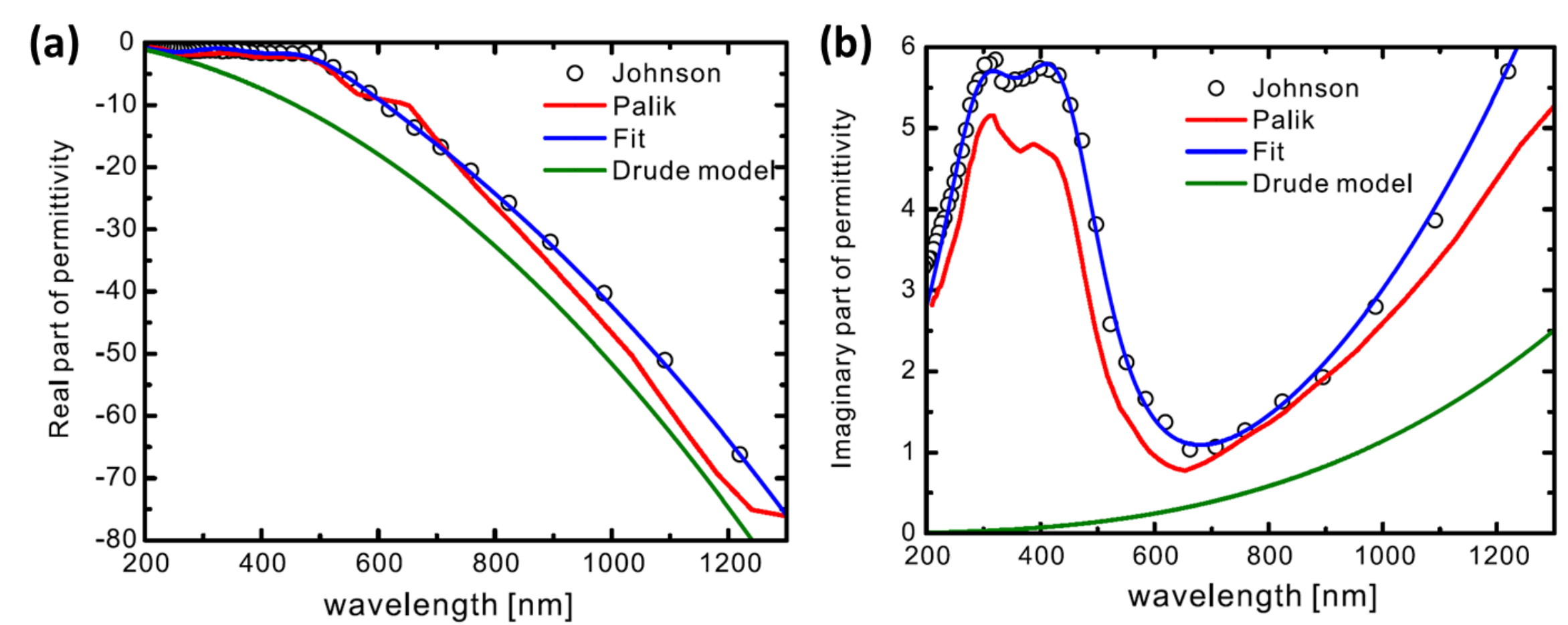}
\caption{(a) Real and (b) imaginary part of permittivity of gold from $200$~nm up to $1300$~nm. The green line is calculated using the Drude model (lowest line). The black circles and red line are the experimental data from Johnson and Palik, respectively. Blue line is the fit of the experimental data. Adapted from \textcite{Lee2012}.}
\label{fig:drude_model}
\end{figure}

It should be noted that for most metals, inter-band transitions (valence – conduction band) also contribute to the absorption and thus for frequencies close to the inter-band transition (usually in the short visible range), the Drude model is not valid anymore. In this review article, we focus only on gold (Au) as it is perhaps the best plasmonic material for the fabrication of hybrid PCFs due to relatively low imaginary part of dielectric function (which is linked to optical absorption of Au) in comparison to other metals. Figure \ref{fig:drude_model} (a) shows the real part of the dielectric permittivity of Au taken from the experimental data of Johnson and Christy (black circle) \cite{Johnson1972} and Palik's Handbook of Optical Constants of Solids (red line) \cite{Palik1997}. The analytic fit of the experimental data (blue line) \cite{Etchegoin2006} and the calculated values based on the Drude model (green line) using Eqs. 17 and 18 are also shown. Similarly, Fig. \ref{fig:drude_model} (b) shows the imaginary part of the permittivity for Au. It is evident that the Drude model (green line) fails to describe the response of Au in the range of $200-600$~nm. This is because the Drude model does not include the absorption arising from the inter-band transitions as we mentioned before \cite{Lee2012}.   

%%%%%%%%%%%%%%%%%%%%%%%%%%%%%%%%%%%%%%%%%%%%%%%%%%%%%%%%%%%%
\subsubsection{\label{fabrication}Fabrication of plasmonic PCF}

Although the integration of glass wires and films inside PCF has been reported using various techniques \cite{Schmidt2008, Markos2012}, the most convenient way of filling molten metal (or semiconductors) inside PCF is perhaps the pressure-assisted method \cite{Tyagi2008, Schmidt2009, Lee2011}. The main challenge of pushing molten metals inside PCF’s holes is to overcome the anti-capillary force introduced from the very high surface tension of the metal which acts negatively and pushes the molten metal out of the holes. The smaller the diameter of the holes, the bigger the threshold pressure required \cite{Lee2011}.

\begin{figure}
\centering
\includegraphics[width=0.98\linewidth]{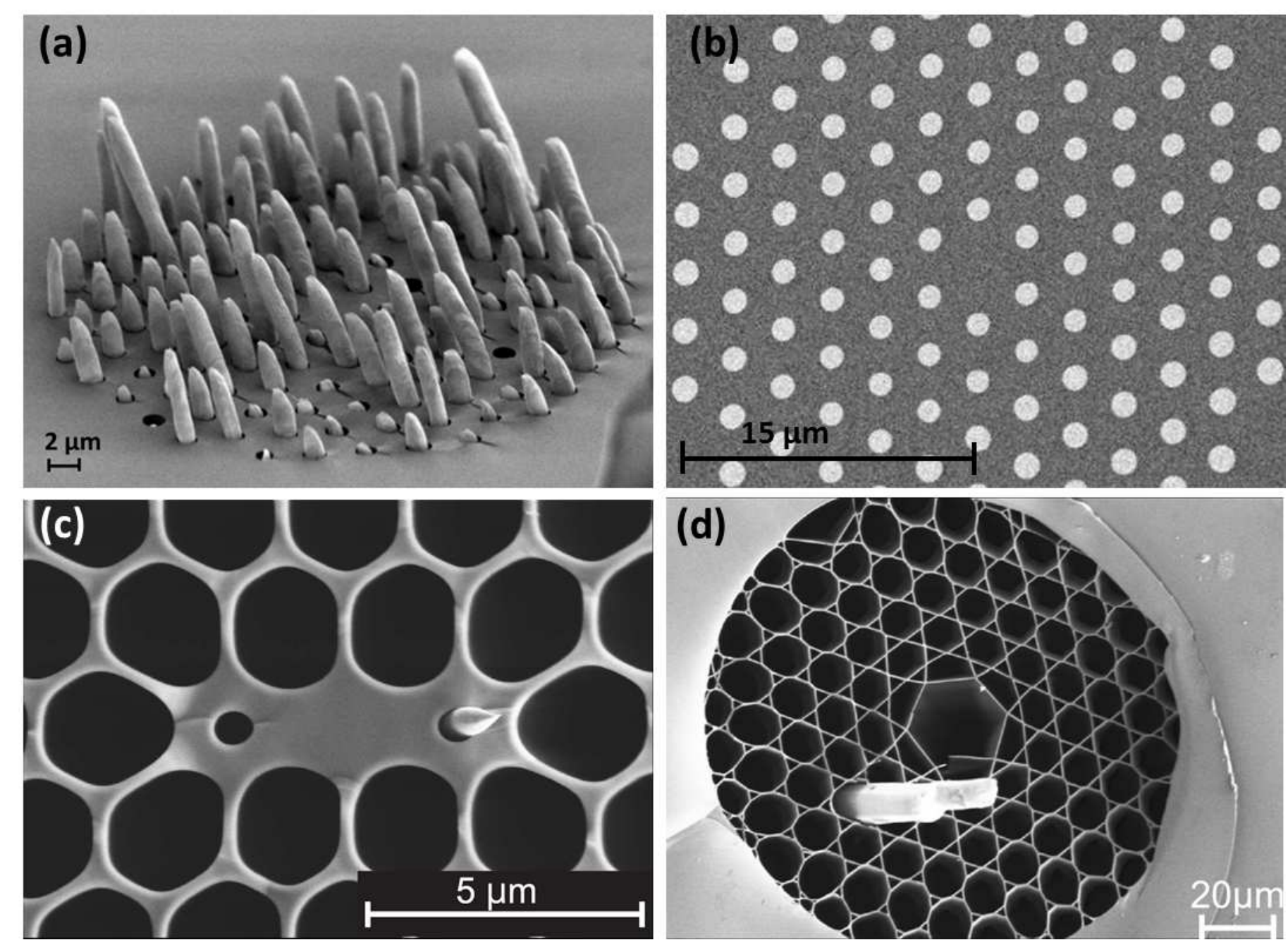}
\caption{SEM images of Au-gold filled (a) solid-core PCF showing the metal wires sticking out (b) after polishing the end-facet of the fiber with a focus ion beam. Selectively Au-filled (c) polarization-maintaining (PM) PCF and (d) Kagom\'{e} hollow-core fiber.Adapted from \textcite{Sempere2010}, \textcite{Lee2012} and \textcite{Uebel2013}.}
\label{fig:sem_selective}
\end{figure}

In 2008, \textcite{Schmidt2008} reported the successful fabrication of both gold and silver metal wire arrays inside a PCF with diameters down to $500$~nm using a high temperature pressure cell. This method was further improved over time and at the moment the integration of metal wires inside PCF can be achieved with a fiber splicer and a pressure cell \cite{Lee2011}, a furnace to liquefy the metals and a pressure system providing pressurized argon up to a few hundreds bar. The main advantage of this technique is that all or only selected air channels (using selective filling masking methods) of a PCF can be filled. Figure \ref{fig:sem_selective} (a) shows how the integrated metal wires into a solid-core PCF are sticking out before polishing the fiber. One method of polishing the end facet of the fiber is using the focused Ga ion beam to ablate small amounts of material and mill the metal \cite{Sempere2010}. However, this method is expensive and time consuming. Figure \ref{fig:sem_selective} (b) shows the clean polished end-facet of the fiber. Figure \ref{fig:sem_selective} (c) and (d) show the selective integration of Au wires in a polarization-maintaining solid-core PCF and a kagom\'{e} hollow-core PCF, respectively. The pressure assisted technique is a versatile method for the development of hybrid PCFs but it also has some drawbacks as mentioned in Section II.A.2. For instance the infused molten material length is restricted to the hot zone of the furnace which is typically around 10s of cm. However, for plasmonic applications, a few centimeters long metal-filled PCF is normally more than enough for efficient excitation of SPPs. 

%%%%%%%%%%%%%%%%%%%%%%%%%%%%%%%%%%%%%%%%%%%%%%%%%%%%%%%%%%%%

\subsubsection{\label{optical}Plasmon resonance-induced polarization effects}

The core guiding mode in a plasmonic metal-filled PCF can be coupled with SPPs when the phase matching conditions are satisfied \cite{Maier2007}. These hybrid fibers have wavelength dependent transmissions because the core guided light couples to leaky SPPs at particular frequencies. Selective filling of individual air holes with metal brings inherently enhanced polarization-dependent transmission which could be eventually used for the development of in-fiber absorbing polarizer or notch filters. 

\begin{figure}
\centering
\includegraphics[width=0.98\linewidth]{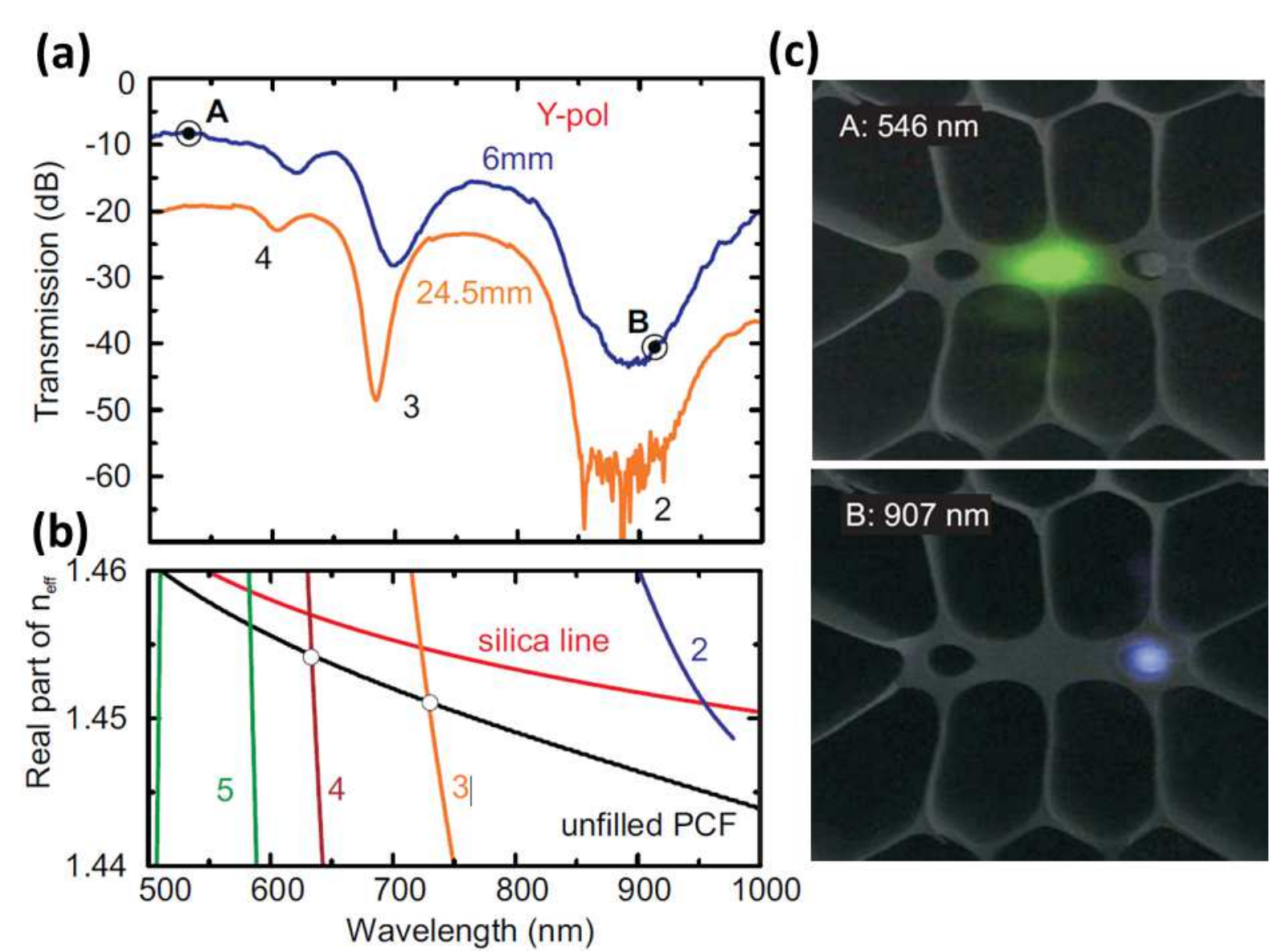}
\caption{(a) Measured transmission spectrum in y-polarization of a 6 mm (blue) and 24.5 mm (orange) length of a PM-PCF. (b) Calculated dispersion of the relevant modes. (Black line) Initial PM-PCF. (colored lines) Modes (m = 2-5), of an isolated wire embedded in silica. (c) Near-field profiles at $546$~nm (point A) and $907$~nm (point B), superimposed on to an SEM of the structure. Adapted from \textcite{Lee2008}.}
\label{fig:plasmon_resonance}
\end{figure}

In 2008, Lee \textit{et al.} demonstrated for the first time how the polarization properties of a highly birefringent PCF can be further improved by incorporating a $\sim$ 6 and $\sim$ 24.5 mm long gold wire in the x-axis of a PM-PCF \cite{Lee2008}. The transmission spectrum of the fiber device shows distinct dips at wavelengths where the core and SPPs phase match and couple as shown in Fig. \ref{fig:plasmon_resonance} (a). The real part of the plasmonic modes (indicated as m=2, 3, 4 and 5) can be calculated analytically by considering an isolated Au nanowire integrated in a silica host. Comparing the experimental with the simulated data (Fig. \ref{fig:plasmon_resonance} (b)), it can be clearly seen that the dips in the transmission spectra correspond to anti-crossings between core mode and plasmonic modes on the wire. Slight blue-shift of the experimental dips are due to narrow air-gaps between the metal wires and the silica glass wall. Figure \ref{fig:plasmon_resonance} (c) shows the near-field profile of the PM-PCF with the integrated metal wire. For the on-resonance state (i.e. the dip in the transmission at $907$~nm), the light is concentrated in the high index metal wire while for the off-resonance state ($546$~nm), the light is confined in the silica core \cite{Lee2012}.  In this work, the authors reported a maximum on-off-resonance ratio of 45 dB measured in a 24.5 mm long sample with a $900$~nm diameter Au wire and maximum resonance loss of  $2.5\ \mathrm{dB/mm}$.  

\begin{figure}
\centering
\includegraphics[width=0.98\linewidth]{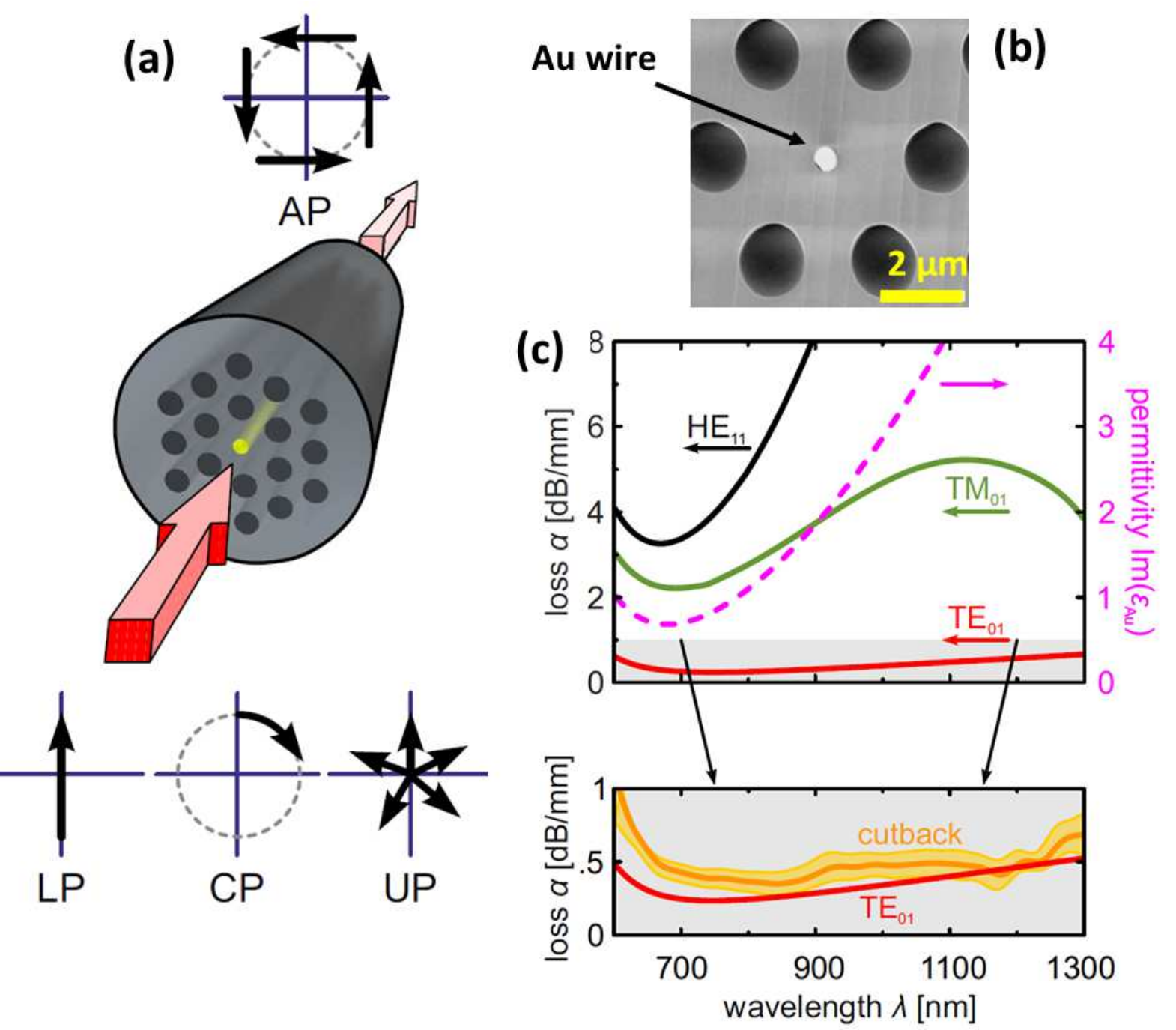}
\caption{(a) Concept of the azimuthal polarization excitation in a metal-filled PCF. (b) SEM image of the PCF with the integrated central Au wire. (c) Calculated loss spectra of the three core modes, together with the imaginary part of the gold (purple dashed line). The gray area is the zoomed section shown the experimental cut-back loss of the azimuthal mode (orange line) compared with theory (red line). The orange-shaded area shows the experimental error range. Adapted from \textcite{Uebel2013}.}
\label{fig:azimuthal_pol}
\end{figure}

In 2011, Uebel \textit{et al.} investigated the polarization properties of the spiral SPPs of metal wires and experimentally demonstrated a novel approach for the generation of azimuthal beams \cite{Uebel2011}. The proposed structure consisted of a gold nanowire running axially through the center of the solid glass PCF core (Fig. \ref{fig:azimuthal_pol} (a) and (b)). This hybrid structure has the property of transmitting only the azimuthally polarized mode, with all other guided modes being strongly absorbed \cite{Uebel2011}. It is worth noting that the output azimuthal polarization was completely independent from the input polarization state as shown Fig. \ref{fig:azimuthal_pol} (a). The polarization extinction ratio was more than 1:10000 over the spectral range from $600$~nm up to $1400$~nm \cite{Uebel2011}. Therefore, it could be used as effective broadband transmission filter for a single, doughnut-shaped mode being azimuthally polarized. The azimuthal polarization properties of the proposed metal-filled PCF are mainly introduced due to the different amount of magnetic field of the three supported modes inside the metal wire. The fundamental and radially polarized modes have about one order of magnitude more field in the gold (and consequently higher loss), only the azimuthal mode thus remains after a certain propagation distance, leading to efficient azimuthal polarization excitation as shown in Fig.\ref{fig:azimuthal_pol} (c). Furthermore the polarization state can be converted from azimuthal to radial by simply using two waveplates \cite{Uebel2011}. 

\textcite{Nagasaki2011} showed numerically that the location of the metal wires in the silica PCF cladding has a crucial effect on the polarization properties of the hybrid structure. They also showed that a large polarization extinction ratio in the PCFs filled with several gold wires aligned parralel to each other can be achieved and they predicted the importance of arranging the metal wires close to each other for high polarization-dependence. Effects arising due to combining more than one metal wire along with the impact of their relative locations with respect to the silica core was then experimentally investigated by \textcite{Lee2012a}. They found that a pair of metal wires being in the close proximity of each other could act as a plasmonic "molecule" where hybridized modes are excited at specific wavelengths by launching light into the glass core. These hybridized modes introduce strong spectral splitting when light couples from the silica core to the nanowire pair mainly due to the hybridization of the plasmonic "orbitals" and the formation of bonding and anti-bonding states \cite{Lee2012a}.\\ 

In conclusion, integration of metal elements inside silica PCFs opened the door to exploitation of novel fundamental effects such as plasmonics.  Furthermore, the use of more than a single metal wire inside PCF could be useful for further investigations such as gap-plasmons, subwavelength interaction of metal nanowire arrays, etc. \cite{Lee2012a}.\footnote{It should be mentioned that interesting investigations have been also recently reported based on step-index fibers with embedded side metal wires or nanotips \cite{Uebel2013a,Tuniz2016}.} 

%%%%%%%%%%%%%%%%%%%%%%%%%%%%%%%%%%%%%%%%%%%%%%%%%%%%%%%%%%%%

\subsection{\label{nonlinear}Nonlinear effects in hybrid solid-core PCF} 

In the previous sections we were focused on how the linear properties of fluids and solids could be efficiently used for the development of sensing and tunable devices. Here, we discuss the nonlinear properties of different liquids that have been combined with silica PCFs for generation of fundamental nonlinear effects such as tunable diffraction and self-defocusing, generation of spatial solitons, nonlinear coupling in an optofluidic fiber and supercontinuum generation. 

\subsubsection{\label{nonlinearities}Nonlinearities in liquids}
\textcite{Shimizu1967} reported for the first time spectral broadening in liquids due to self-phase modulation. The main reason of using liquids for the investigation of nonlinear effects was that the
Kerr nonlinearity can be several times higher than silica (see Table 1). Until a few years ago, the experiments with liquids were mainly performed in liquid cells or capillaries of a few centimeters long \cite{Bridges1982,He1990}. New horizons appeared with the advent of PCFs, have made the interaction of intense light pulses of high peak powers with matter over interaction lengths of hundreds of meters a reality.

The retarded response of any nonlinear material can be described by a function $R(t)$ in the generalized nonlinear Schr\"{o}dinger equation (NLGSE). A number of different physical effects contribute to the response $R(t)$ function and therefore it can be rewritten as \cite{Hellwarth1971, Etchepare1985}: 
\begin{equation}
R(t)=f_{e}\delta (t)+f_{R}h_{R}(t)+f_{m}h_{m}(t)
\end{equation}
The first contribution is due to the electron motion arising from the incoming light. The tightly bound electrons lead to a response time of a few femtoseconds. For a laser pulse with a duration of several hundred femtoseconds, the electron motion is considered instantaneous and thus the temporal dependence can be written as $\delta-$function. The second contribution is coming from the Raman effects. The function $h_{R} (t)$ describes the temporal response \cite{Pricking2011}. A general form of $h_{R} (t)$ has been thoroughly described by \textcite{Pricking2011} in which they show that the fission length of a launched higher-order soliton dramatically increases if the characteristic time of the retarded response is close to the input pulse duration. They also investigate the effect of the retarded response on the soliton self-frequency shift. The final contribution is arising by the molecular re-orientation \cite{Shen2003,Sutherland2003, Vieweg2012}. This effect occurs only in gases and liquids. The molecules are aligned by the incoming light and then provide a significant polarizability through a sequence of delayed full and partial alignment revivals \cite{Shen2003,Sutherland2003}. Depending on the liquid, the delay time can be from femtosecond regime up to a few picoseconds \cite{Levine1975,Kajzar1985, Mcmorrow1988}. This dramatic influence of molecular re-orientation can lead to high nonlinearity of some liquids compared to solids where the main effects are coming only from the electronic part.  The influence of the retarded response of liquids in fluid-filled fibers to the soliton dynamics has been also thoroughly investigated in \cite{conti2010highly, Pricking2011}. 

For the observation of nonlinear effects in liquids, one of the most crucial parameters is their nonlinear refractive index. Furthermore, the absorption of the liquid must be relatively low in order for the nonlinearity to take place after pumping the liquid. For example a material with one of the highest nonlinear refractive indices, is Salol ($\mathrm{C_{13} H_{10} O_{3}}$) (see Table 1). However, it has several O-H groups and therefore it has strong absorption peaks in the near-infrared (NIR), which makes this liquid not suitable for optical experiments, at least for NIR. $\mathrm{CCl_{4}}$ and $\mathrm{CS_{2}}$ on the other hand their nonlinearity is not that high as Salol but they have relatively high transparency in the visible and NIR region. Some of the most popular liquids that have been used for nonlinear experiments are summarized in Table 1.\footnote{The data extracted from\cite{stolen_self-phase-modulation_1978,Ho1979,couris2003,Itoh2004,Vieweg2012}.}

\begin{table}[H]
\caption{\label{Table 1}Nonlinear refractive index of various liquids compared to fused silica.}
\begin{ruledtabular}
\begin{tabular}{cc}
Name & \textit{$n_{2}$} ($\times 10 ^{-20} m^{2}/W)$)\\
\hline

Water ($H_{2}O$)  & 1.3 \\
Ethanol ($C_{2}H_{6}$) & 7.6 \\
Toluene ($C_{6}H_{5}CH_{3}$) & 170 \\
Carbon disulfid ($CS_{2}$) & 514 \\
Salol ($C_{13}H_{10}O_{3}$) & 1540 \\
Chloroform ($CHCl_{3}$) & 31\\
Carbon tetrachloride ($CCl_{4}$) & 15 \\
Fused silica ($SiO_{2}$) & 2.6 \\
\end{tabular}
\end{ruledtabular}
\end{table}

\subsubsection{\label{devices}Nonlinear fiber devices}

\textcite {Rosberg2007} experimentally demonstrated that infiltration of a high index fluid (castor oil) inside the holes of the PCFs can act as thermally tunable waveguides themselves. By increasing the temperature, they showed that the thermal nonlinearity introduce dramatic enhancement of the discrete diffraction or self-defocusing in the 2-D microstructure of the fiber cladding having thus a dominant effect in the output power \cite{Rosberg2007}. The spatial control of light was enabled by the combined effects of discreteness, strong material tunability and nonlinearity, and was independent on any architectural light guiding core defects as in the case of conventional PCF structures \cite{Rosberg2007}. The latter can be used to build a tunable all-optical power limiter \cite{Rosberg2007}. Discreteness of waveguide lattices or  "discrete optics" was first introduced by Christodoulides \textit{et al.} dealing with all devices with more than two coupled waveguides \cite{Christodoulides1988}, which were mostly linearly arranged \cite{Christodoulides2003,Vieweg2012}.

\begin{figure}
\centering
\includegraphics[width=0.98\linewidth]{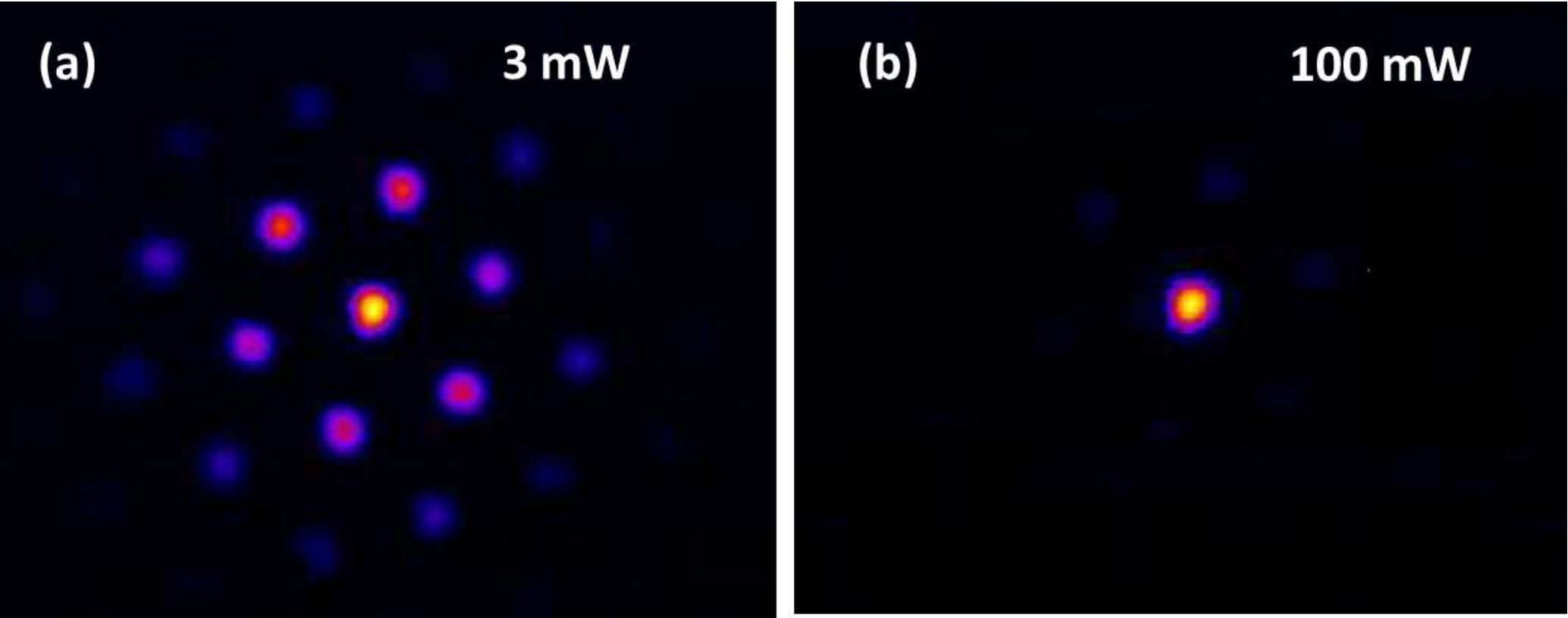}
\caption{Experimentally observed output diffraction pattern and soliton localization in a liquid-infiltrated PCF at (a) low and (b) high input power. Adapted from \textcite{Rasmussen2009}.}
\label{fig:soliton_localization}
\end{figure}

In a similar liquid-filled PCF configuration, \textcite{Rasmussen2009} studied the nonlocal nonlinear interaction between the hexagonal lattice sites and demonstrated for the first time the formation of nonlocal gap solitons in a hybrid PCF. The nonlinearity in their system was defocusing and had thermal origin due to the negative thermo-optic coefficient of the infused liquid, while the nonlocality introduced by the diffusive nature of heat transfer. The authors experimentally verified their predictions by using a commercially available PCF filled with a high index fluid. At a constant temperature of 76$^{\circ}$C, the authors observed a linear diffraction at low input power 3 mW (Fig. \ref{fig:soliton_localization} (a)) corresponding to propagation of approximately three diffraction lengths. Interestingly, at high laser power of 100 mW a clearly nonlinear self-localization observed to almost a single lattice site as shown in Fig.\ref{fig:soliton_localization} (b) \cite{Rasmussen2009}.

In 2012, Vieweg \textit{et al.} reported the formation and power dependent of spatial solitons in an optofluidic PCF in which a waveguide array of 5 strands were selectively filled with the nonlinear $CCl_{4}$ \cite{Vieweg2012a}. The authors observed power dependent spatial soliton formation in this selectively liquid-filled PCF with enhanced nonlinear optical properties \cite{Vieweg2012a}. Moreover, they also demonstrated how the spatial solitons can be influenced and hence controlled by the thermo-optical effect provided by the infused liquid. 

Even though the thermal nonlinearity of liquid-filled PCFs offered the possibility for some interesting investigations within the discrete optics field, the most crucial factor for enhanced nonlinear effects is still the nonlinear or Kerr coefficient of a material. An all-optical switch using the material Kerr nonlinearity to control the switching behavior was first proposed by \textcite{Jensen1982}. In the linear regime, it is well-known that the control of the output power ratio for a fixed length cannot be controlled optically but only by applying an external perturbation. The case is different if the two waveguides being in close proximity exhibit strong Kerr nonlinearity. In conventional fused silica couplers or dual-core fibers \cite{Friberg1987a,Friberg1988}, the nonlinear switching is significantly limited due to the low $n_{2}$. However, the use of highly nonlinear liquids overcomes this limitation offering higher levels of nonlinearity while the host material can be still the well-grounded and mature silica. 

\begin{figure}
\centering
\includegraphics[width=0.98\linewidth]{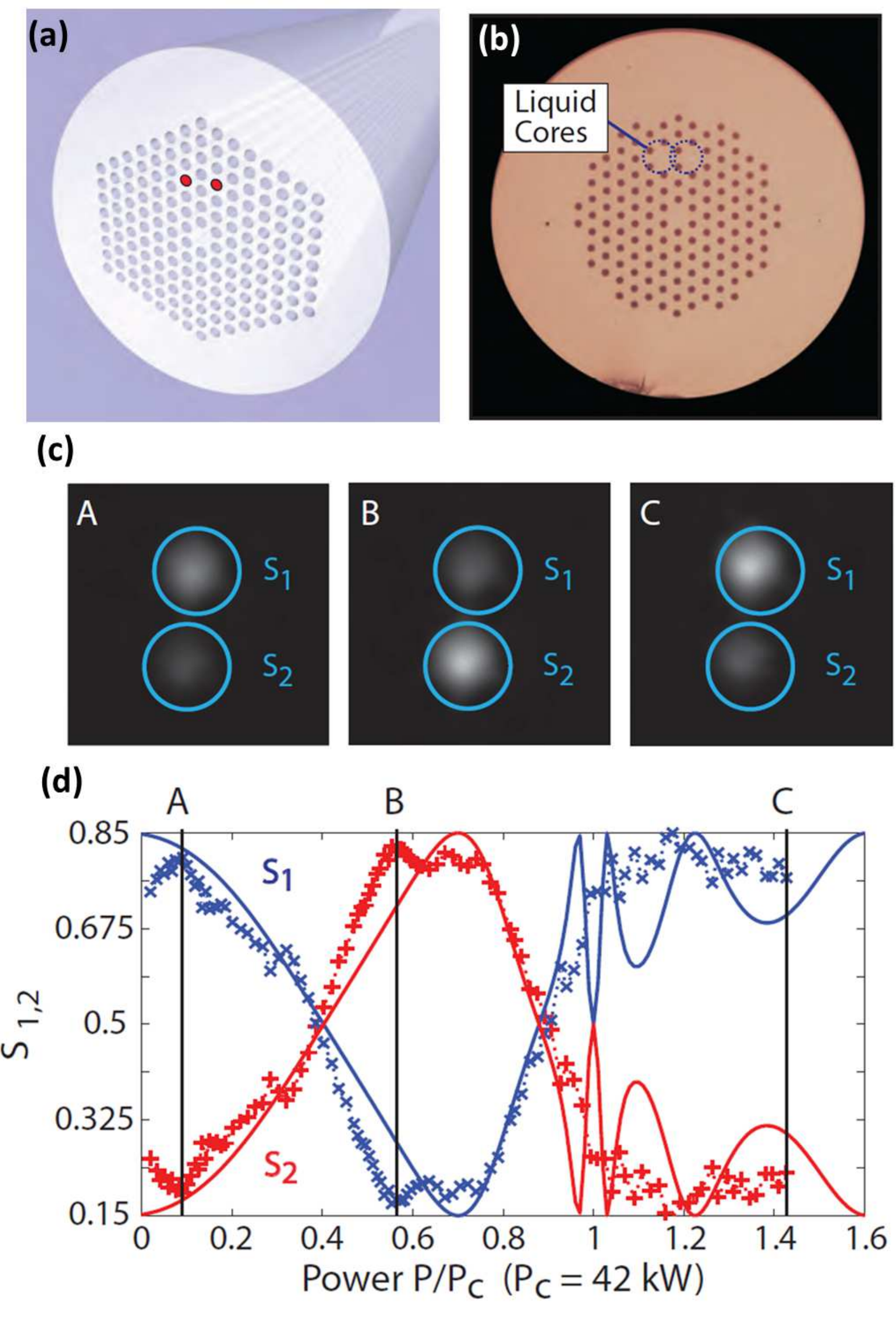}
\caption{(a) Schematic representation of the optofludic coupler. (b) Optical images of the actual filled strands of PCF with $CCl_{4}$. (c) Measured intensity distributions at the output of the nonlinear coupler at different powers: (A) 3.5kW, (B) 24.2kW, and (C) 58.7kW. (d) Experimentally measured and normalized power distribution between the two channels S1 (blue) and S2 (red) of the optofluidics coupler versus input power. (Solid lines) theoretically calculated dependencies of the power distribution. Adapted from \textcite{Vieweg2012}.}
\label{fig:nonlinear_coupler}
\end{figure}
In 2012, Vieweg \textit{et al.} demonstrated a novel optofluidic nonlinear coupler fabricated by selective filling of two strands of a silica PCF with the liquid $CCl_{4}$ as shown in Fig. \ref{fig:nonlinear_coupler} (a) and (b) \cite{Vieweg2012b}. As it can be seen from Table 1, $CCl_{4}$ has a relatively much higher nonlinear refractive index compared to silica ($\sim n_{2} =2.6\times10^{−20} m^{2}/W$) \cite{stolen_self-phase-modulation_1978} and consequently a large ultra-fast Kerr nonlinearity.  By pumping one of the strands with laser pulses ($1040$~nm at 20 MHz repetition rate, 180 fs pulse duration), the authors demonstrated a clear power dependent coupling between two channels (indicated as S1 and S2). The recorded near-field intensity patterns of the two liquid filled holes at different input powers as well as the normalized power distribution of the coupler, is shown in Fig. \ref{fig:nonlinear_coupler} (c) and (d), respectively \cite{Vieweg2012b,Vieweg2012}.

\subsubsection{\label{nonlinear prop.}Nonlinear propagation and supercontinuum generation}

\begin{figure}
\centering
\includegraphics[width=0.98\linewidth]{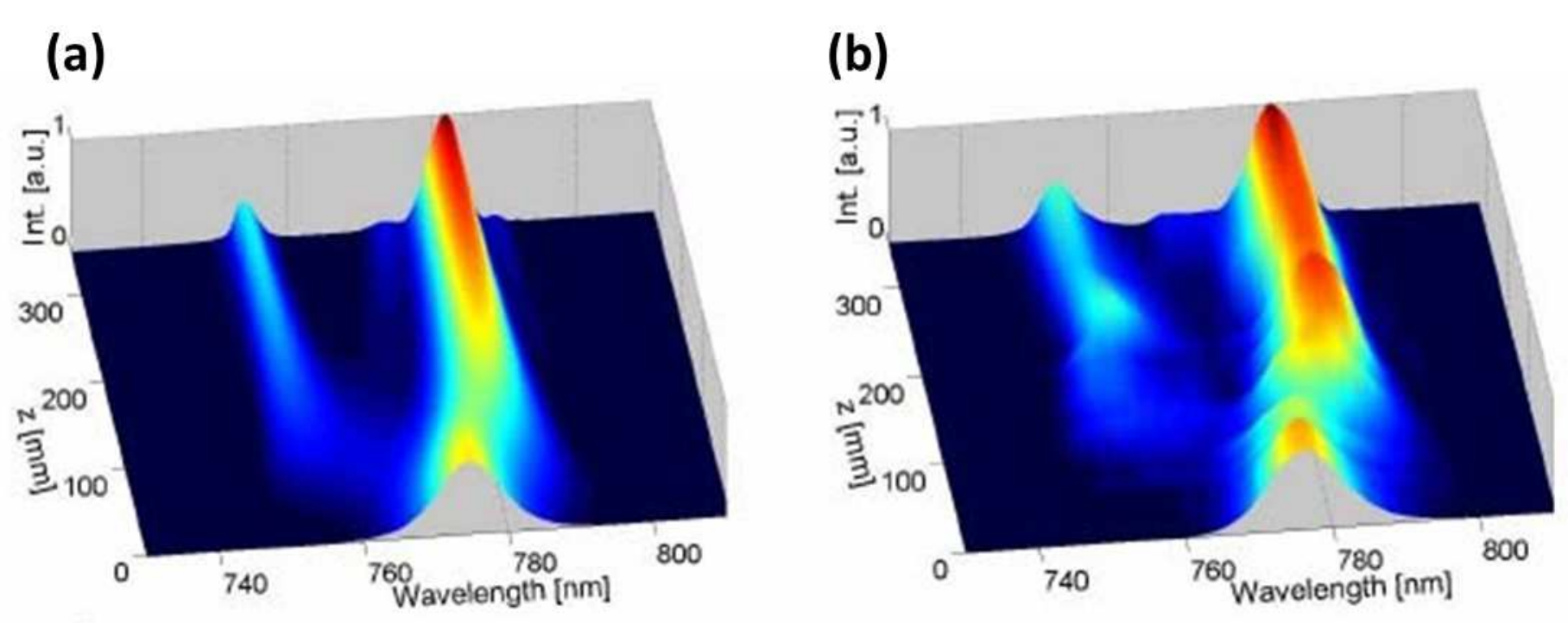}
\caption{(a) Simulated and (b) measured spectral evolution vs. increasing input power of the high-index filled PCF showing the red-shifting generated soliton and the dispersive wave when pumped close to its ZDW. The output was recorded using a Frequency Resolved Optical Gating (FROG) system. Adapted from \textcite{Fuerbach:05}.}
\label{fig:nonlinear_propagation}
\end{figure}

Group velocity dispersion (GVD) has a crucial role in waveguides for the efficient generation of nonlinear effects as it will be also discussed in Section III.D \cite{Ranka2000, dudley_supercontinuum_2006}. For hybrid PCFs, the GVD is a combination of the infiltrated, host material and the waveguide dispersion. \textcite{Fuerbach05} characterized for the first time the nonlinear propagation of ultra-short pulses ($\sim$ 70 fs) from a Ti:Sapphire laser through a high-index fluid-filled PCF. The infused PCF had a zero dispersion wavelength (ZDW) at $\sim$ $770$~nm and therefore soliton propagation at around $780$~nm was observed well below the material’s ZDW. The authors demonstrated, by tuning the laser line around the location of the ZDW, soliton propagation and red shifting, as well as the formation of dispersive waves in the normal dispersion regime \cite{Fuerbach:05}. With increasing power however, self-phase modulation (SPM) compensated the dispersive broadening and therefore a soliton appears. The length of the hybrid PCF was fixed at 180 mm. Figure \ref{fig:nonlinear_propagation} (a) and (b) shows the simulated and measured spectral evolution with respect the input power pumped close to the ZDW, respectively.  A red-shifted soliton appears when the fiber was pumped into the anomalous dispersion regime. The soliton shifts towards longer wavelength with increasing powers, due to the Raman soliton self-frequency shift. On the other side of the ZDW, dispersive waves were also generated \cite{Fuerbach:05}. It should be emphasized that the location of the ZDW wavelength can be controlled based on the infiltrated liquids. Furthermore, the ability to control the refractive index of the infused material using thermal, electrical or optical ways enables new highways for adiabatic soliton compression or optimized supercontinuum generation \cite{Travers2005, Kudlinski2006}. 

The idea of obtaining supercontinuum generation in a PCF whose core is filled with highly nonlinear liquids ($CS_{2}$ and nitrobenzene) was first theoretically proposed by \textcite{Zhang2006}. They predicted that efficient continuum from $700$~nm to more than $2500$~nm can be generated by pumping the hybrid liquid-filled PCF in the anomalous dispersion regime with sub-picosecond pulses at $1550$~nm. Interestingly, \textcite{Bozolan2008} reported first that supercontinuum generation can be even achieved in distilled water. By selectively filling the central hole of a HC-PBG, the guiding mechanism was converted from bandgap to index-guiding \cite{Bozolan2008}. The authors reported spectral broadening from $\sim$600 to $\sim$1140 nm in the water-filled PCF pumped close to the ZDW of the fiber with pulses having $\sim$1.45 MW peak power \cite{Bozolan2008}. 

Two years later in 2012, \textcite{Bethge2010} further investigated the possibility of using water as nonlinear medium and reported two-octave spectral coverage from $410$ to $1640$~nm. The main advantage of the proposed water filled PCF was that compared to continua spectra generated in glass core PCFs, the liquid core supercontinua show a much higher degree of coherence. Moreover, the larger mode field area and the higher damage threshold of the water core enabled significantly higher pulse energies, ranging up to 0.39 $\mu$J \cite{Bethge2010}. Similarly, \textcite{PhanHuy2010} investigated both numerically and experimentally the self-phase modulation in liquid-filled PCFs. 

\begin{figure}
\centering
\includegraphics[width=0.98\linewidth]{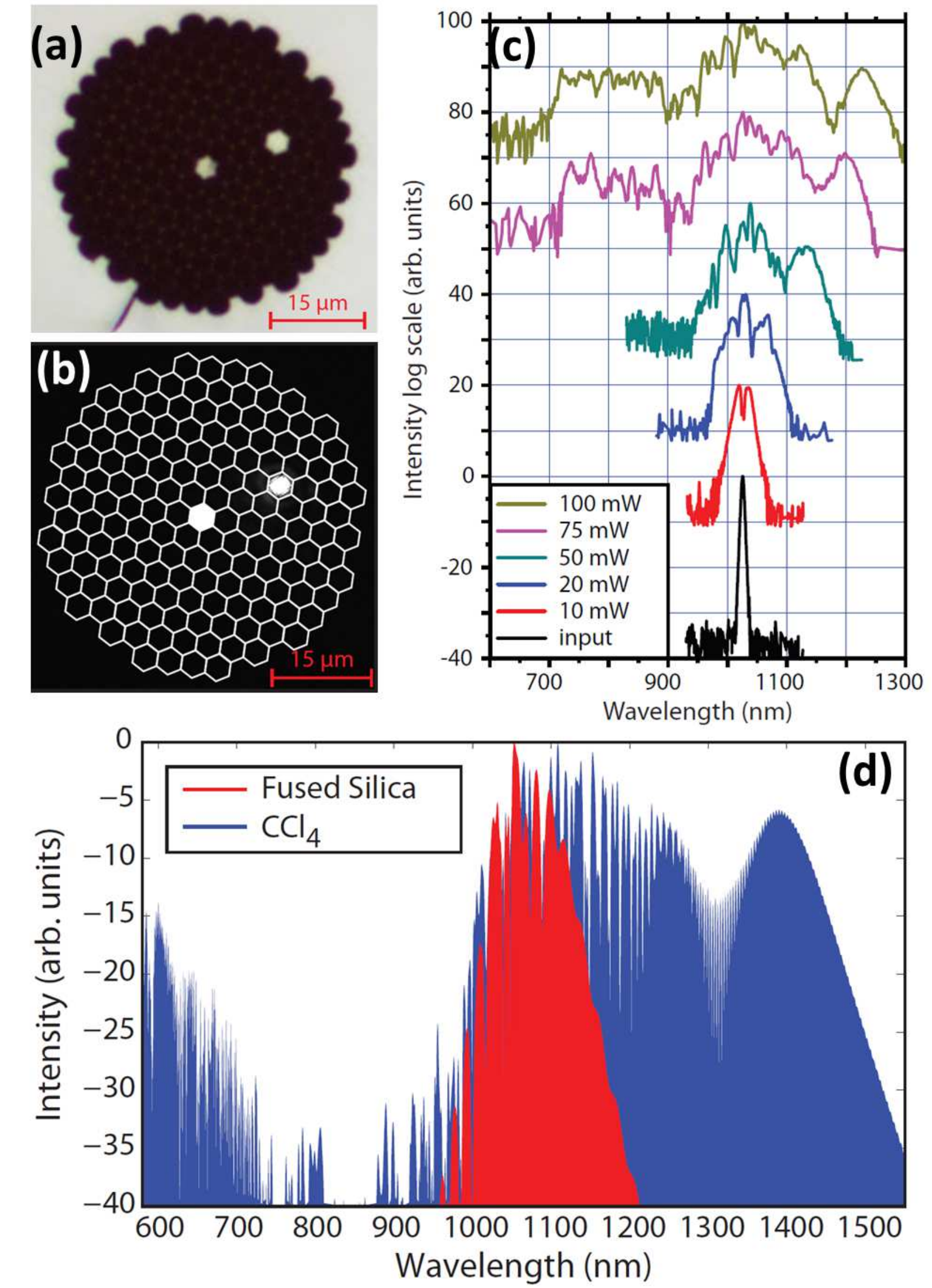}
\caption{(a) Optical image of the selectively single strand filled PCF with $CCl_{4}$. (b) Near field profile superimposed in cross section of the fiber. (c) Spectral evolution for different input powers of a 26 cm long fiber with a single strand filled with and diameter of 2.5 μm. (d) Spectral broadening of the same structure with a glass core (red curve - light grey) and a liquid $CCl_{4}$ core (blue curve - dark grey). For comparison purposes, the linear refractive index and the dispersion properties are deliberately set to be identical. Adapted from \textcite{Vieweg2012}.}
\label{fig:spectral_broadening_liquid}
\end{figure}

\textcite{Vieweg2010a} used $CCl_{4}$ instead of water as the active nonlinear medium for supercontinuum generation. By selectively filling a single strand of a commercial PCF (Fig. \ref{fig:spectral_broadening_liquid} (a) and (b)), the authors used an Yb:KGW oscillator with a repetition rate of 44 MHz ($\sim$ 210 fs long pulses at $1026$~nm) to pump the formed liquid waveguide.  Figure \ref{fig:spectral_broadening_liquid} (c) shows the spectral evolution in a 26 cm long filled single strand with respect the incoming power. For low input powers, SPM takes places while as the power increases then there is soliton formation and the Raman-shift of  solitons become clearly observable \cite{Vieweg2010a}. At full power of 100 mW, the authors obtained a spectral broadening of $600$~nm, covering the spectral region from $700$~nm to $1300$~nm. The spectrum revealed the red-shifting soliton as well as dispersive waves in the blue-edge region \cite{Vieweg2010a}. This spectral broadening could not be achieved simply by using a silica PCF with the same diameter due to lack of high nonlinearity \cite{Vieweg2012}. Using numerical simulations, the authors showed the tremendous influence of the nonlinear refractive index on the generated spectrum. Figure \ref{fig:spectral_broadening_liquid} (d) provides a direct comparison of how the supercontinuum spectrum is significantly reduced if the nonlinear refractive index of silica (red) was used instead of $CCl_{4}$ (blue) keeping all the other parameters the same \cite{Vieweg2012}.

The most important common advantage of all the aforementioned reports is that the ZDW can be tuned accordingly by appropriately choosing the core diameter of the host fiber as well as the active liquid material. This provides a great degree of flexibility to move the ZDW in wavelengths where high power and low-cost laser sources exist. In general, liquid-filled PCFs (including the selectively filled fibers) offer the basis of a plethora of new reconfigurable and versatile nonlinear optical fiber devices with unprecedented performance. There is still plenty of room for improvements as well as future investigations of utilizing for example novel functional nanomaterials or even solution-processed highly nonlinear materials for enhanced nonlinear applications \cite{Kuhlmey2009, Schmidt2016}. 

\section{\label{3}Gas-filled Hollow-core PCF}
\subsection{Introduction}
The study of the interaction of laser light and gases and plasmas is of major importance to science. Section II was focused on the most important advances on hybrid PCFs combined with solid and liquid materials. This section of the article on the other hand summarizes achievements using gas-filled hollow-core photonic crystal fibers (gas-filled \hc{s}). Gases and partial plasmas offer the greatest transparency range of any nonlinear active media and hence enable the largest spectral extent for nonlinear experiments. They also offer access to the nonlinear response of atoms and molecules in their most pure form, often without significant influence from neighboring molecules, a fact highlighted by \textcite{new_optical_1967} in the first report on third harmonic generation in gases. This enables direct tests of quantum mechanical calculations of nonlinear coefficients, simplifies understanding of complex dynamics, and provides a means for nonlinear spectroscopy of basic molecular processes. Gases and plasmas also have much higher damage thresholds than other nonlinear media and are, in a sense, self-healing (i.e. if we ionize the gas it eventually recombines again, unlike solids which become permanently damaged); this allows energy and intensity scaling of nonlinear devices to their most extreme limits. Finally, many molecules of spectroscopic interest, either in the atmosphere or of importance to biology, can be sensed in gaseous form.

A waveguide geometry offers a number of clear advantages. Firstly, there is the usual argument for fiber optics, that we can significantly increase the light-matter interaction, characterized by the intensity-length product $IxL$, compared to any focusing geometry. This enables much lower thresholds and higher efficiencies for what can often be weak nonlinear processes. Secondly, in the correct fibers, the waveguide dispersion can be used to qualitatively alter laser pulse propagation dynamics by introducing anomalous dispersion. Gas media are almost exclusively normally dispersive at optical frequencies. An additional ``control knob'' on the dispersion enables a whole new range of dynamics--most remarkably soliton interactions with gases and plasmas; something not possible in a bulk geometry.

For these reasons, hollow capillary fibers (HCF), made from a single hollow core inside a solid glass cladding have proved extremely successful as hosts for gases and liquids in optical experiments. \textcite{ippen_low-power_1970} already used them for nonlinear optics, and they currently form the basis of intense few-cycle laser pulse compressors in ultrafast science (see Section~\ref{sec:hcfcomp}). By introducing a microstructure to the cladding further control over the propagation losses and dispersion properties of the waveguide can be achieved as described in Section~\ref{B}. The general class of gas-filled \hc{s} has dramatically improved our ability to tailor the context of light-gas interaction. The primary difference to hybrid PCF filled with solids or liquids is that gases are compressible, and hence the gas number density is directly related to, and controlled by, the applied gas pressure. Consequentially the nonlinearity and dispersion of the waveguide system can be tuned almost real-time during an experiment, and perhaps more significantly, without additional fabrication steps.

The potential for interesting devices and effects to be observed in gas-filled \hc{s}, was noted in the first report of such fibers by \textcite{cregan_single-mode_1999}. Since then a very large number of results have been obtained. Here we attempt to provide a holistic review, with most of our focus on nonlinear optics. The reader is additionally directed to existing partial reviews of the topic\footnote{\textcite{benabid_hollow-core_2006,bhagwat_nonlinear_2008,dudley_ten_2009,benabid_linear_2011,travers_ultrafast_2011,russell_hollow-core_2014,saleh_soliton_2016}.}.

\subsection{Types of gas-filled HC-PCF}
There are two primary types of hollow-core PCFs used for gas-based optical devices and experiments, as outlined in Section~\ref{B}: those based on the photonic band-gap effect, which we label as \pgpcf{}, and those based on anti-resonance or inhibited coupling guidance, such as \kpcf{} or single-ring negative-curvature \hc{}, which we refer to in what follows as \bhc{}. This latter class includes a wide range of structures. The most prominent among them for gas-based optics has been the \kpcf{} \hc{}, perhaps with negative curvature in the core (also referred to as hypocycloid core-contour design \cite{Wang2011}). But it must be stressed that in terms of guidance properties, especially dispersion and nonlinear properties, all of the \bhc{s} are extremely similar, and results observed in one kind of fiber will almost certainly be observed in another for similar parameters.

\subsection{Canonical experimental setup}
\begin{figure}
    \centering
    \includegraphics[width=0.98\linewidth]{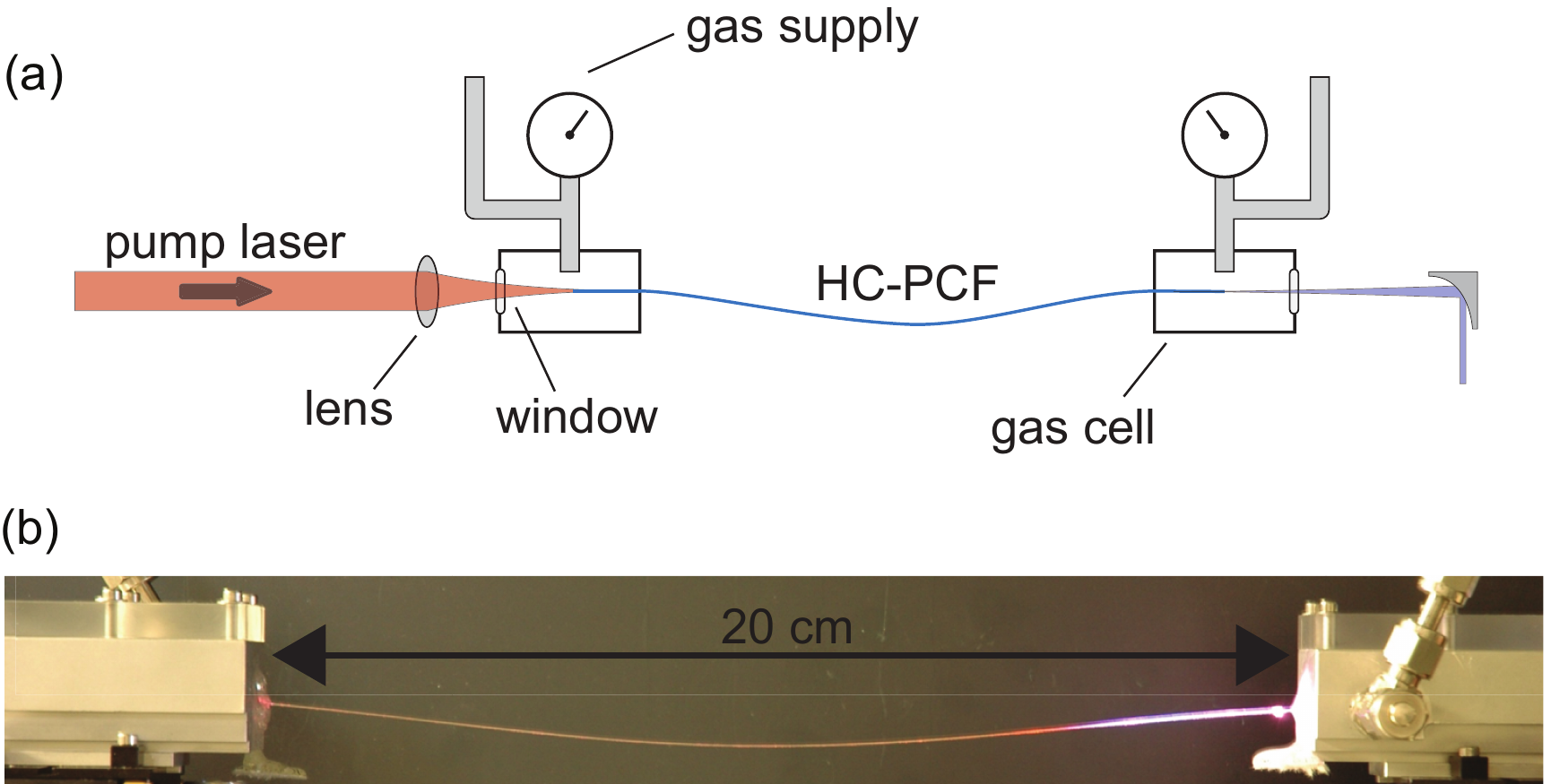}
    \caption{(a) Example of the usual setup for free-space coupling into a gas-filled HC-PCF, allowing for vacuum propagation, and pressure gradients between the cells. (b) Photograph of one realization of the setup, showing side-scattering at the point of UV emission (as described in Section~\ref{sec:dwave_kgm}).}
    \label{fig:gas_setup}
\end{figure}
Fig.~\ref{fig:gas_setup} shows the canonical arrangement for investigating optics in gas-filled HC-PCF. The fiber is held between two gas cells which form a gas and vacuum tight seal between the core and microstructure of the fiber and the ambient atmosphere. Through these gas cells the fiber can be both evacuated and filled with different gases or gas mixtures\footnote{The dynamics of the gas filling have been considered in a number of works, such as \textcite{henningsen_dynamics_2008, dicaire_analytical_2010, ermolov_low_2013, wynne_gas-filling_2015}.}. Pressures over 50~bar can be routinely administered. Additionally, as the only path between the two gas cells is through the fiber itself, a pressure (and hence gas density) gradient can be formed between them, either increasing or decreasing. This allows the optical properties of the fiber as a whole, i.e. the nonlinearity or dispersion, to be modified axially through the fiber, in a similar way to a tapered solid-core fiber. The resulting pressure distribution inside the fiber can be obtained from the standard equations for pressure-volume flow and the conductance of a circular bore to be
\begin{equation}
\label{eqn:gas_pgradient}
p(z)=\sqrt{p_0^2-\frac{z}{L}\left (p_0^2-p_1^2\right )},
\end{equation}
where $z$ is the axial fiber coordinate, $L$ is the fiber length, $p_0$ is the pressure at the input and $p_1$ is the pressure at the output.
If a pressure gradient is not required, the fiber can be placed inside a tube that connects the gas cells. In this way it was shown by \textcite{azhar_raman-free_2013} that extremely high pressures can be reached, exceeding 150 bar. \textcite{lynch-klarup_supercritical_2013, azhar_nonlinear_2013} both demonstrated that HC-PCF can even be filled with Xenon (Xe) in the supercritical state, at where its density increases dramatically (see Fig.~\ref{fig:gas_prop}(a) below). Taking the drive for increased density to a further extreme, \textcite{azhar_nonlinear_2012} demonstrated nonlinear optical interaction inside a HC-PCF filled with liquid Argon (Ar). Although in that initial result, throughput was limited due to the formation of menisci at the gas-liquid transitions, one can imagine ways to circumvent that issue.

In the usual system, shown in Fig.~\ref{fig:gas_setup}, light is coupled in and out from the gas cell through windows, carefully selected and positioned to avoid nonlinear effects of the focusing beam and absorption for the spectral range under interest. Typically, few-mm silica windows are sufficient, but for example, thinner MgF$_2$ windows can be used for VUV applications.

Generalizations of the setup in Fig.~\ref{fig:gas_setup} include delivering the output directly to a differentially pumped vacuum chamber for VUV or XUV experiments \cite{ermolov_low_2013, fan_integrated_2014, tani_wavelength-tunable_2017}, connection to a VUV spectrometer \cite{belli_vacuum-ultraviolet_2015,ermolov_supercontinuum_2015} or avoidance of gas-cells altogether by splicing the HC-PCF directly to solid-core fibers, as shown by \textcite{benabid_compact_2005}.

\subsection{Properties of gas-filled HC-PCF}
\subsubsection{Attenuation}
The key advantage that HC-PCF has over conventional HCF is low intrinsic attenuation and bend loss even for very small core sizes (e.g. core radius $a<10$~\um{}). The record low loss of an HC-PBG is around 1.7~dB/km (Mangan et al. 2004), and for \bhc{} it is usually closer to 0.1~dB/m with recent reports demonstrating losses as low as 7.7 dB/km \cite{Debord:17}. While this latter value is still high compared to conventional fibers, it is sufficient for nonlinear experiments that use a few meters or even just a few centimetres; still providing a very large advantage over free-space focusing. While the low loss allows experiments and devices to use small core sizes, the low bend loss also enables flexible fiber delivery of laser light and probing of difficult to access locations, such as vacuum, hazardous locations (nuclear reactors) or the human body. In contrast HCF has very high bend loss and an attenuation constant that scales approximately (for the HE$_{nm}$ mode) as $\alpha_{HCF}\approx 3u_{nm}^2/k^2a^3$, where $k$ is the wavenumber and $u_{nm}$ is the $m^{\mathrm{th}}$ zero of the Bessel function $J_{n-1}$ \cite{marcatili_hollow_1964}. For a 30~\um{} core diameter, as commonly used with \bhc{} with $\sim 0.1$~dB/m, a HCF will have a loss value of 44~dB/m. Therefore, while HCF is used very successfully for guiding intense (mJ-scale) laser pulses over lengths of a few meters in large cores ($a>100$~\um{}), considerable new nonlinear optical physics and applications can be opened up by using smaller-core HC-PCF.

\subsubsection{Mode propagation constant\label{sec:gas_disp}}
The propagation constant and dispersion of narrowband guiding HC-PCF such as those based on a photonic band-gap (\pgpcf{}) or other narrow-guidance \hc{}, does not have a simple analytical expression. In general one must solve finite element (FEM) simulations or similar to obtain the mode field patterns and dispersion curves.

Broadband-guiding \hc{s} in contrast, such as \kpcf{} \hc{} or single-ring negative-curvature \hc{}, have a modal dispersion that is very well approximated by that of HCF (away from resonances). Following \textcite{stratton_julius_adams_electromagnetic_1941}, a simple analytical expression for the axial mode propagation constant of the HE$_{nm}$ mode of HCF was found by \textcite{marcatili_hollow_1964},
\begin{equation}
\label{eqn:gas_prop_const}
\beta=n_{\mathrm{eff}}k_0=\sqrt{k_0^2n_\mathrm{gas}^2-\frac{u_{nm}^2}{a^2}},
\end{equation}
where $n_{\mathrm{eff}}$ is the effective mode index, $k_0$ is the vacuum wavenumber and $n_\mathrm{gas}$ is the gas density dependent refractive index of the filling gas. Eq.~\ref{eqn:gas_prop_const} is derived under several approximations, including that the core size $a\gg\lambda$, where $\lambda$ is the wavelength under consideration.

The accuracy of Eq.~\ref{eqn:gas_prop_const} for modelling \bhc{} has been verified by multiple FEM analyses \cite{im_guiding_2009,travers_ultrafast_2011} and by a number of direct experiments in the near-infrared (IR), visible and ultraviolet (UV) \cite{nold_pressure-controlled_2010,finger_accuracy_2014}. As HC-PCFs rarely have circular cores there is some ambiguity about the value to assign to $a$ in Eq.~\ref{eqn:gas_prop_const}. \textcite{travers_ultrafast_2011} found that for the hexagonal core shapes of conventional \kpcf{} \hc{} the so called area-preserving core radius (defined so that the resulting circle has the same area as the actual fiber core) gives better agreement for the propagation constant. 

At longer wavelengths, $\lambda > 1$~\um{}, with usual core diameters (i.e. $\sim30$~\um{}) Eq.~\ref{eqn:gas_prop_const} becomes less appropriate as the condition that $a\gg\lambda$ is violated. In that case, \textcite{finger_accuracy_2014} noted that the dispersion estimated using Eq.~\ref{eqn:gas_prop_const} significantly diverges from both measurement and FEM analysis. They introduced a frequency dependent definition of the core size to partially correct for the error. In practice this works well, but its advantage compared to full FEM simulations is somewhat reduced by the fact that, to be trustworthy, it requires some empirical factors that themselves usually require a number of data points from FEM calculations for the specific fiber structure. In addition, it is more usual to use larger core \bhc{} at longer wavelengths \cite{yu_negative_2016}, which mitigates the deviation from Eq.~\ref{eqn:gas_prop_const}.

\subsubsection{Pressure tunable dispersion}
\begin{figure}
    \centering
    \includegraphics[width=0.98\linewidth]{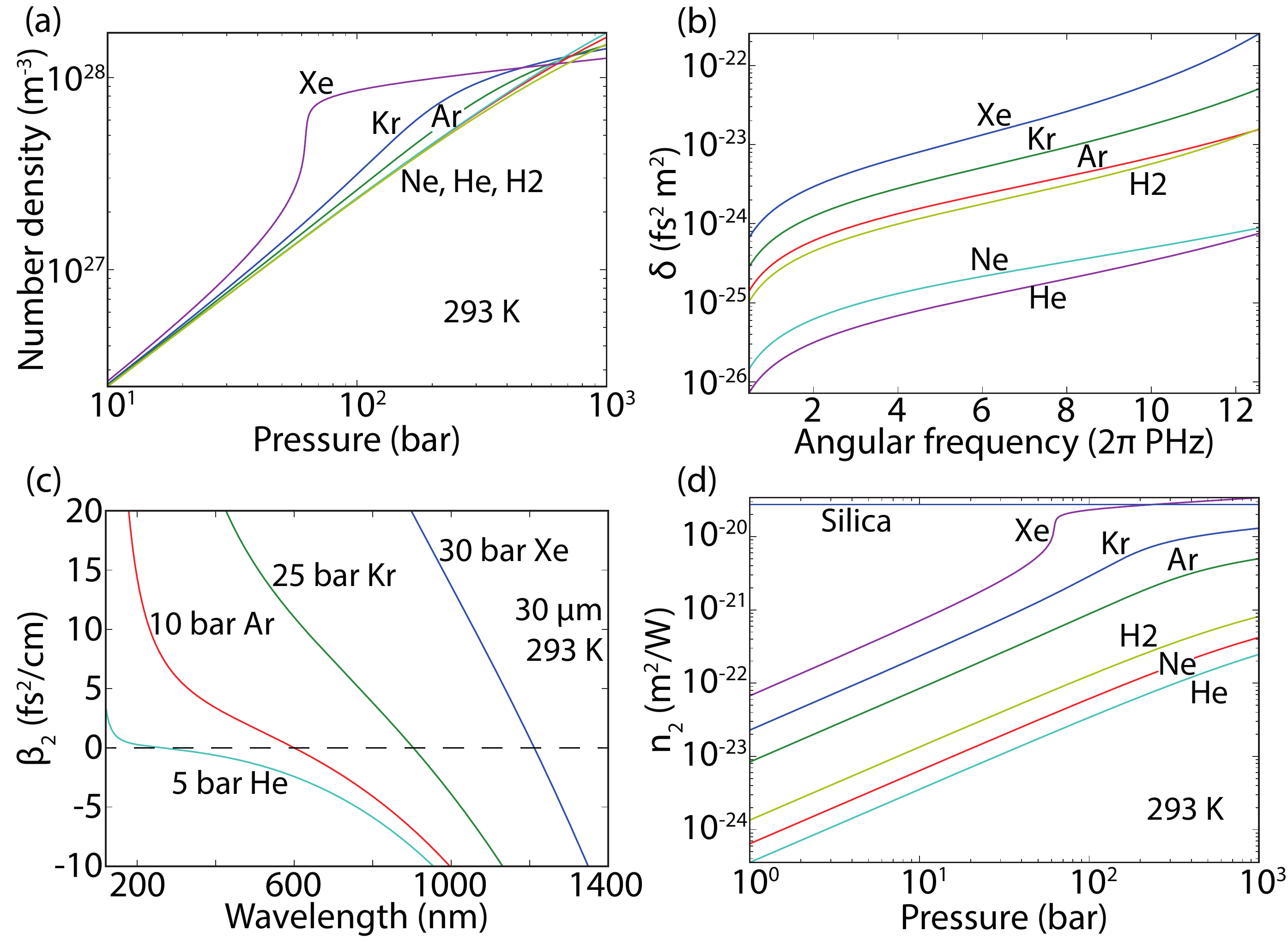}
    \caption{The properties of gases and gas-filled HC-PCF. (a) The variation of number density as a function of pressure, at 293~K, for gases commonly used to fill HC-PCF. He and H$_2$ are close to ideal gases, whereas Xe is far from an ideal gas at pressures over 10~bar. Calculated using REFPROP \cite{lemmon_nist_2013}. (b) Eq.~\ref{eqn:gas_dispersivity} as a function of angular frequency (equivalent wavelength range: 200~nm to 4000~nm) for commonly used gases. (c) GVD curves for various filling gases and pressures at 293~K, in a 30~\um{} core diameter \bhc{}, illustrating tuning of the zero dispersion point from $\sim300$~nm to $\sim1200$~nm. (d) The nonlinear refractive index of common gases as a function of pressure (silica glass is added as a reference line).}
    \label{fig:gas_prop}
\end{figure}

Even for a fixed waveguide structure, Eq.~\ref{eqn:gas_prop_const} allows for easy tuning of the frequency dependent propagation constant, and hence dispersive properties of the fiber, simply by changing the gas pressure. The total refractive index of a filling gas made from multiple species is
\begin{equation}
\label{eqn:gas_polarizability}
n_\mathrm{gas}^2-1=\frac{1}{\varepsilon_0}\sum_tN_t\alpha_t,
\end{equation}
where $N_t$ is the number density of species $t$ and $\alpha_t$ is its frequency dependent linear polarizability, which can be obtained from published Sellmeier equations\footnote{See e.g. \cite{borzsonyi_dispersion_2008,peck_refractivity_1977,shelton_refractive-index_1986}. Note that there is an error in the $C_1$ coefficient for Xe in \cite{borzsonyi_dispersion_2008}, it should be $12.75\times10^{-3}$ rather than $12.75\times10^{-6}$.}. For ideal gases one can replace $N_t$ in Eq.~\ref{eqn:gas_polarizability} with the relative pressure multiplied by a reference density, $N_t=N_{t,0}p_tT_0/p_0T_t$ where $p_t$ and $T_t$ are the partial pressure and temperature of species $t$, and $N_{t,0}$, $p_0$ and $T_0$ are for the given standard conditions. However, care must be taken when working with gases that are not ideal. In particular, Xe and Kr at room temperature and above 10 bar behave far from ideal, as shown in Fig.~\ref{fig:gas_prop}(a).

The modal GVD is given by $\beta_2=\partial^2\beta/\partial\omega^2$, using Eq.~\ref{eqn:gas_prop_const} we find that
\begin{equation}
\label{eqn:gas_GVD}
\beta_2=\sum_tN_t\delta_t-\frac{u_{nm}^2c}{\omega^3a^2},
\end{equation}
where $\omega$ is the angular frequency, $c$ is the vacuum speed of light, and
\begin{equation}
\label{eqn:gas_dispersivity}
\delta_t=\frac{1}{\varepsilon_0c}\left(\frac{\partial\alpha_t}{\partial\omega}+\frac{\omega}{2}\frac{\partial^2\alpha_t}{\partial\omega^2}\right)
\end{equation}
is a quantity that depends on the gas species (through the molecular polarizability and its frequency dependence) only, and not the gas pressure. It is plotted in Fig.~\ref{fig:gas_prop}(b) for a number of common gases.
%  \begin{figure}
%  \centering
%  \includegraphics[width=0.98\linewidth]{gas_dispersivity}
%  \caption{Eq.~\ref{eqn:gas_dispersivity} as a function of angular frequency (equivalent wavelength range: 200~nm to 4000~nm) for commonly used gases.}
%  \label{fig:gas_dispersivity}
%  \end{figure}

The second term of Eq.~\ref{eqn:gas_GVD} shows that the total GVD of a gas-filled HC-PCF has an anomalous (negative) component that increases strongly for small core sizes and low frequencies. From Fig.~\ref{fig:gas_prop}(b) we see that $\delta_t$ is always normal (positive) for optical frequencies. Therefore, by balancing the core size and gas pressure (through $N_t$) one can tune the zero dispersion point, $\omega_{\mathrm{ZD}}$ given by $\beta_2(\omega_{\mathrm{ZD}})=0$, of the overall mode dispersion. As an example, Fig.~\ref{fig:gas_prop}(c) shows dispersion curves for some common fiber parameters. The zero dispersion point is easily tunable from $\sim300$~nm to $\sim1200$~nm, and the dispersion values are of the order 10~fs$^2$/cm over a wide range. These parameters are ideal for ultrafast nonlinear dynamics, and this facility for tuning the dispersion landscape is of key importance in tuning and controlling the nonlinear dynamics that occur inside the fiber, as we shall see in subsequent sections.
%  \begin{figure}
%  \centering
%  \includegraphics[width=0.98\linewidth]{gas_GVD}
%  \caption{GVD curves for various filling gases and pressures at 293~K, in a 30~\um{} diameter \arpcf{}, illustrating tuning of the zero dispersion point from $\sim300$~nm to $\sim1200$~nm.}
%  \label{fig:gas_gvd_total}
%  \end{figure}

\subsubsection{Third order nonlinearity of the filling gas}
The instantaneous third order nonlinear susceptibility of a gas $\chi^{(3)}$, directly scales with its number density \cite{butcher_elements_1991}, thus the total $\chi^{(3)}$ of a gas mixture is simply the weighted sum of its constituents
\begin{equation}
\chi^{(3)}=\sum_t\frac{N_t}{N_{t,0}}\chi_0^{(3)},
\end{equation}
where $\chi_0^{(3)}$ and $N_{t,0}$ are known for each gas for some standard conditions\footnote{See e.g. \textcite{lehmeier_nonresonant_1985,shelton_nonlinear-optical_1990,wahlstrand_high_2012}.}. We can define an associated nonlinear refractive index, as usual through $n_2=3\chi^{(3)}/4\varepsilon_0 cn^2$.

%\begin{figure}
%\centering
%\includegraphics[width=0.98\linewidth]{gas_n2}
%\caption{The nonlinear refractive index of common gases as a function of pressure.}
%\label{fig:gas_n2}
%\end{figure}
The number densities shown in Fig.~\ref{fig:gas_prop}(a) scale $\chi^{(3)}$, so for Xe the increase in nonlinearity is significantly super-linear with gas pressure. Fig.~\ref{fig:gas_prop}(d) shows $n_2$ for common gases as a function of pressure. Note that supercritical Xe can have a nonlinear Kerr coefficient exceeding that of fused silica, but without the Raman contribution. This can be useful for quantum optics applications which require high nonlinearity for e.g. parametric generation, but for which Raman is a source of noise photons (see Section~\ref{sec:gas_mi}).

\subsubsection{The effective mode area}
The effective mode area of a mixed material PCF needs to take care of the differing light-field intensities in different fiber regions, and the associated differences in $n_2$. One approach is to use a material specific effective area
\begin{equation}
\label{eqn:gas_aeff}
A_{\mathrm{eff}}^\nu=\frac{\left(\iint \mathbf{e}_j^\perp\cdot \mathbf{e}_j^\perp d\mathbf{r}_\perp\right )^2}{\iint_\nu \left( \mathbf{e}_j^\perp\cdot \mathbf{e}_j^\perp \right)^2d\mathbf{r}_\perp},
\end{equation}
with the lower area integral over regions of different materials labelled by $\nu$. In Eq.~\ref{eqn:gas_aeff} $\mathbf{e}_j^\perp$ are the transverse electric fields of the mode $j$, and $\mathbf{r}$ describes the transverse coordinate.

For \hc{s} or other \bhc{s} where the light is almost exclusively guided in the filling-gas\footnote{Typically, the peak intensity in the glass is more than three orders of magnitude lower, and the total area of the light-glass overlap is five orders of magnitude smaller \cite{travers_ultrafast_2011}.}, the integrals are effectively over the core-region with the filling gas alone, and a very good approximation for effective area of the HE$_{11}$ mode can be found from the analytic mode functions of HCF \cite{marcatili_hollow_1964}, to be $A_{\mathrm{eff}}\approx 1.5a^2$. For a typical \bhc{} core radius of 15~\um{}, the effective area is $\sim340$~\um{}$^2$. This should be compared to solid-core PCF which can have effective areas smaller than 1~\um{}$^2$.

For \pgpcf{} fibers the integrated intensity over the glass region of the fiber can become quite large, and the glass nonlinearity can become non-negligible. \textcite{luan_femtosecond_2004} first noticed this and found that the nonlinear contributions from the two regions can be almost equal. Therefore there are usually two contributions for \pgpcf{} fiber, one from the filling gas, as in \bhc{}, and the other from the glass material making up the fiber structure. The approach of Eq.~\ref{eqn:gas_aeff} was first used to assess the total nonlinearity on \pgpcf{} fiber by \textcite{luan_femtosecond_2004} and has also been used in numerical studies by \textcite{fedotov_ionization-induced_2007} and \textcite{laegsgaard_dispersive_2008}. \textcite{hensley_silica-glass_2007} experimentally showed that the contribution of glass to the nonlinearity of a \pgpcf{} can vary by nearly an order of magnitude depending on the structure, between an order of magnitude or just a factor of two smaller than the air in the core.

\subsubsection{Nonlinear coefficient}
The nonlinear coefficient $\gamma$ conveniently describes the nonlinear phase acquired as a function of modal power through\footnote{The sign of this phase-modulation is ambiguous and depends on the definitions used when driving the propagation equations. Here we follow the common definition in the fiber optics community.}
\begin{equation}
\label{eqn:phase} 
\phi_\mathrm{NL}(t)=\gamma P(t)L_{\mathrm{eff}},
\end{equation}
where $P(t)$ is the instantaneous pulse power and $L_{\mathrm{eff}}=[1-\exp(-\alpha L)]/\alpha$ is the effective medium length (with $L$ the physical length) accounting for linear attenuation $\alpha$.

Using the above definition for the effective area, the total nonlinear coefficient is given by
\begin{equation}
\gamma= \sum_\nu\gamma_\nu=k_0\sum_\nu \frac{n_2^\nu}{A_{\mathrm{eff}}^\nu}.
\end{equation}

Compared to conventional fiber and solid-core PCF, $A_{\mathrm{eff}}$ is much larger\footnote{There are exceptions, such as very large mode area solid-core PCF.} and $n_2$ is usually lower\footnote{Except for example, supercritical Xe.}, and so the nonlinear coefficient is usually substantially smaller. The consequence is that HC-PCF can be used either for high-power linear pulse delivery, or for extremely high power nonlinear optics. Whereas nonlinear optics in solid-core fiber usually operates around the 10~kW level (or indeed much lower), in \hc{} 100~MW is more usual, with some experiments exceeding 1~GW.  

\subsection{Pulse delivery and power handling in \hc{}\label{sec:delivery}}
One of the first uses of the low nonlinearity of HC-PCF was for laser pulse delivery. \textcite{konorov_laser_2003} delivered trains of $\sim$20-40~ps pulses (separated by 8~ns) with a total energy up to 1~mJ through a \pgpcf{} with 14~\um{} core diameter, albeit with low transmission. Subsequently \textcite{shephard_high_2004,shephard_improved_2005} delivered 0.5~mJ, 65~ns pulses through \pgpcf{} for laser micro machining applications.

Recently it has become clear, somewhat surprisingly, that \bhc{} can be made with extremely low loss in the infrared spectral region, far beyond the silica transmission window \cite{yu_spectral_2013,yu_negative_2016}. Capitalising on this \textcite{urich_delivery_2012} delivered 14~mJ pulses at the Er:YAG wavelength of 2.94~\um{}, well beyond the normal transmission band of bulk silica.

\textcite{humbert_hollow_2004} noted that larger core (19-cell) \pgpcf{}, while handling larger optical powers, have narrower transmission bandwidths than 7-cell designs, and hence are less suitable for guiding ultrashort pulses. In fact, the broad bandwidths supported by \bhc{} are naturally more suited for ultrafast beam delivery, and recently continuous ps pulse delivery in the green spectral region was demonstrated by \textcite{debord_ultra_2014}. They delivered $\sim$300~nJ, 27~ps pulses at 515~nm over 2~m. \textcite{jaworski_high_2015} further improved on this, delivering 0.57~mJ ns pulses and 30~$\muup$J, 6~ps pulses in the green spectral region. Further work on average power scaling of ps pulse delivery is ongoing \cite{michieletto_hollow-core_2016}.

At even shorter wavelengths, continuous-wave delivery has been demonstrated at 280~nm in the deep-UV \cite{gebert_damage-free_2014}, a region that is more difficult for solid-core fibers due to photo-induced material damage. In addition a large number of experiments generating ultrashort pulses in the deep and even vacuum UV (down to 113~nm) have been reported (see Section~\ref{sec:dwave_kgm}), indicating that delivery beyond the bulk material transmission window is possible at both spectral extremes. 

High energy femtosecond pulse delivery was recently reported by \textcite{debord_multi-meter_2014}, who transmitted $\sim$600~$\muup$J of energy ($\sim$1 mJ input) in 600~fs pulses over 10~m of \bhc{} filled with 3~bar He. Even larger core fibers have been predicted to handle even larger pulse energies \cite{Debord:15}.

It should also be noted that all of the nonlinear experiments reviewed below also constitute a form of laser pulse delivery, with pulse durations $< 5$~fs and peak powers even exceeding 2~GW demonstrated.

In terms of average power handling, there have been many reports of operating \bhc{} at the level of multiple 10's of Watts, and recently, \textcite{hadrich_scalability_2016} established that \bhc{} could handle killowatt average power levels. 

\subsection{Self-phase modulation and pulse compression}
Instead of high power pulse delivery, the relatively low (compared to solid-core fiber) nonlinearity of HC-PCF can be used for even higher power nonlinear optics as described in the many following sections. Of these, high energy ultrafast pulse compression is perhaps the most successful application of HC-PCF so far.

First demonstrated by \textcite{fisher_subpicosecond_1969}, the compression of laser pulses using self-phase modulation (SPM) to broaden the spectrum with an approximately linear chirp, followed by chirp compensation using a delay line, such as a grating pair \cite{treacy_compression_1968}, has become a central tool in ultrafast laser systems. The inherent advantages of using an optical waveguide geometry, which enables one to increase the intense nonlinear interaction length while staying below the intensities for spatial effects such as self-focusing, where recognized by \textcite{ippen_selfphase_1974} and first utilized for compression by \textcite{nakatsuka_nonlinear_1981} and \textcite{shank_compression_1982}.

The maximum of the phase shift acquired by an optical pulse in a fiber is determined from Eq.~\ref{eqn:phase} to be $\phi_m=\gamma P_0L_{\mathrm{eff}}=L_{\mathrm{eff}}/L_{\mathrm{NL}}$
where $P_0$ is the peak power and the second equality defines $L_{\mathrm{NL}}$ to be the length after which $\phi_m=1$~rad. \textcite{pinault_frequency_1985} estimated\footnote{Following a number of earlier results, such as \textcite{fisher_subpicosecond_1969} and \textcite{stolen_self-phase-modulation_1978}.} that as the phase follows the pump intensity, the spectral broadening factor is $\Delta\omega/\Delta\omega_0\approx[1+(0.88\phi_m )^2 ]^{1/2}$. Following \textcite{fisher_subpicosecond_1969}, this implies, assuming complete phase compensation, a temporal pulse compression factor of
\begin{equation}
\label{(eqn:gas_spm_comp}
F_c\approx0.88\phi_m=0.88\frac{L_{\mathrm{eff}}}{L_{\mathrm{NL}}},
\end{equation}
where we have assumed $\phi_m\gg1$.

At high pulse energies the use of glass-core optical fibers for ultrafast SPM based pulse compression becomes impossible, as the associated peak powers exceed those for optical damage, or the critical power required for self-focusing and spatial degradation of the beam, $P_{\mathrm{cr}}\approx2\pi/k_0^2n_0n_2$ \cite{fibich_critical_2000}. For silica this limit is $\sim4$~MW at 800~nm. This limit depends on power, not intensity, so simply increasing the core size does not overcome it.

\subsubsection{Compression using hollow capillary fibers for SPM}
A major advance in ultrafast laser science occurred when \textcite{nisoli_generation_1996} demonstrated that using gas-filled hollow capillary fibers (HCF), with bore sizes $>150$~\um{} diameter, for SPM, enables pulse compression of mJ scale pulses to $\sim10$~fs. In a further refinement \textcite{nisoli_compression_1997} demonstrated compression to $<5$~fs. Subsequently many results followed, and such short pulses have been achieved at over 5~mJ (Bohman et al. 2010). These pulses are the primary drivers of intense ultrafast laser experiments, including most attosecond experiments\footnote{\textcite{brabec_intense_2000,krausz_attosecond_2009}.}.

However, for compressing at lower peak powers, just above what is possible with solid-core fibers, HCF suffers from a major limitation, in that its propagation loss scales very unfavorably for the smaller cores required to obtain sufficient nonlinear phase shift. The record low power experimentally demonstrated to date is 100~MW \cite{mansour_generation_2006}.

\subsubsection{Compression using HC-PCF for SPM \label{sec:hcfcomp}}
\begin{figure}
    \centering
    \includegraphics[width=0.98\linewidth]{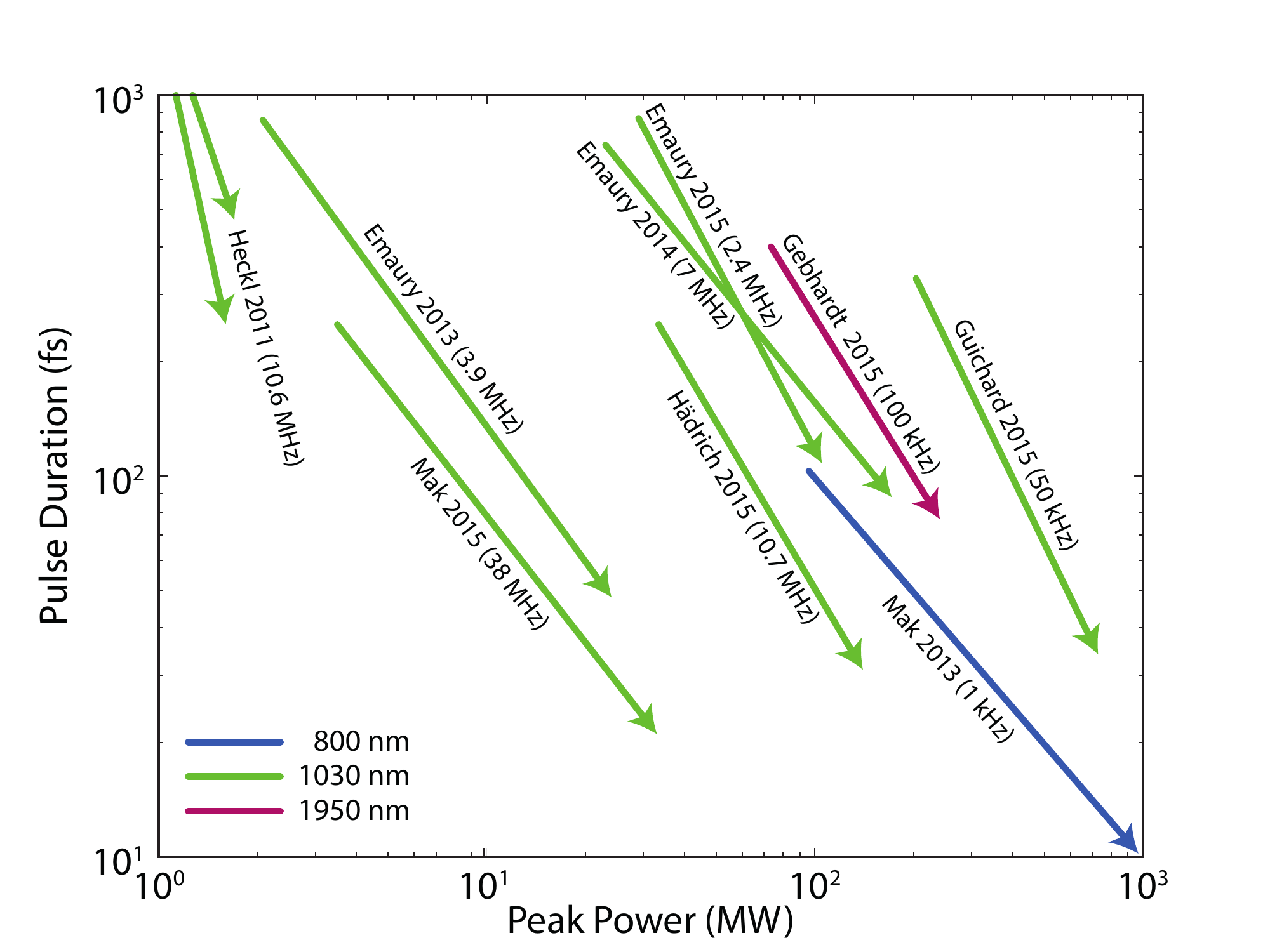}
    \caption{Pulse compression results in gas-filled \bhc{} or similar fibers using SPM spectral broadening and chirped mirrors for phase compensation. The arrows start at the input pulse duration and peak power and point to the final compressed pulse duration and peak power. The references and repetition rates are indicated beside the arrows.}
    \label{fig:gas_spm_comp}
\end{figure}
In contrast, the smaller propagation losses of HC-PCF for small core sizes allow for the use of much lower pressures and shorter fiber lengths, in addition HC-PCF has the significant advantage of considerably lower bend loss. They have therefore been widely used for pulse compression in the intermediate regime between solid-core fiber or HCF, with input peak powers between 1~MW to several 100~MW, as summarized in Fig.~\ref{fig:gas_spm_comp}. This power range is of growing importance due to the increasing availability of high repetition rate, fiber or thin-disk based, laser sources in this region. These sources tend to provide pulses in the range of several 100~fs, and hence must be compressed for many ultrafast applications that require few-cycle pulses.

SPM of sub-$\mu$J pulses was first observed in a PBG by \textcite{konorov_self-phase_2004}, broadening 100~fs pulses with an 8~nm bandwidth to more than 35~nm. However, the output was not compressed. Broadband-guiding HC-PCF, such as \kpcf{} or single-ring negative-curvature \hc{}, is much more suited to pulse compression as it supports the bandwidths necessary for short pulse compression.

The first demonstration using \bhc{} was by \textcite{heckl_temporal_2011}, who compressed 1~ps pulses with $\sim1$~MW at 10~MHz, down to 250~fs in a 36~\um{} diameter fiber filled with 8~bar Xe. The same group subsequently published several papers with similar pulse durations, but increasing peak power, and over 100~W average power \cite{emaury_beam_2013,emaury_efficient_2014}. \textcite{mak_two_2013} demonstrated compression, at 800~nm, from $\sim100$~fs to 10~fs at $\sim10$~$\muup$J--the shortest pulses and highest peak power ($\sim1$~GW) so far achieved using SPM-mirror compression.  \textcite{mak_compressing_2015} further demonstrated two stage compression, with the result of an SPM-mirror compressor (21~fs, 33~MW peak power, 28~W average power) as input to a soliton-effect compressor (to be described in Section~\ref{sec:fsscomp}). \textcite{guichard_nonlinear_2015} used simply an air-filled \kpcf{} \hc{} for convenient spectral broadening, and compression to 0.7~GW, 34~fs pulses.

\textcite{kottig2017generation} compressed 20.5~$\mu$J, 329~fs pulses from a fiber laser system operating at 1030~nm, to 24~fs with 17.4~$\mu$J. At the 1.92~MHz repetition rate, this corresponded to 33.5~W. In that work the pump pulses were converted to circular polarization before the \bhc{} to reduce the effects of ionization and also the strength of the Kerr nonlinearity. This reduction was used so that a higher gas pressure could be used for the same spectral broadening. In this way the total group velocity dispersion at the pump wavelength can be reduced, improving the quality of the compressed pulses after conversion back to linear polarization and phase compensation.

\textcite{gebhardt_nonlinear_2015} proved that \bhc{} can also be used for compression at 1950~nm in the thulium-fiber band and \textcite{murari_kagome-fiber-based_2016} demonstrated pulse compression of 1.8~ps pulses at 2050~nm from an Ho:YLF amplifier to 300~fs.

Finally, both \textcite{hadrich_exploring_2015,emaury_compact_2015}, used the results of compressing high repetition rate sources, with \bhc{}, for the generation of high-harmonics, improving the photon flux in the extreme UV.

\subsubsection{Using \pgpcf{} as the linear dispersive element}
While the dominant role of HC-PCF in pulse compression at present is as the spectral broadening component, its first use for pulse compression was in fact as the linear delay line to compensate the phase of pulses broadened through SPM in solid core fibers. \textcite{de_matos_all-fiber_2003,de_matos_all-fiber_2004,limpert_all_2003}, demonstrated that the strong anomalous dispersion of \pgpcf{} fibers, and low nonlinearity compared to solid-core fibers, meant that they could be used instead of gratings or chirped mirrors, allowing an all-fiber setup for compressing fiber laser systems and all-fiber chirped pulse amplifiers.

\subsection{Soliton propagation in HC-PCF}
The waveguide contribution to the dispersion of HC-PCF modes enables the GVD to be anomalous, even though most of the light is propagating in gas or vacuum. This has profound consequences for the resulting nonlinear propagation dynamics, as it allows for optical soliton propagation, as first discovered by \textcite{zakharov_exact_1972}. Soliton propagation in solid-core fibers, predicted by \textcite{hasegawa_transmission_1973} and demonstrated by \textcite{mollenauer_experimental_1980}, have been very widely explored \cite{taylor_optical_2005}, leading, amongst many other processes to the formation of bright white-light supercontinua \cite{dudley_supercontinuum_2006}. Until 2003, no temporal optical solitons had been observed in gases, as they are mostly normally dispersive at common optical laser frequencies. Filling HC-PCF with gases removes this limitation, and has enabled a host of new nonlinear soliton dynamics in gases, combining many of the dynamics previously explored in solid-core fibers, such as soliton-effect self-compression and dispersive-wave emission, with the qualitatively different material responses of gases; in particular the molecular response and the fact that photoionization and plasma-effects can be explored without permanent material or waveguide damage.

Soliton dynamics can occur in the anomalous dispersion regime\footnote{When considering bright optical solitons in a medium with positive Kerr nonlinearity--which is most gases at visible-NIR optical frequencies.} when the soliton order $N > 1/2$. $N$ is given by\footnote{This is precise only for input pulses with a $\mathrm{sech}^2(t/\tau_0)$ pulse shape, but is a reasonable approximation in other cases.}
\begin{equation}
\label{eqn:gas_sol_order}
N^2=\frac{L_\mathrm{D}}{L_{\mathrm{NL}}}=\frac{\gamma P_0\tau_0^2}{|\beta_2|}
\end{equation}
where $\tau_0=\tau_{FWHM}/1.763$ is the duration of the input pulse, and $L_\mathrm{D}=\tau_0^2/|\beta_2|$ is the length over which linear dispersive effects alter the phase by 1~rad.

Tuning the gas pressure changes $\beta_2$ and $n_2$, whereas tuning the core size changes $\beta_2$ and $A_{\mathrm{eff}}$. Thus significant control over the dispersion, soliton order and the power required for soliton dynamics can be obtained simply by tuning the core size and gas pressure.

\subsubsection{Solitons in \pgpcf{} \label{sec:solsinpbf}}
The much smaller $\gamma$ in HC-PCF, compared to solid-core fibers, means that soliton dynamics occur at much higher peak powers. In \pgpcf{} fiber the dispersion is also very large (i.e. -30~ps$^2$/m) and anomalous over most of its transmission band, further enhancing this scaling.
This was exploited by \textcite{ouzounov_generation_2003}, who first demonstrated soliton propagation in gas-filled \pgpcf{} fiber using either air or Xe as the filling gas, and pumping around 1425 nm. The key result was the observation that the output pulses in Xe are almost unchanged from the 75~fs pulses at the input, despite propagating though 1.7~m of fiber with a peak power of 5.5~MW.

In air, a red-shift of the solitons was observed, indicating a soliton self-frequency red-shift (SSFS) reminiscent of that observed in solid-core fibers \cite{dianov_stimulated-raman_1985,mitschke_discovery_1986,gordon_theory_1986}. The effect occurs for fundamental optical solitons that have a sufficient coherent bandwidth for impulsive Raman scattering which in turn leads to a red-shift of the pulse, as explained in Section~\ref{sec:coherentraman}. The stability of the soliton effect maintains the integrity of the pulse shape as it downshifts in frequency.

These early results were confirmed over a longer, 5 m, length by \textcite{luan_femtosecond_2004}, who also noted the soliton shift. Subsequently \textcite{ivanov_frequency-shifted_2006} used frequency shifted MW solitons around 620~nm as the Stokes pulse in two-color CARS experiments (see Section~\ref{sec:cars}).

A dramatic example of the soliton self-frequency shift in air-filled \pgpcf{} fiber was the work by \textcite{gerome_high_2008}, who demonstrated a Raman-soliton self-frequency shift from $\sim$1070 to 1100~nm in an 8~m \pgpcf{} fiber. 

\subsubsection{Soliton self-compression in \pgpcf{}}
After observations of soliton propagation and self-frequency shifting, the next key result was the observation of soliton self-compression. \textcite{konorov_self-compression_2005} observed a slight self-compression of 100~MW, 270~fs pulses in a 9~cm, 50~\um{} core diameter HC-PBG fiber filled with air. \textcite{ouzounov_soliton_2005} observed compression of 100~fs pulses at 1450~nm to 50~fs at 225~nJ energy after propagating through 24~cm of an \pgpcf{} fiber filled with 4.5~bar Xe. A number of numerical papers followed (\textcite{bessonov_temporal_2005,bessonov_pulse_2006}, illustrating some of the dynamics that are possible.

Soliton self-compression occurs when the soliton order $N > 1$. From Eq.~\ref{eqn:gas_sol_order} we see that this means that initially $L_{NL} < L_D$, and thus nonlinear broadening through SPM is initially dominant. As the spectrum expands, the role of dispersion increases and, being anomalous, compensates the nonlinear SPM phase, leading to temporal pulse compression inside the fiber. This in turn further enhances the SPM driven broadening, and the process continues. Soliton self-compression was first experimentally confirmed in conventional fibers by \textcite{mollenauer_experimental_1980, mollenauer_extreme_1983}. In the ideal case the maximum compression factor has been numerically determined to be\footnote{See e.g. \cite{dianov_optimal_1986,chen_nonlinear_2002,voronin_soliton-number_2008}.}
\begin{equation}
F_{sc}  =\frac{\tau_0}{\tau_c}\approx4.5N.
\end{equation}
\textcite{voronin_soliton-number_2008} numerically demonstrated that this is of course limited by higher-order effects, including high-order dispersion, self-steepening and Raman; all of which can cause soliton fission, as discussed in much earlier works\footnote{\cite{golovchenko_decay_1985, beaud_ultrashort_1987, kodama_nonlinear_1987}.}.  \textcite{ouzounov_soliton_2005} noted that the large dispersion slopes characteristic of \pgpcf{} fibers limit the high-ratio pulse compression to the picosecond regime. Further refinements include the works by \textcite{mosley_ultrashort_2010}, who used a short pulse source in the green to demonstrate soliton-effect compression from $\sim300$~fs to $\sim100$~fs in a 1~m PBG at 532~nm; and \textcite{peng_high_2011}, who reported the delivery and compression of 1~ps, 5~$\mu$J pulses at 1550~nm at 100~kHz (0.5~W).

Overcoming the limitations of \pgpcf{} in terms of bandwidth handling (and hence the shortest pulses achievable) requires the use of \bhc{}, and this has led to many exciting results on soliton dynamics, as we describe in Section~\ref{sec:fsscomp}.

\subsubsection{Adiabatic soliton compression in \pgpcf{}}
Instead of tuning the parameters such that $N > 1$, adiabatic compression\footnote{For original results in conventional fiber see \cite{kuehl_solitons_1988,chernikov_femtosecond_1991}.} is based around keeping $N \approx 1$, but changing the fiber and pulse parameters such that the pulse duration decreases smoothly, through
\begin{equation}
\label{eqn:sol_adiabatic}
\tau_0=\frac{2|\beta_2|}{\gamma E_{\mathrm{sol}}},
\end{equation}
where $E_{\mathrm{sol}}=2P_0\tau_0$ is the soliton energy. Clearly loss reduces $E_{\mathrm{sol}}$ and hence temporally broadens the soliton, whereas gain, decreasing dispersion or effective area, or increasing nonlinear refractive index (through say a gas pressure gradient) compresses the soliton.

\textcite{gerome_delivery_2007}, demonstrated adiabatic soliton compression in an 8~m tapered \pgpcf{} fiber, generating 90~fs, 70~nJ solitons. The taper only reduced the core diameter from 7.2~\um{} to 6.8~\um{} but the GVD at 800~nm from close to 80 to nearly 0~ps/nm/km. \textcite{laegsgaard_theory_2009} numerically studied adiabatic compression by means of a pressure gradient, but this has not been experimentally verified. \textcite{welch_solitons_2009} further demonstrated adiabatic and soliton effect compression, using dispersion variation in a 35~m tapered \pgpcf{} fiber, with larger compression ratios, of 2.5~ps to 215~fs and 1.2~ps to 175~fs at an output energy of 5~nJ and 9.4~nJ respectively.

\subsubsection{Solitons in \bhc{}\label{sec:fsscomp}}
In a series of pioneering papers \textcite{im_guiding_2009,im_high-power_2010,im_soliton_2010} studied the properties of gas-filled \bhc{} and its perfect suitability for soliton dynamics. \textcite{travers_ultrafast_2011} also performed detailed and systematic numerical simulations of a wide range of ultrafast nonlinear optics in \bhc{}. Together these works predicted high power soliton propagation, self-compression to few-cycle pulses, supercontinuum formation and dispersive-wave generation to the vacuum UV--all results that would be confirmed experimentally and form the basis of a new sub-field of nonlinear fiber optics, as described in the following sections. In addition to the relatively low value of $\gamma$, common to almost all HC-PCF, \bhc{} also exhibits very wide transmission windows and regions with weak, flat and pressure tunable anomalous dispersion (see Fig.~\ref{fig:gas_prop}(c)). These are the key features that make \bhc{} so suited to ultrafast nonlinear optics and soliton dynamics in particular. 

\textcite{joly_bright_2011} provided the first experimental evidence of soliton propagation and self-compression in \bhc{}, by compressing 30~fs pulses to 9~fs at the output of the fiber. Numerical simulations suggested that 4~fs pulses must have occurred inside the fiber, and also highlighted the importance of self-steepening and shock-formation \cite{demartini_self-steepening_1967} on the formation of such short pulses. In the case of self-compression, \textcite{travers_ultrafast_2011} predicted sub-cycle, 0.5~fs pulse generation at 800~nm. Subsequently, sub 2~fs pulses were reliably inferred by \textcite{holzer_femtosecond_2011} and \textcite{ermolov_supercontinuum_2015}, with corresponding peak powers close to 2~GW. The inference was made through comparison to rigorous numerical simulations and the requirement for such short pulse durations to occur in order to drive the physical mechanism of the formation spectral features identifiable in the experiments (see the dispersive-wave and plasma sections below). True confirmation of these results is awaiting sufficiently sophisticated pulse characterization measurements. \textcite{mak_two_2013} measured soliton self-compression to 6.8~fs from 24~fs pump pulses at 800~nm. The corresponding peak power was estimated to be 0.9~GW inside of the fiber core. Subsequently, \textcite{balciunas_strong-field_2015} compressed 35~$\mu$J pulses to 4.5~fs at 1.8~\um{}, which is sub-cycle. The corresponding peak power of the output pulses was 2~GW. These pulses were fully characterized and were also proved to be CEP stable and of sub-cycle duration through stereo-ATI measurements. This source is useful for attosecond pulse generation and initial high-harmonic generation experiments have been reported \cite{fan_integrated_2014}, (see also Section~\ref{sec:hhg}).

\begin{figure}
    \centering
    \includegraphics[width=0.98\linewidth]{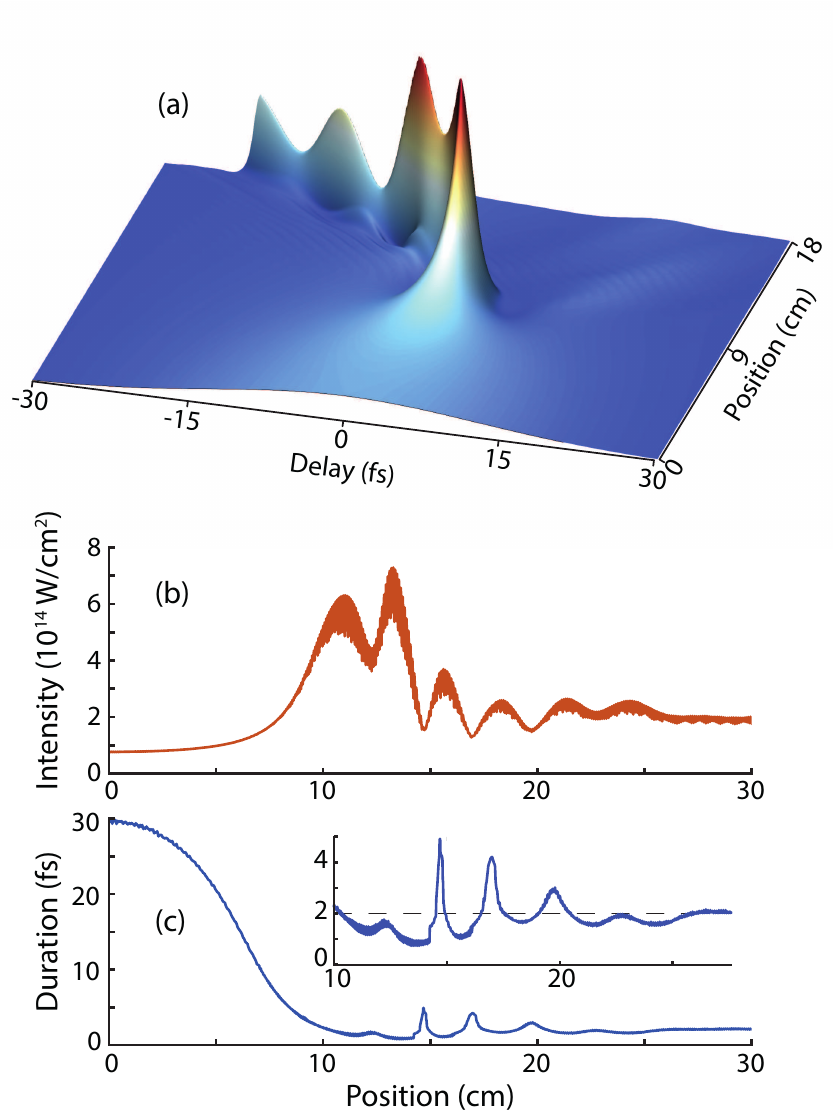}
    \caption{(a) Temporal intensity evolution through the fiber for soliton-effect self-compression from 30~fs to $\sim1$~fs in a 26~\um{} core diameter \bhc{} filled with 28~bar He. The initial pulse has an energy of 3.54~$\mu$J, corresponding to an $N = 3.5$ order soliton. (b) Corresponding peak intensity. (c) Corresponding full-width at half-maximum (FWHM) pulse duration.}
    \label{fig:gas_solcomp}
\end{figure}
As an example of the pulse-compression dynamics, the results of numerical simulation\footnote{Note that the dynamics in this figure are extreme in terms of pulse duration ($\sim1$~fs) and a high peak intensity, which leads to strong ionization of the gas. The model used, described by \textcite{tani_multimode_2014}, rigorously accounts for the high order effects resulting from such propagation, including self-focusing, plasma formation and shock formation.} of a 30~fs pulse at 800~nm propagating through a 26~\um{} core diameter \bhc{} filled with 28~bar He ($\lambda_{\mathrm{ZD}}=2\pi/\omega_{\mathrm{ZD}}=342$~nm), are shown in Fig.~\ref{fig:gas_solcomp}. The soliton order was set at $N=3.5$, which required 3.54~$\mu$J pulse energy. Fig.~\ref{fig:gas_solcomp}(a) illustrates the dramatic pulse compression and enhancement of the peak intensity as the pulse propagates through the fiber. From Fig.~\ref{fig:gas_solcomp}(b) we note that the intensity increases 7-fold from $\sim1$ to $\sim7\times10^{14}$~W/cm$^2$. The corresponding pulse durations are shown in Fig.~\ref{fig:gas_solcomp}(c), decreasing from 30~fs to $\sim1$~fs in the first 10~cm. Most extraordinarily, the pulse duration stays constant at around 2~fs throughout the rest of the fiber propagation, over 20~cm, despite the fact that the dispersion length is $\sim2$~mm. Both the compression and the stable pulse duration are evidence of the existence of soliton dynamics. Inspection of the numerical results indicates that the self-compression process is assisted by self-steepening, leading to shock formation on the trailing edge, and ionization and plasma formation, leading to a shock on the leading edge (the full plasma dynamics are reviewed in detail in Section~\ref{sec:plasma}).

\subsubsection{Resonant dispersive-wave emission in \bhc{}\label{sec:dwave_kgm}}
Soliton propagation in the presence of higher order dispersion can lead to the emission of radiation at frequencies shifted from the center of the soliton spectrum through a process known as resonant dispersive-wave emission (RDW)\footnote{Sometimes also referred to as resonant radiation, dispersive-wave generation or fiber Cherenkov radiation.}\footnote{It has actually been shown that solitons are not required \cite{roger_high-energy_2013,webb_generalized_2013}, but we keep the focus here on soliton emitted RDW as that is the regime for all of the HC-PCF experiments to date.}, first discussed in detail by \textcite{wai_nonlinear_1986}, and subsequently widely explored in solid-core optical fibers. This process has been harnessed in broadband-guiding gas-filled \hc{} for the efficient generation of tunable ultrafast pulses between 113~nm and 550~nm, conveniently spanning a spectral region with a wide range of applications, especially in spectroscopy.

The frequency of emission is determined by a phase-matching condition which can be derived from the generalized nonlinear Schr\"odinger equation under the approximation that the dispersive-waves do not themselves interact nonlinearly with the field\footnote{In some circumstances, under which cross-phase modulation ought to be important, it is somewhat surprising that this approximation works. The usual explanation is that the group velocity walk-off between the soliton and the dispersive-wave is large, and hence they have limited time to interact. But this does not always adequately explain why the phase-matching is not altered.}. The resulting condition is simply\footnote{See e.g. \textcite{skryabin_theory_2005} for a derivation.} 
\begin{equation}
\label{eqn:dwave_gen}
\Delta\beta(\omega)=\beta(\omega)-\beta_{\mathrm{NL}}(\omega)=0
\end{equation}
where $\beta_{\mathrm{NL}}(\omega)$ is the propagation constant at frequency $\omega$ including any terms causing a nonlinear phase-shift, and $\beta(\omega)$ is the usual linear propagation constant. \textcite{karpman_radiation_1993} explained RDW emission as tunneling from inside the self-trapped potential of the soliton to free-propagating radiation, and this interpretation is clear from Eq.~\ref{eqn:dwave_gen}--the emission occurs when the nonlinearly propagating waves are phase-matched to the free-propagating linear waves. \textcite{akhmediev_cherenkov_1995} explained the process as an analogue of Cherenkov radiation emitted by the solitons. \textcite{erkintalo_cascaded_2012} provided a further refinement to our understanding with a frequency domain perspective, explaining RDW emission through cascaded four-wave mixing (FWM). In particular, this picture predicts the spectral recoil of solitons in the opposite direction from the emitted frequency, observed during RDW emission, as the accumulation of Stokes or anti-Stokes in the FWM process. It also describes RDW emission when the pump pulses are not solitons \cite{roger_high-energy_2013,webb_generalized_2013}.

In the case that the input pulse consists of a single fundamental soliton, then $\beta_{\mathrm{NL}}(\omega)$ can be written explicitly, so that the phase-matching condition becomes
\begin{equation}
\label{eqn:dwave_sol}
\Delta\beta(\omega)=\beta(\omega)-\beta(\omega_p)-\beta_1(\omega_p)[\omega-\omega_p]-\frac{\omega}{\omega_p}\gamma P_0=0,
\end{equation}
where $\omega_p$ is the central frequency of the soliton. The shock term, $\omega/\omega_p$, in front of the nonlinear phase was introduced by \textcite{roger_high-energy_2013}.

Eq.~\ref{eqn:dwave_sol} generalizes to the case of pumping with high-order solitons undergoing self-compression by replacing $P_0$ with $P_{\mathrm{sc}}$ the total power of the self-compressed pulse. This simply reflects the fact that the nonlinear propagation constant $\beta_{\mathrm{NL}}(\omega)$ is related to the total field intensity interacting with the RDW, not any specific soliton component. $P_{\mathrm{sc}}$ is difficult to estimate analytically, but some numerical rules were established by \textcite{dianov_optimal_1986} and \textcite{chen_nonlinear_2002}, giving $P_{\mathrm{sc}}\approx4.5P_0N$. A more accurate way of establishing the phase-matched frequencies was introduced by \textcite{austin_dispersive_2006}, in which the full field envelope is used for the nonlinear phase-shift. The downside to this approach is that it requires numerical propagation simulations, at which point one can just numerically measure the emission frequency. Nevertheless, this analysis does help identify the gain dynamics, especially with respect to self-steepening \cite{travers_phase-matching_2011}.

\begin{figure}
    \centering
    \includegraphics[width=0.98\linewidth]{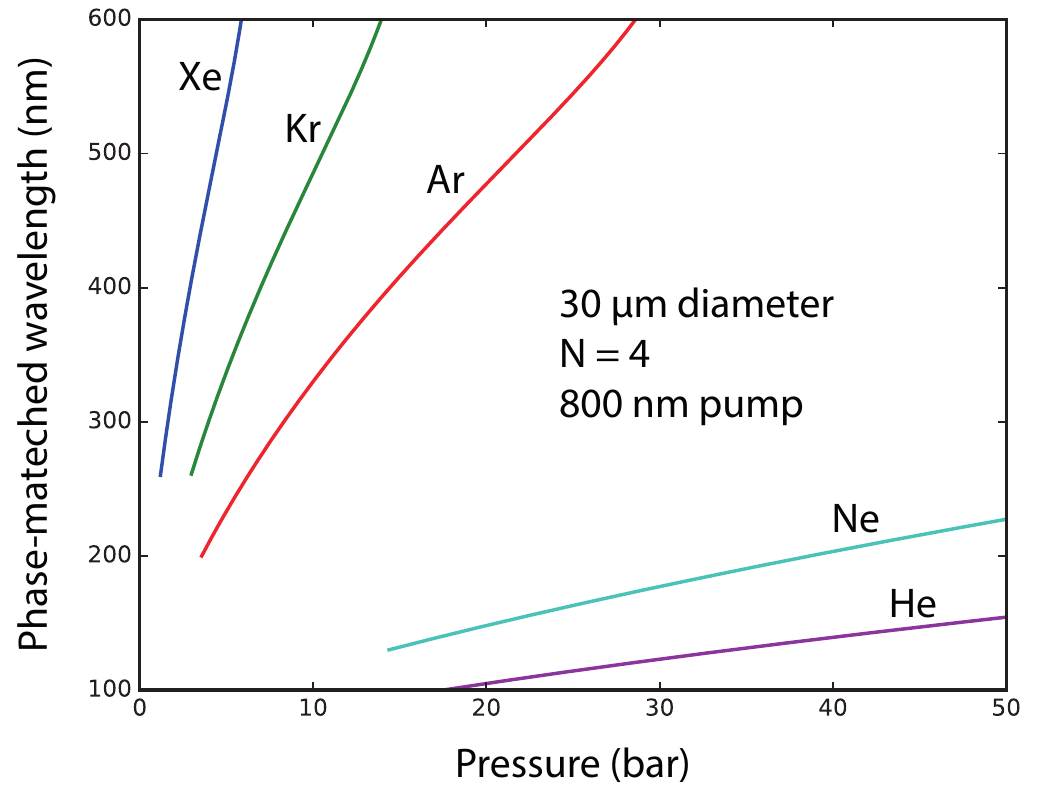}
    \caption{Phase-matched RDW wavelengths as a function of gas-filling pressure in a 30~\um{} core diameter \bhc{}, for an 800~nm, 30~fs pump pulse with energy tuned for an input soliton order of $N = 4$.}
    \label{fig:gas_dwavepm}
\end{figure}

The use of \bhc{} for RDW emission was first proposed by \textcite{im_high-power_2010}. Broadband-guiding \hc{} offers a number of advantages for the generation of high frequency RDW:
\begin{enumerate}
    \item It is possible to tune the zero dispersion frequency (wavelength) very high (low), which can be shown (from Eq.~\ref{eqn:dwave_sol}) to consequently shift the phase-matched RDW frequency higher, extending to the vacuum-UV.
    \item The low dispersion and broadband guidance enable the extreme pulse compression and shock dynamics discussed previously, which enhances the conversion efficiency to highly shifted phase-matched frequencies.
    \item The ability to photo-ionize the gas without damaging the waveguide can further shorten the pump pulse, frequency shift it, and also alter the dispersion landscape (these effects are discussed in Section~\ref{sec:plasma}).
\end{enumerate}
Fig.~\ref{fig:gas_dwavepm} illustrates the opportunities for phase-matching dispersive-waves in \bhc{} pumped at 800~nm. The phase-matched RDW points range from $\sim100$~nm with He, up to beyond 600~nm with Ar, Kr and Xe. Of course, shifting to longer pump wavelengths also enables the emission of infrared RDW.

\begin{figure}
    \centering
    \includegraphics[width=0.98\linewidth]{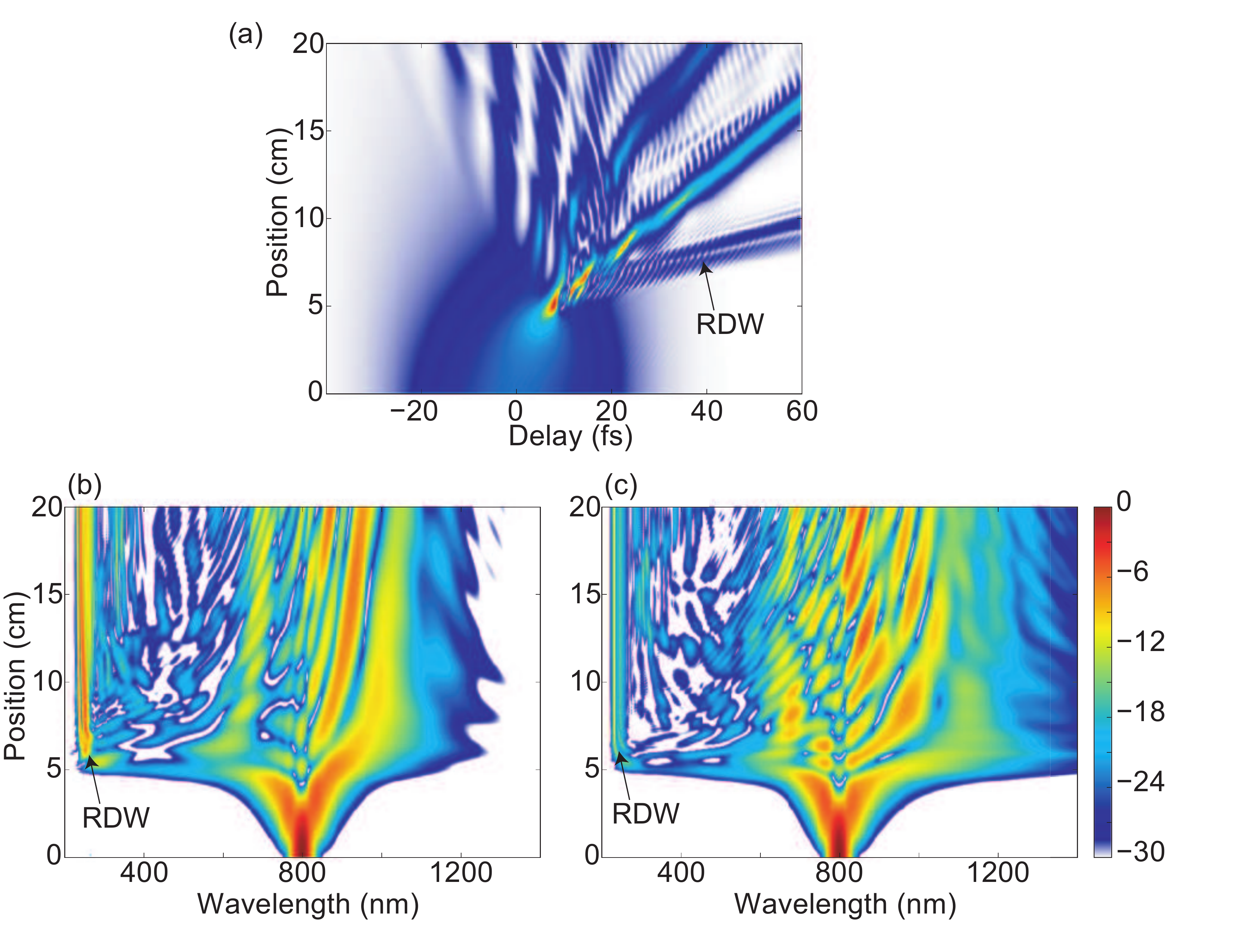}
    \caption{Pulse propagation simulations of RDW emission for parameters similar to \textcite{joly_bright_2011} (see text). (a) The temporal intensity evolution along the fiber length. (b) The spectral intensity evolution along the fiber length, including self-steepening. (c) As for (b) but without the self-steepening term; in this case the RDW is much weaker.}
    \label{fig:gas_dwave_shock}
\end{figure}

\paragraph{Experiments.}
RDW emission in \bhc{} was first demonstrated experimentally by \textcite{joly_bright_2011}. In that experiment, a $\sim30$~\um{} core diameter \bhc{} was filled with 0 to 10~bar Ar and pumped with 30~fs, $\sim1$~$\mu$J pulses. Wavelength tunable dispersive-wave emission between 200~nm and 320~nm was achieved and numerical simulations clearly identified that the self-steepening term, leading to shock formation, increased the conversion efficiency by roughly an order of magnitude, to 8\%.
\textcite{biancalana_theory_2004} had shown that the amplitude of the RDW is directly related to that of the driving wave at the emission frequency, and noted that this tends to decay exponentially with increasing frequency shift. Therefore, when pumping with high order solitons, the resulting self-compression and shock-formation significantly enhances the RDW emission, by increasing the spectral overlap between the nonlinear driving field and the phase-matched RDW point. Fig.~\ref{fig:gas_dwave_shock} illustrates the pulse propagation for parameters similar to \textcite{joly_bright_2011}. In the first 5~cm the pulse undergoes extreme soliton self-compression. At the compression point, the spectrum has a large spectral asymmetry towards the blue side due to shock formation, which disappears upon further propagation. A large amount of energy is, however, deposited at an extreme wavelength, and this is seen as a week scattered wave in the time-domain picture. This is the RDW. In Fig.~\ref{fig:gas_dwave_shock}(c) the spectrum evolution is shown without shock. At the compression point, the spectrum is more symmetric and there is less energy seeding the RDW emission, which is consequently weaker.

\begin{figure}
    \centering
    \includegraphics[width=0.98\linewidth]{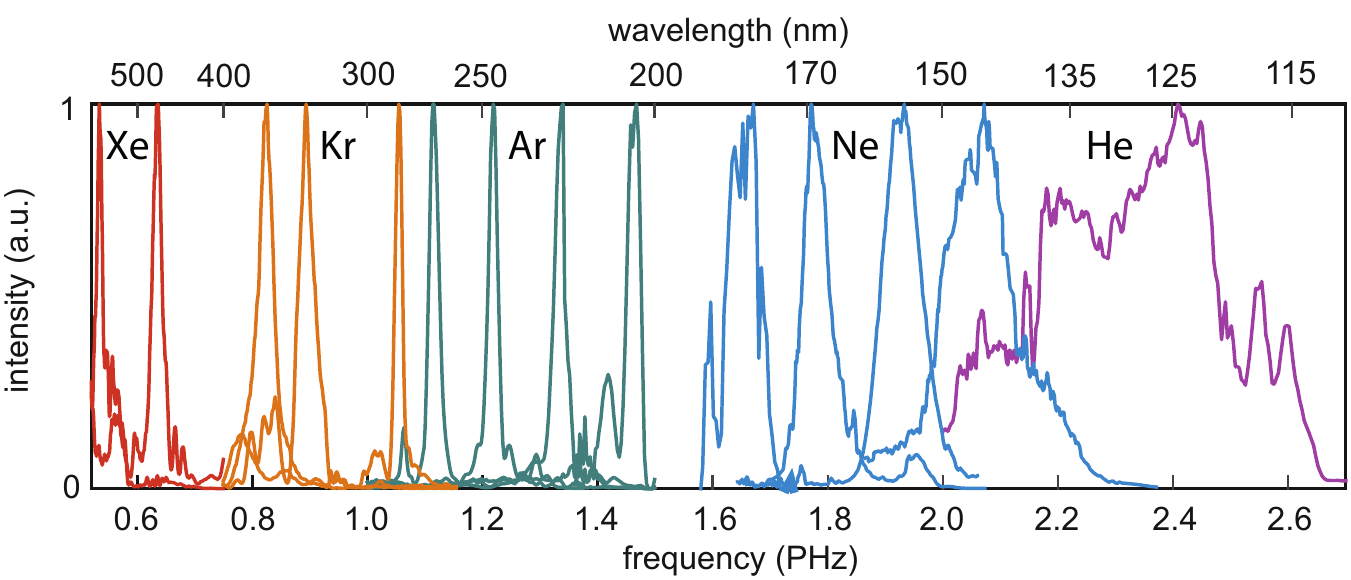}
    \caption{Experimental spectra of RDW emission from kagom\'{e} HC-PCF collated from \textcite{mak_tunable_2013} and \textcite{ermolov_supercontinuum_2015}.}
    \label{fig:gas_dwave_exp_tuning}
\end{figure}
By tuning the dispersion landscape through both the gas species and gas pressure, it was confirmed by \textcite{mak_tunable_2013} that RDW between 180~nm and 550~nm can be achieved, as shown in Fig.~\ref{fig:gas_dwave_exp_tuning}. Furthermore, it was found that there are optimal conditions of fiber length and pulse energy in order to achieve the cleanest emission in terms of spectral shape and bandwidth; fractional bandwidths $\sim0.03$ were observed. Numerical simulations were used to confirm that the whole spectrum was coherent, when the soliton order was sufficiently low. At higher soliton orders, plasma effects degraded the coherence.

Following the first demonstration of VUV generation and guidance in \bhc{} by \textcite{belli_vacuum-ultraviolet_2015} (to be discussed later in Section~\ref{sec:vuvraman}), \textcite{ermolov_supercontinuum_2015} extended the RDW tuning down to 113~nm in Ne and He gas, limited by the absorption of the MgF$_2$ windows used on the gas cells, as also shown in Fig.~\ref{fig:gas_dwave_exp_tuning}. In that result it was also noted that ionization can further enhance the efficiency of dispersive-wave emission due to the soliton self-frequency blue-shift (this is discussed further in Section~\ref{sec:plasma_dwave}).

\paragraph{Power-scaling and deep-UV pulse duration.}
\textcite{kottig2017generation} significantly scaled the repetition rate of RDW emission by pumping with a ytterbium fiber laser. The 329~fs pulses were first compressed using an SPM-mirror system, based on Ar-filled \bhc{}, to 24~fs with $\sim20$~$\mu$J. These were then coupled into the same fiber, but with an HF-etched core wall, and filled with Ne or He, to generate deep-UV RDW emission. At a 100~kHz repetition rate, RDW emission at 205~nm was achieved with over 1~$\mu$J of deep-UV pulse energy. At 1.92~MHz repetition rate over 0.5~$\mu$J was achieved at 275~nm, corresponding to a deep-UV average power of over 1~W. In a closely related work \textcite{kottig_high_2015} characterized the pulse duration of the high power RDW emission, determining that the deep-UV pulses  were $< 10$~fs at the fiber exit, although significantly broadened due to the air dispersion and gas-cell windows on route to the two photon autocorrelator used for pulse duration measurements. More recently \textcite{ermolov_multi-shot_2016} used a transient-grating cross-correlation frequency-resolved optical gating (FROG) system to verify that the RDW emission at 270~nm (at 1~kHz, pumped by a 800~nm Ti:Sapphire laser) was just 5~fs in duration.

\paragraph{Pressure gradients and higher order modes.}
\textcite{mak_tunable_2013} also demonstrated RDW emission in fibers with both positive and negative pressure gradients. The use of a pressure gradient alters the generation dynamics by changing the phase-matching condition continuously along the fiber. Therefore, by tuning the pump energy, and hence the location of the compression point, different emission wavelengths and bandwidths are generated. \textcite{tani_multimode_2014} demonstrated dispersive-wave emission to higher order modes. In that case, due to the stronger dispersion of the higher order modes, the dispersive waves are emitted at larger and larger frequency shifts. This plays a role in the VUV supercontinuum described in Section~\ref{sec:vuvscg}, and \citeauthor{tani_multimode_2014} described how the transfer of energy to higher order modes through RDW emission can constitute a new form of self-focusing. Finally, photoionization and plasma formation can alter both the emission dynamics of dispersive-waves, and lead to phase-matching and emission and new frequencies. These are described in Section~\ref{sec:plasma_dwave}.

\paragraph{Infrared pumping for RDW emission.}
Several works have shown RDW emission from longer pump wavelengths, generated with an optical parametric amplifier (OPA). \textcite{Cassataro:17} tuned the pump wavelength between 1.6~\um{} and 1.8~\um{}, showing both supercontinuum formation in the infrared from 270~nm to 3.1~\um{} and RDW emission around 300~nm. This was the first result in gas-filled \bhc{} to show tuning of the RDW emission with pump wavelength. Previous work tuned the gas pressure and pump energy. \textcite{Meng:17} also recently showed RDW emission from an OPA tuned between 1.3~\um{} and 1.5~\um{} in an Ar-filled \bhc{}. As the pump wavelength in this case was much closer to the zero dispersion wavelength, the RDW emission was shifted to around 1000~nm. The closeness of the pump to the phase-matched emission wavelength increased the conversion efficiency to around 16\%, the highest yet reported.

\paragraph{Applications.}
An ultrafast pulse source ($< 10$~fs) with 10~nJ to $> 100$~nJ, tunable between 113~nm and 550~nm, can be expected to be of interest for a range of applications, particularly in ultrafast spectroscopy of molecular systems which have many excitation pathways in the deep and vacuum UV region. So far there have only been two works on applications of RDW emission. \textcite{joly_new_2011} made some numerical estimates about seeding free electron lasers to improve their coherence properties, but no experimental demonstration has followed.

\textcite{bromberger_angle-resolved_2015} successfully used RDW emission for angle-resolved photoemission spectroscopy (ARPES) at 9~eV. The wide tunability throughout the VUV (the most important photon energy range for ARPES) and the short pulse duration of the RDW emission make this source highly interesting for time-resolved ARPES measurements. The ability to scale the repetition-rate of the emission to the MHz range will also improve the rate of data collection and the statistical analysis of the relatively low photo-electron yields from such experiments.

\subsection{Plasma effects in \bhc{}\label{sec:plasma}}
At sufficiently high intensities the filling-gas in an HC-PCF can be ionized by the light field. This can have a strong effect on the optical pulse propagation, even if $< 1$\% of the gas is ionized. Combined with the ability of HC-PCF to guide optical solitons in gas, this has opened a new subfield of nonlinear fiber optics: soliton-plasma interactions.

The linear and nonlinear optical properties of the ions produced by photoionization can be expected to follow the general trend of atoms with higher ionization potentials, and exhibit weaker dispersion and nonlinearity than the neutral atoms. Taking the extreme limit, that they become completely negligible, the reduction in linear and nonlinear susceptibility of the filling gas will be at most the same as the ionization fraction, i.e. $< 1$\%. The same is not true of the electrons that result from photoionization. The nonlinear response from free electrons can be neglected as it requires relativistic intensities, beyond the range so far considered in HC-PCF (around $10^{18}$~W/cm$^2$). So the only contribution from the electrons is their linear response. Two factors make this very significant: (i) the polarizability of free electrons is very large (see next section); (ii) the free electrons are themselves produced through a highly nonlinear process, causing them to accumulate exceedingly rapidly in the optical pulse at the peaks of the electric field cycles. So while the plasma optical response is linear, the combined photoionization and plasma response is highly nonlinear.

\subsubsection{Free electron polarizability, plasma refractive index and dispersion}
The polarizability of a free electron is $\alpha_e=-\mathrm{e}^2/\mathrm{m}_e\omega^2$, where $\mathrm{e}$ and $\mathrm{m}_e$ are the electron charge and mass. Comparing this to the polarizability of neutral atoms shows that the polarizability of free electrons is much larger and of opposite sign. For example at a wavelength of 800 nm, $\alpha_e=-5.08\times10^{-39}$~F m$^2$, and the polarizability of Ar is $\alpha_{\mathrm{Ar}}=1.82\times10^{-40}$~Fm$^2$. Thus the dispersive effect of the plasma would equal that of the gas at an ionization fraction of just 3.6\%. Even for much smaller fractions, the perturbation to the refractive index can be much stronger than that due to the Kerr effect.

\begin{figure}
    \centering
    \includegraphics[width=0.98\linewidth]{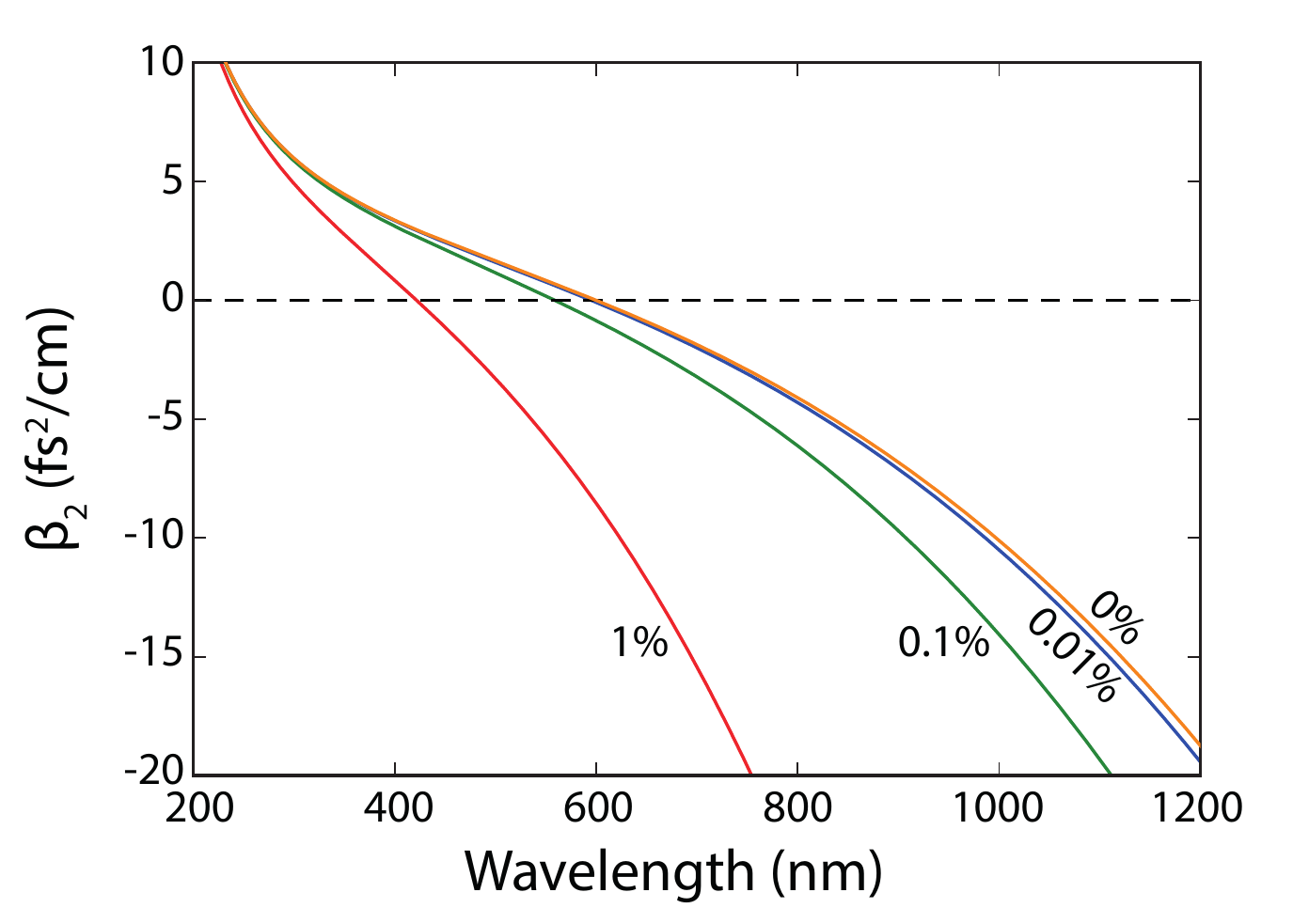}
    \caption{Group velocity dispersion of 10~bar Ar in a 30~\um{} core diameter \bhc{} with the indicated ionization fractions.}
    \label{fig:gas_plasma_gvd}
\end{figure}

The refractive index of the plasma can be defined from the polarizability in the usual way
\begin{equation}
n_{\mathrm{pl}}(\omega)=\sqrt{1-\frac{N_e\alpha_e}{\varepsilon_0}}\approx1-\frac{N_e}{2N_{\mathrm{crit}}}=1-\frac{\omega_\mathrm{pl}^2}{2\omega^2},
\end{equation}
where the plasma frequency $\omega_\mathrm{pl}^2=N_e\mathrm{e}^2/\varepsilon_0\mathrm{m}_e$ is the frequency of the electron oscillations and the critical density $N_\mathrm{crit}=\varepsilon_0\mathrm{m}_e\omega^2/\mathrm{e}^2$ is the density at which $\omega_\mathrm{pl}=\omega$. From this we can determine the GVD induced by the plasma
\begin{equation}
\beta_2^{(\mathrm{pl})}(\omega)\approx-\frac{\omega_\mathrm{pl}^2}{c\omega^3}.
\end{equation}
Thus the plasma GVD is anomalous and becomes very large in the infrared. Fig.~\ref{fig:gas_plasma_gvd} shows the GVD of a 30~\um{} core diameter \bhc{} filled with 10~bar Ar at various degrees of ionization. The zero dispersion wavelength can be dramatically altered for 1\% ionization, and the anomalous GVD at longer wavelengths becomes significantly stronger.

\subsubsection{Ionization rates}
The liberation of an electron from an atom can be considered through a multitude of phenomenological processes. The two dominant ones for the intensities so far achieved in HC-PCF (up to $10^{15}$~W/cm$^2$) are multiphoton ionization, where an electron absorbs sufficient photons from the driving laser to gain enough energy to escape the atomic potential; and tunnel ionization, where the electric field of the driving laser distorts the atomic potential so that the electron is more likely to tunnel through the barrier. The typical way to distinguish these regimes is to consider the Keldysh parameter \cite{keldysh_ionization_1965}
\begin{equation}
\gamma_K=\sqrt{\frac{U_I}{2U_p}},
\end{equation}
where $U_I$ is the ionization potential energy, and $U_p=\mathrm{e}^2E^2/4\mathrm{m}_e\omega^2$ is the ponderomotive energy\footnote{The cycle-averaged kinetic energy of an electron in a laser field.} for field strength $E$. In a semi-classical picture $\gamma_K$ is the time taken for an electron to tunnel through the Coulomb barrier of the atom in units of the electric-field cycle period. It is usually stressed, though somewhat controversially, that for $\gamma_K<1$ tunnel ionization occurs, and for $\gamma_K\gg1$ multiphoton ionization occurs. The parameters used in \bhc{} usually correspond to $\gamma_K\approx1$, preventing such a clean distinction. The so-called Perelomov, Popov and Terent’ev (PPT) rate \cite{perelomov_ionization_1966}, accurately reproduces both regimes, and interpolates between them, so we use it for the following analysis\footnote{The widely used ADK rate \cite{ammosov_tunnel_1986} is the tunnel limit of PPT with generalized coefficients to handle more complex initial electron states; the ADK coefficients should be integrated into the PPT model. The PPT rate can also be extended to include the non-adiabatic field following of the ionization rate, i.e. when the field strength changes within the tunneling time \cite{yudin_nonadiabatic_2001}. The power law ionization rates specialized for just multiphoton ionization are inappropriate for $\gamma_K\approx1$, and are easily derived from PPT in the limit $\gamma_K\gg1$.}.

\begin{figure}
    \centering
    \includegraphics[width=0.98\linewidth]{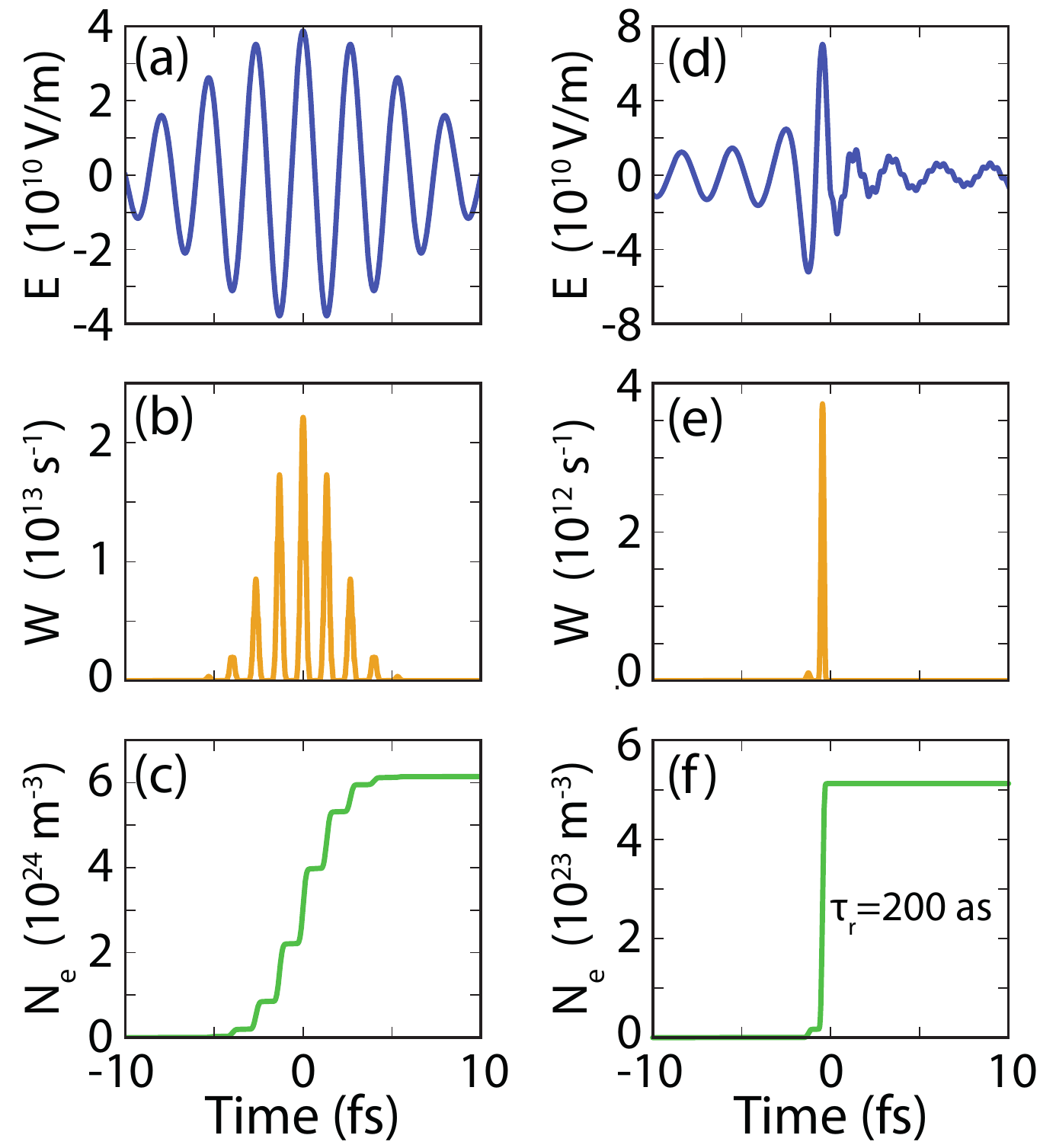}
    \caption{(a-c) Ionization of 10~bar Ar with a 10~fs Gaussian pulse with peak intensity $2\times10^{14}$~W/cm$^2$. (a) The electric field; (b) the ionization rate; (c) the accumulated free electron density. (d-f) Ionization of 28~bar He with the self-compressed pulse from Fig.~\ref{fig:gas_solcomp} (1.2~fs duration, $6.6\times10^{14}$~W/cm$^2$). In (f) the rise time of the free electron density step is 200~as.}
    \label{fig:gas_rate}
\end{figure}
For a single ionization level the free electron density is simply calculated through\footnote{Here we neglect avalanche ionization, which is usually, although not always, negligible for the gas pressures and free electron densities obtained in HC-PCF.}
\begin{equation}
\partial_tN_e=\left(N_a-N_e\right)W\left[E(t)\right],
\label{eqn:ne}
\end{equation}
where $N_a$ is the original neutral atom density and $W[E(t)]$ is the electric field dependent ionization rate. Fig.~\ref{fig:gas_rate}(b) shows the calculated ionization rate for a 10~fs Gaussian pulse with $2\times10^{14}$~W/cm$^2$ peak intensity (Fig.~\ref{fig:gas_rate}(a)), propagating in 10~bar Ar. Note that the rate is strongly peaked at the peaks of the electric field. Fig.~\ref{fig:gas_rate}(c) shows the corresponding free electron density accumulating step-wise with time, up to an ionization fraction of $\sim2.5$\%.

Fig.~\ref{fig:gas_rate}(d-f) shows the same plots, but for the self-compressed pulse from Fig.~\ref{fig:gas_solcomp}, which has 1.2~fs duration, and an intensity of $6.6\times10^{14}$~W/cm$^2$. In this case the rate shows a single strong peak and the free electron density increases in a step of just 200 attoseconds\footnote{It was recently established by \textcite{serebryannikov_quantum_2014} that the PPT model is still valid in this extreme sub-cycle regime.}. The step change in Fig.~\ref{fig:gas_rate}(f) corresponds to a refractive index modulation $\Delta n \sim 0.1$\%, comparable to the core-cladding step in a step-index solid-core fiber. This fast refractive index modulation drives the soltion-plasma effects considered in Section~\ref{sec:plasmasfbs}.

The refractive index step moves with the pulse, i.e. at the speed of light, causing phase modulation and acting as a scattering center, and also possibly useful as an event horizon in analogue gravity experiments, as proposed (using a different nonlinearity) by \textcite{philbin_fiber-optical_2008}.

\subsubsection{Nonlinear contribution of plasma creation}
The nonlinearity arising due to ionization arises from the fast accumulation of free electrons shown in Fig.~\ref{fig:gas_rate}(c). The large difference in dispersion of the free electrons means that this fast accumulation of plasma imparts a strong phase-modulation back on the laser pulse\footnote{The sign of this phase-modulation is ambiguous and depends on the definitions used when driving the propagation equations. However, more important is the frequency shift, which has a non-ambiguous opposite sense to the refractive index shift.}. The corresponding frequency shift is
\begin{equation}
\label{eqn:plasma_blue}
\Delta\omega(t)=-k_0\frac{\partial n_\mathrm{pl}}{\partial t} L_{\mathrm{eff}}\approx k_0  \frac{N_a}{2N_{\mathrm{crit}}}W[E(t)]L_{\mathrm{eff}},
\end{equation}
where we have assumed low ionization, such that $N_a-N_e\approx≈N_a$.

The recombination of free electrons is a very slow process compared to ultrafast laser pulses, and so the phase and frequency shifts noted above are one-sided, leading to a frequency upshift of the light field. This effect was first explored in free space by \textcite{bloembergen_influence_1973,yablonovitch_self-phase_1974,corkum_amplification_1985}; and then refined in a large series of papers\footnote{See e.g. the papers by \textcite{wood_measurement_1991,rae_detailed_1992,le_blanc_spectral_1993}.}. \textcite{tempea_nonlinear_1998} proposed plasma blue-shifting in HCF as a spectral broadening mechanism for pulse compression, and a clear demonstration of plasma blue shifting in HCF was performed by \textcite{babin_ionization_2002}.

Although useful for estimating the plasma density, Eq.~\ref{eqn:plasma_blue} is insufficient for modelling the full plasma dynamics in HC-PCF, as it neglects the interaction with the Kerr effect and anomalous dispersion which enable soliton-plasma solutions. A more complete model can be obtained by including the ionization and plasma phase modulation and absorption dynamics in the propagation equations by co-solving the coupled equations describing the plasma current\footnote{See \textcite{geissler_light_1999} for a self-consistent derivation and \textcite{rae_detailed_1992} for an early approach.}. The model introduced by \textcite{tani_multimode_2014} and used throughout this part of this review is one example.

\subsubsection{Soliton-plasma effects in HC-PCF\label{sec:plasmasfbs}}
\textcite{fedotov_ionization-induced_2007} showed both theoretically and experimentally that ionization in an air-filled HC-PBG fiber can cause a blue-shift of the input pulse. The pump at 807~nm with 60~fs, 2.3~$\muup$J pulses ($5\times10^{13}$~W/cm$^{2}$) was shifted over 10~THz towards the blue, and this agreed fully with numerical simulations. The shift was described as being a soliton self-frequency blue-shift, although no experimental or theoretical evidence was presented to confirm this. \textcite{fedotov_ionization-induced_2007}, also noted that the blue-shifting effect had to overcome the opposite frequency-shift due to the Raman effect in air (see Section~\ref{sec:solsinpbf}).

\begin{figure}
    \centering
    \includegraphics[width=0.98\linewidth]{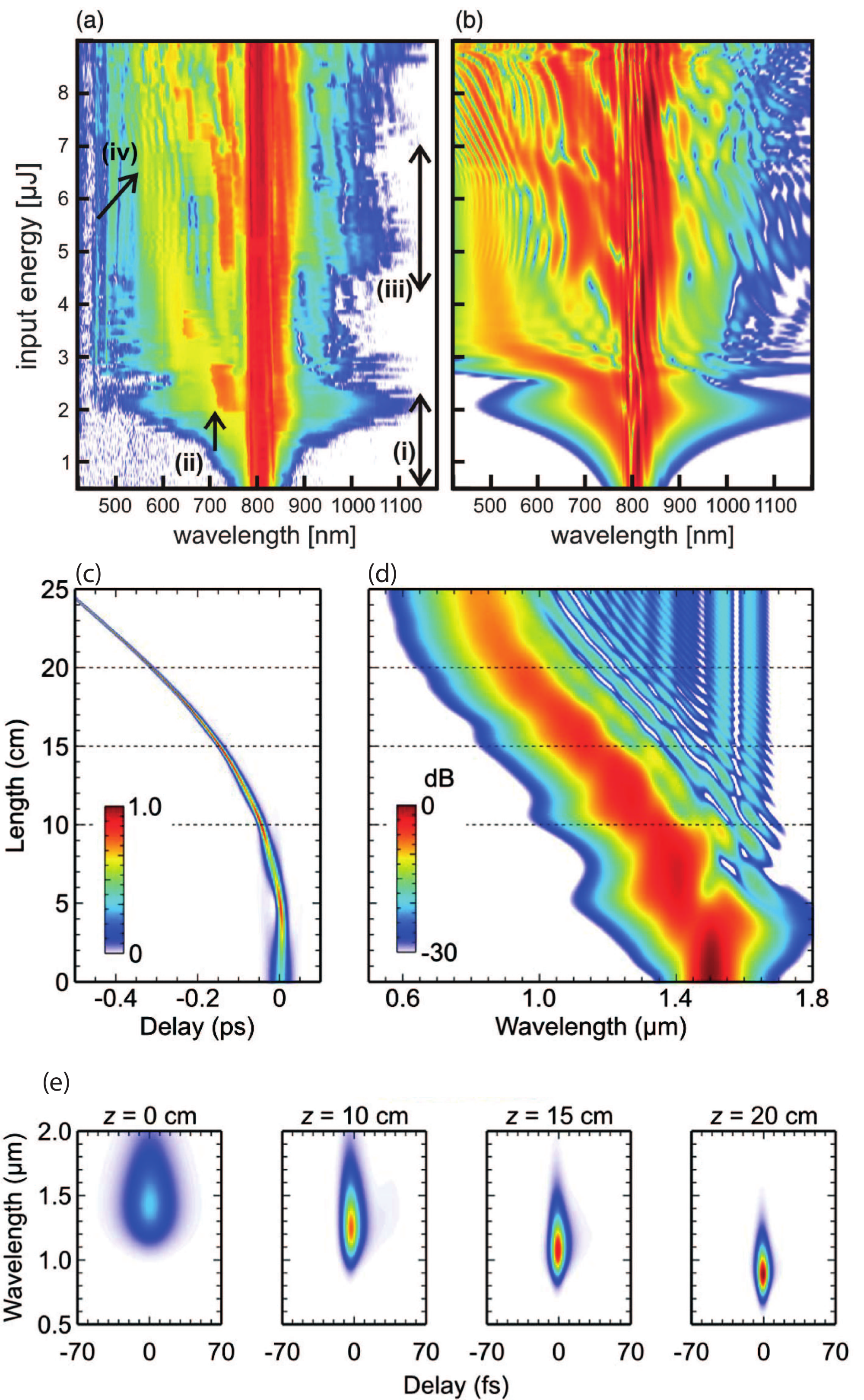}
    \caption{(a,b) Adapted from \textcite{holzer_femtosecond_2011}: (a) Experimental and (b) numerical output spectra of a \bhc{} (26~\um{} core diameter) filled with 1.7~bar of argon as a function of input pulse energy (65~fs, 800~nm). In region (i) there is no ionization, at (ii) a blue shoulder emerges, moving to higher frequencies as the pulse energy increases; at (iii) a second ionization stage
        starts; and at (iv) spectral interference between the two blue-shifting pulses is apparent. (c-e) Adapted from \textcite{chang_combined_2013}: (c) temporal and (d) spectral evolution during fundamental soliton self-frequency blue-shift; (e) Cross-correlation frequency-resolved optical gating traces (at the positions marked with dashed lines in (c,d)) showing the pulse compression and frequency shifting.}
    \label{fig:gas_plasma}
\end{figure}
\textcite{holzer_femtosecond_2011} demonstrated much larger pulse shifts in \bhc{}. Pumping a 26~\um{} core diameter Ar-filled \bhc{} with 800~nm, 65~fs, 1~$\muup$J to 9~$\muup$J led to the observation of a series of blue-shifting pulses emitted from the pump after a characteristic soliton self-compression stage. The frequency shift was in excess of 125~THz, as shown in Fig.~\ref{fig:gas_plasma}(a,b). Numerical simulations and analysis clearly evidenced that the blue-shifting mechanism was an ionization effect and that the blue-shifting pulses were solitons. A simultaneously published paper by \textcite{saleh_theory_2011} detailed an analytic model for the soliton effects in the ionization regime, and through perturbation theory, explained the soliton blue-shift in a way directly analogous to the Raman self-frequency shift. They also found further interesting effects such as non-local soliton clustering (both temporal and spectral) mediated by the plasma\footnote{See also \textcite{saleh_understanding_2011}.}. In addition they studied the interaction between plasma creation and Raman scattering.

The work of \textcite{holzer_femtosecond_2011} and \textcite{saleh_theory_2011} primarily considered an effective soliton fission at a single soliton self-compression point due to plasma induced phase-modulation. In this case the initial pump pulse is insufficiently intense to ionize the gas significantly, but due to self-compression its intensity upon propagation is increased such that strong plasma generation does occur. After this point the pump pulse breaks up into a series of blue-shifting fundamental solitons. Recently \textcite{kottig2017phz} have shown that at higher pump energies the plasma phase-modulation becomes strong before the soliton self-compression point is reached. Instead, the pulse experiences asymmetric spectral broadening \cite{SalehMI2012}, caused by the combined effects of the optical Kerr nonlinearity and plasma-induced phasemodulation. Combined with the anomalous group-velocity dispersion, this eventually leads to a coherent pulse splitting. The resulting sub-pulses then undergo further self-compression and and fission. Recent numerical simulations have also shown that under the correct parameter regimes the pump pulse can undergo multiple stages of pulse compression and soliton fission \cite{habib2017multiple, SelimHabib:17}. Something hinted at in earlier work, but not fully elucidated.

For high energy long pump pulses, \textcite{SalehMI2012} showed theoretically and numerically, that a new kind of modulational instability can arise due to the additional plasma nonlinearity in gas-filled \bhc{}.

At the other extreme \textcite{chang_combined_2013} predicted that fundamental solitons can also continuously blue shift---something unexpected, as the photoionization process removes energy from the soliton, so one would expect it to broaden and loose sufficient intensity for further blue-shifting. A fortuitous combination of physical parameters enable this through adiabatic soliton compression, as indicated in Fig.~\ref{fig:gas_plasma}(c-e). The input soliton has sufficient intensity to ionize the gas and hence blue-shift. In the process it loses some energy, and so would be expected to broaden and have reduced intensity. However, the blue-shift moved the soliton to a frequency with both a higher nonlinear coefficient (through the frequency dependence of the nonlinear phase shift) and to a region with a lower magnitude of GVD. Hence, by Eq.~\ref{eqn:sol_adiabatic} it compresses, and regains sufficient intensity to shift further. This process repeats, and \textcite{chang_combined_2013} numerically demonstrated a shift from 1500~nm to 800~nm while simultaneously compressing the pulse from 30~fs to 4~fs, with a 30\% conversion efficiency. The use of pressure gradients was shown to give good control of the frequency shift.

Recently \textcite{kim_extreme_2015} proposed that by mixing gases one could independently control the Kerr and ionization based nonlinearities, using a mixture of two gases with significantly different ionization potentials. In this way one gas maintain the Kerr driven soliton dynamics, while the other gives rise to a separately controllable blue-shift.

\subsubsection{ Interaction between ionization and RDW emission\label{sec:plasma_dwave}}
The interaction between photo-ionization and RDW emission was studied numerically by \textcite{chang_influence_2011}. The main conclusion was that soliton dynamics initially dominate at higher gas pressures, where the Kerr nonlinearity and GVD are significant. In that case the soliton self-compression and RDW emission dynamics occur as described in Section~\ref{sec:dwave_kgm}. As the pressure is lowered, the Kerr effect and gas GVD become less signifiant in comparison to the phase-modulation due to the creation of free electrons. The leads to an accelerating self-compressed pulse as opposed to the delayed one that is usual due to the recoil effect in RDW emission, and an altering of the dispersion landscape such that the RDW emission point is shifted. 

\paragraph{VUV supercontinuum generation\label{sec:vuvscg}}
\begin{figure}
    \centering
    \includegraphics[width=0.98\linewidth]{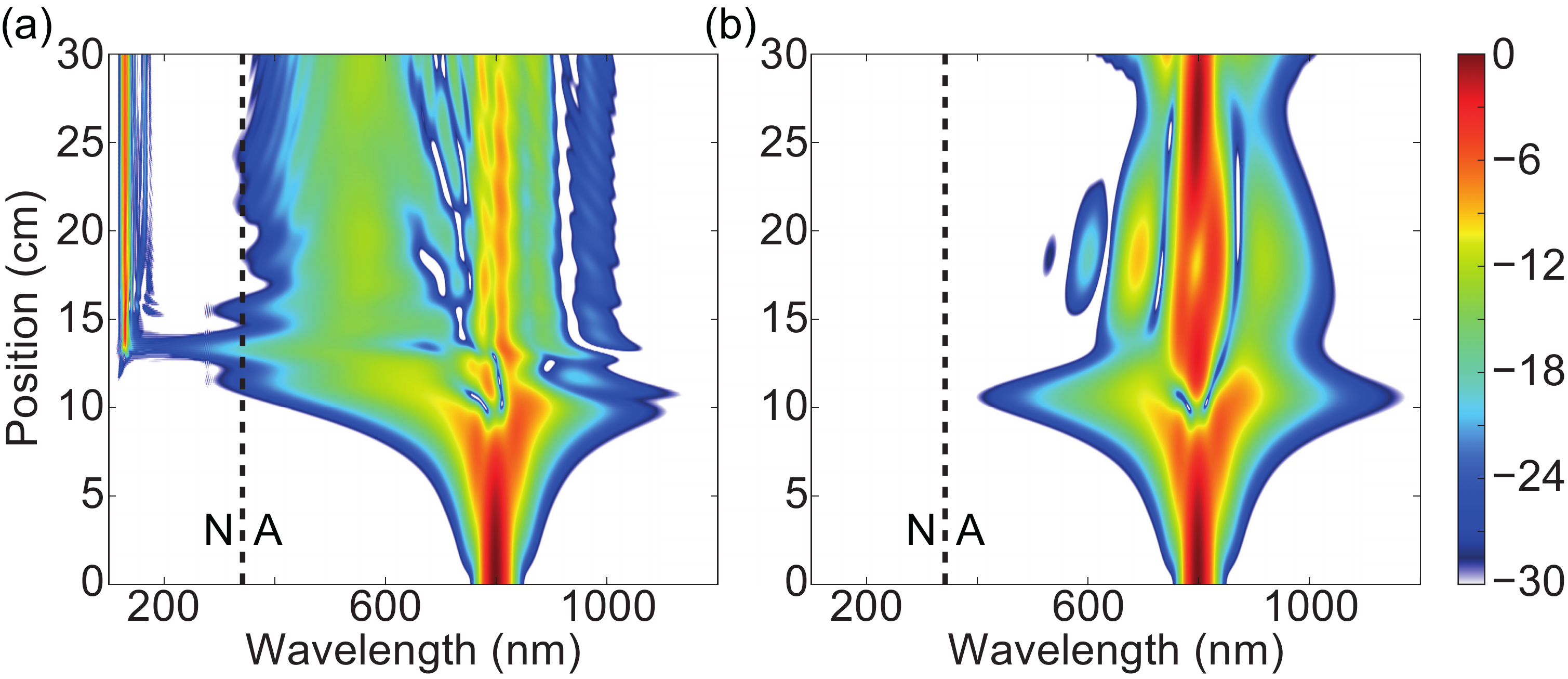}
    \caption{The spectra corresponding to Fig.~\ref{fig:gas_solcomp} are shown with (a) and without (b) the influence of ionization. These parameters correspond to the experimentally observed VUV RDW emission shown in Fig.~\ref{fig:gas_dwave_exp_tuning}. The dashed black lines mark the zero dispersion wavelength (340 nm) and N and A indicate normal and anomalous dispersion. Without ionization the VUV RDW is not emitted. Adapted from \textcite{ermolov_supercontinuum_2015}.}
    \label{fig:RDWplasmaenhance}
\end{figure}
The first demonstration of vacuum-ultraviolet supercontinuum generation was by \textcite{belli_vacuum-ultraviolet_2015}, who used the Raman response in a hydrogen-filled \bhc{} to broaden RDW radiation down to 124~nm. This is discussed further in Section~\ref{sec:vuvraman}.

Subsequently, \textcite{ermolov_supercontinuum_2015} found that blue-shifting solitons strongly enhance VUV dispersive-wave emission in noble gases, as predicted by \textcite{saleh_understanding_2011}. This occurs through two mechanisms. Firstly, the blue-shift pushes the soliton closer to the zero dispersion wavelength, increasing the spectral overlap between the soliton tail and the RDW frequency. Secondly, the pulse compression associated with blue-shifting solitons further enhances this spectral overlap. Both of these exponentially increase the generated dispersive-wave signal. This feature is illustrated in Fig.~\ref{fig:RDWplasmaenhance}, where the spectra corresponding to Fig.~\ref{fig:gas_solcomp} are shown with and without the influence of ionization. With ionization a clear blue-shifting soliton appears around 8~cm. It shifts towards the zero dispersion wavelength and emits a strong dispersive-wave around 130~nm before recoiling. Without ionization, the soliton compression around 800 nm is not sufficient to excite a RDW at the phase-matched point in the VUV, because of a lack of spectral overlap.

\begin{figure}
    \centering
    \includegraphics[width=0.98\linewidth]{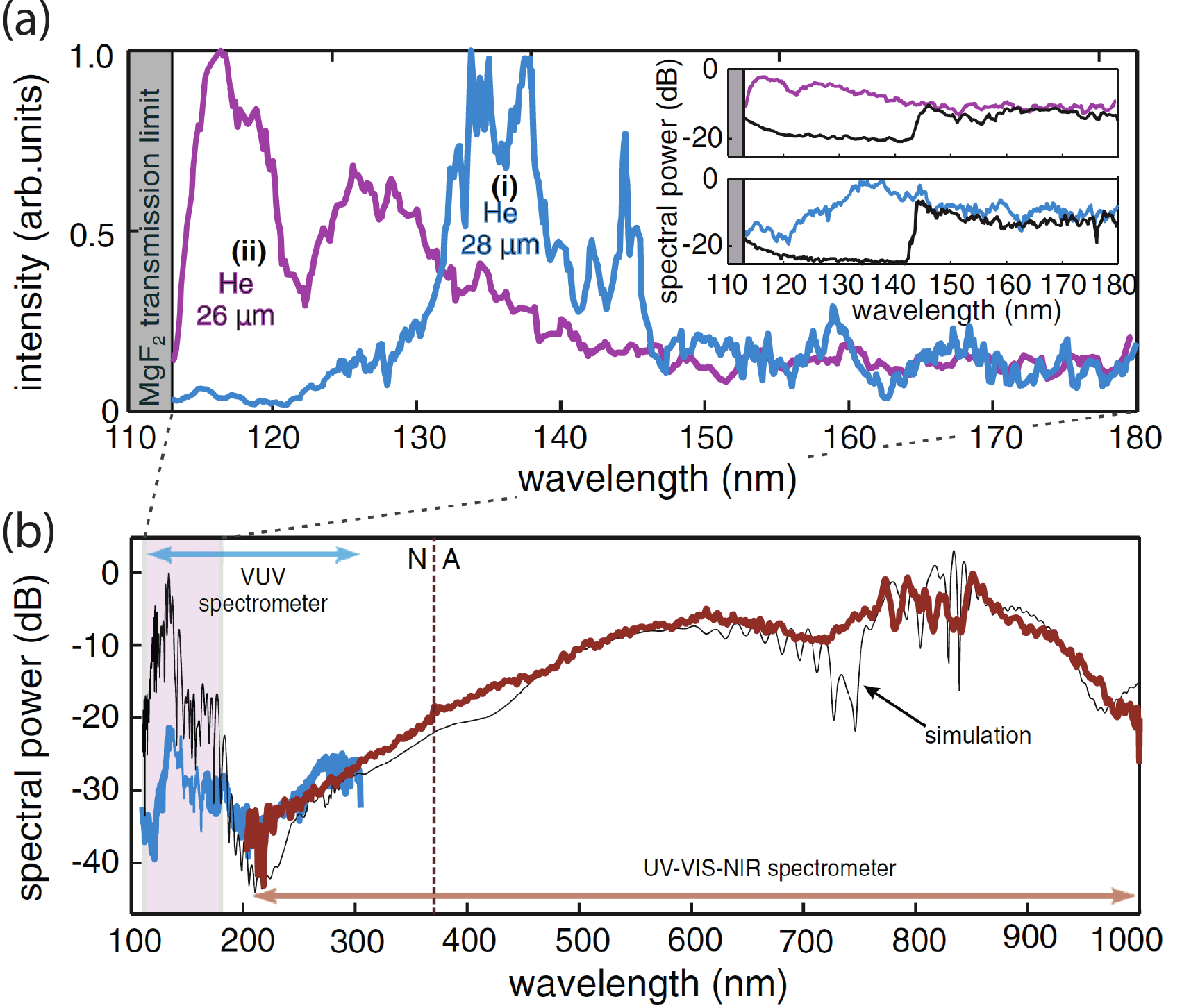}
    \caption{Experimental VUV supercontinuum generation results from \textcite{ermolov_supercontinuum_2015}, in He-filled \bhc{}. (a) (linear scale) for core diameters of (i) 28~\um{} and (ii) 26~\um{}; (inset) the same data on a logarithmic scale, including the spectra recorded through a sapphire filter (black lines). (b) The full supercontinuum corresponding to (a)(i). The solid black line is the simulated spectrum. The dashed vertical line marks the ZDW (N = normal, A = anomalous GVD).}
    \label{fig:gas_vuvscg}
\end{figure}
This mechanism forms the basis of a different VUV supercontinuum generation mechanism in \bhc{}, where \textcite{ermolov_supercontinuum_2015} demonstrated that the combination of VUV RDW emission, plasma soliton blue-shift and four-wave mixing can lead to the formation of a supercontinuum spanning from 113~nm to beyond 1100~nm, shown in Fig.~\ref{fig:gas_vuvscg}.

In addition to the core mechanism of enhanced RDW emission due to blue-shifting solitons, four additional ingredients added to the supercontinuum. (i) After emitting a RDW the soliton spectrally recoils to lower frequencies, where it recompresses and begins to blue-shift once again, emitting a secondary RDW at a slightly lower frequency than the first. This process can repeat several times, such that multiple waves are emitted. (ii) As first discussed by \textcite{tani_multimode_2014}, RDWs can be emitted in higher order modes at higher frequencies. (iii) The RDWs in different modes interact by four-wave mixing (FWM) further filling in gaps in the continuum. (iv) By tuning the parameters such that a soliton compression event occurs at the output of the fiber, other gaps in the spectrum can be filled in. Combined these processes enable the emission of a bright, spatially coherent, continuous spectrum from the VUV to 1100~nm. \textcite{ermolov_supercontinuum_2015} measured down to 113 nm, which was the transmission limit of the MgF2 window used. Numerical results suggest that this continuum is also fully temporally coherent, raising the possibility of compression to sub-femtosecond durations in the optical region.

\paragraph{Generation of Mid-IR RDW\label{sec:dwave_mir}}
An additional interaction between ionization and RDW emission was discovered by \textcite{novoa_photoionization-induced_2015}. The refractive index modulation induced by the fast creation of plasma can lead to additional phase-matched frequencies in the mid-infrared. An additional term was introduced into the nonlinear propagation constant to account for the plasma dispersion, i.e. Eq.~\ref{eqn:dwave_sol} becomes
\begin{multline}
\label{eqn:dwave_ion}
\Delta\beta(\omega)=\beta(\omega)-\beta(\omega_p)-\beta_1(\omega_p)[\omega -\omega_p]-\frac{\omega}{\omega_p}\gamma P_\mathrm{sc}\\
+\frac{\omega_p}{\omega}\frac{\omega_pN_e}{2n_0cN_\mathrm{crit}}=0. 
\end{multline}

The last term is the additional shift in propagation constant due to plasma. It is opposite in sign to the other terms, and so can give rise to new phase-matching solutions. Without it, RDW emission in gas-filled \bhc{} can only occur in the normal dispersion region at up-shifted frequencies. The plasma term in Eq.~\ref{eqn:dwave_ion} provides two additional solutions, one of which lies in the MIR spectral region and shifts to longer wavelengths with increasing free-electron density.  In addition, the shallow dispersion curves of the \bhc{} system ensure extended coherence lengths over a broadband spectral region. The frequency correction factor for the plasma term is also of the opposite sense compared to the Kerr factor---it provides enhancement at lower frequencies with respect to the pump.

Using this scheme, \textcite{novoa_photoionization-induced_2015} numerically predicted emission of a RDW at 4.2~\um{} from a 1.4~\um{} pump wavelength. \textcite{kottig2017novel} experimentally verified this prediction, observing plasma-induced RDW between 3.2~\um{} and 3.8~\um{}, pumped at 1030~nm. This radiation then forms part of a 4.7 octave supercontinuum that spans from 180~nm to 4.7~\um{}, with up to 1.7~W of average power, at 151 kHz repetition rate.

\subsubsection{Microwave induced plasma in \bhc{}\label{sec:micro}}
In the previous discussion the plasma inside the fiber was caused by photo-ionization of the filling-gas by the laser pulse itself. An alternative means of exciting a plasma inside an HC-PCF was introduced by \textcite{debord_generation_2013}. They used an external continuous-wave surfatron microwave coupler as an excitation source to obtain a 6~cm long plasma in 1~mbar of Ar inside a 60~\um{} core diameter HC-PCF. In a further refinement \textcite{vial_generation_2016} miniaturized the setup by using a microstrip split-ring resonator for the microwave excitation. They obtained a very high microwave coupling efficiency and only 10~W of microwave power was required to generate a plasma in the \bhc{} core. These schemes are useful for creating emission sources or perhaps even \bhc{-based} gas laser devices, however, the relatively low gas pressures and electron densities ($10^{15}$~cm$^{-3}$) compared to the laser pulse induced ionization ($10^{18}$~cm$^{-3}$) restricts their use for nonlinear optics applications.

\subsection{Raman effect}
The advent of low-loss step-index fibers and subsequent birth of the field nonlinear fiber optics triggered a revolution in our view of the Raman effect \cite{stolen_raman_1972,stolen_early_2008,clesca_raman_2015}. The advent of \hc{s} has led to a similarly radical transformation. The first report by \textcite{benabid_stimulated_2002} marked the beginning of gas-phase nonlinear optics in \hc{s}. Using a hydrogen-filled \bhc{} \textcite{benabid_stimulated_2002} reported a two orders of magnitude reduction (compared to single-pass Raman gas cells) in the threshold required for the generation of two frequency down- and up-shifted Stokes and anti-Stokes Raman sidebands of a 532~nm pump laser. This changed the view of stimulated Raman scattering (SRS) from a nonlinear process demanding high-energy ($\sim$1~mJ) laser systems, often accompanied by competing nonlinear effects, e.g. self-focusing and/or SPM, to a process with a high-level of controllability, easily achievable using compact laser sources with moderate micro-Joule energies. Driven by the possibility of a meters-long interaction length in a low-loss, gas-filled \pgpcf{} fiber, the pioneering work by \textcite{benabid_stimulated_2002} was followed by a series of studies which resulted in further reduction in the required threshold for SRS and the realization of sub-Watt, continuous-wave-pumped gas-Raman laser \cite{benabid_ultrahigh_2004,benabid_compact_2005, couny_subwatt_2007}.

\subsubsection{Spontaneous Raman Scattering}
Raman scattering \cite{raman_new_1928,long_raman_2002} is a two-photon inelastic scattering process with an intrinsically quantum mechanical origin \cite{berestetskii_quantum_1982,sakurai_advanced_1967,baym_lectures_1969,long_raman_2002}. In the so-called energy transfer model, a photon of frequency $\omega$ incident on a molecule in an initial state $\left|i\right\rangle $ is inelastically scattered off the electron cloud resulting in a photon of frequency $\omega\prime$, while leaving the molecule in a final state $\left|f\right\rangle$. The energy of the scattered photon is given by $\hbar\omega'=\hbar\omega-\hbar\omega_{fi}$ where $\hbar\omega_{fi}=\hbar(\omega_f-\omega_i)$ is the quantum defect. If the final state is an excited state, i.e. $\omega_f>\omega_i$, then $\omega_{fi}$ is positive and the scattered photon is a red-shifted Stokes photon. If the initial state is an excited state, i.e. $\omega_i>\omega_f$ then the scattered photon is a blue-shifted anti-Stokes photon. The coupling between the electromagnetic field and the molecule is mediated by the change in the molecular polarizability tensor, $\boldsymbol{\mathrm{\alphaup }}$ as the molecule oscillates, i.e. rotates or vibrates. Considering an incident power of $P_{\mathrm{inc}}$ over a sample volume $V$ with an area $A$, a length $\mathrm{d}z$, and a number density of the scattering centers, $N$, then the average power of light at Stokes frequency scattered into a unit solid angle in angular direction $\left(\theta ,\ \phi \right)$ from an incident beam of power $P_{\mathrm{inc}}$ is given by $P_{\mathrm{scat}}=N_tP_{\mathrm{inc}}\left({\omega }_{\mathrm{s}}/{\omega }_{\mathrm{p}}\right){\left(\mathrm{d}\sigma /\mathrm{d}\mathit{\Omega}\right)}_{\left(\theta ,\phi \right)}\ \mathrm{d}z$. Here ${\left(\mathrm{d}\sigma /\mathrm{d}\mathit{\Omega}\right)}_{\left(\theta ,\phi \right)}$ is the Raman scattering cross-section per molecule \cite{yariv_quantum_1989}.

Spontaneous Raman scattering is an incoherent process \cite{wang_theory_1969}. This incoherent nature allows for the accumulation of the spontaneous Raman signal from different scattering centers with an increase in the interaction length, $L$, i.e. $P_{\mathrm{scat}}\propto LP_{\mathrm{inc}}$. Long interaction lengths are offered by low-loss waveguides such as HC-PCFs, which makes them very attractive for Raman spectroscopy and sensing applications. It should be noted, however, that for the purpose of performance studies one must carefully take into account the waveguide loss which limits the effective interaction length, and the collection efficiency of the scattered radiation by the waveguide \cite{holtz_small-volume_1999,eftekhari_comparative_2011,buric_enhanced_2008}.

\paragraph{Spontaneous Raman scattering: gas phase}

The differential Raman cross-section is small, typically in the range of $10^{-34}$~m$^{2}$molecule$^{-1}$sr$^{-1}$. Consequently spontaneous Raman scattering is weak, which is particularly problematic in the case of Raman spectroscopy and detection of minute sample volumes and trace gases. Given low propagation losses, increasing the interaction length and the spatial overlap between the incident laser radiation and the specimen loaded in the micron-size core of a HC-PCF, significantly enhances the Raman signal. In general these characteristics make a HC-PCF an ideal platform for a multitude of optical sensing applications. In a spectacular demonstration \textcite{millot_frequency-agile_2015} combined advanced concepts in telecommunication and supercontinuum photonics with a long-interaction length of $\sim$50~m in a \pgpcf{}, to demonstrate highly-sensitive absorption spectroscopy without using any mode-locked lasers or high-finesse multi-pass gas cells.

Although ideas regarding the use of hollow waveguides and photonic crystal fibers as fiber sensors have a long history \cite{sudo_optical_1990,saito_infrared_1997,monro_developing_1999,monro_sensing_2001}, there has been a strong interest in using HC-PCFs for (Raman) sensing application, partly because of their low-propagation losses in the visible and near infrared spectral region for small core diameters ($\sim$10-50~\um{}). \textcite{ritari_gas_2004} validated the feasibility of using \pgpcf{s} for sensing applications, demonstrating the potential of these fibers for the detection of minute sample volumes as well as their compatibility with standard fiber-optics components. A major step in this direction was made by \textcite{benabid_compact_2005} who successfully realized an all-fiber gas cell consisting of a \pgpcf{} filled with gas and spliced hermetically at both ends to standard single-mode optical fibers.

Fiber optics probes are instrumental in modern Raman spectrometers for in vitro measurements of analytes in aqueous environments \cite{santos_fiber-optic_2005,motz_optical_2004}. A major problem in using solid-core fiber probes is the strong Raman background from the fused silica. This mainly comes from the fiber's core and hinders the detection of the weak Raman signal from the sample under study. \textcite{may_fiber_1996,konorov_hollow-core_2006} neatly solved this problem by using a 19-cell \pgpcf{} fiber as a Raman probe for transmitting an intense excitation pulse at 532~nm to the sample. The \pgpcf{} fiber was bundled with three larger fused-silica solid core fibers (core diameter of 400~\um{}) for collecting and transferring the scattered Raman signal from the sample back to the spectrometer. In spite of this they showed an order of magnitude reduction in the Raman background signal, with the remaining background noise coming mainly from the collection fibers.

The relatively large size of the collection optics in the work of \textcite{konorov_hollow-core_2006} may pose a limitation for some applications requiring fine probes. \textcite{brustlein_double-clad_2011} used a double-clad \pgpcf{} fiber as a compact, alternative collection strategy. By incorporating a second cladding as a high numerical aperture multimode waveguide channel (see \textcite{myaing_enhanced_2003,fu_nonlinear_2005}) they could greatly enhance the detection sensitivity. Using this fiber they demonstrated fiber-based, label-free imaging using coherent anti-Stokes Raman scattering (CARS) microscopy (see Section~\ref{sec:cars}). An order of magnitude reduction in the background noise over the results of \textcite{konorov_hollow-core_2006} was reported by \textcite{ghenuche_kagome_2012} using a large-core \bhc{}. In their all-HC-PCF excitation-detection scheme, \textcite{ghenuche_kagome_2012} demonstrated that due to the wide transmission window of the \bhc{}, as compared to \pgpcf{} fiber, both the excitation pulse and the Raman signal could be respectively delivered and collected in the large core of the fiber. Interesting enough, the background noise, mainly originating from the tiny silica bridges between the hollow channels, appeared in a different spatial pattern distribution than the core mode which carried the main Raman signal. This allowed them to reject the background noise simply by spatial filtering of the output of the fiber using a 20~\um{} pinhole conjugated to the fiber's output tip.

In an attempt for a real-time measurement of (trace) gas concentration with generally low Raman cross-section, \textcite{buric_enhanced_2008} demonstrated the gas-phase Raman spectroscopy in a 1.5~m long \pgpcf{} (core diameter 4.9~\um{}) with a signal improvement factor of $\sim$500 over the conventional free space configuration. They conclude that, compared to an optimum collection geometry in free space, this enhancement is mainly due to the long interaction length offered by a HC-PCF. Moreover, \pgpcf{} showed also a better improvement when compared to hollow waveguides ($\sim$1~m long) which showed enhancement factors of 30 and 50 over free space for normal and resonance Raman scattering, respectively \cite{schwab_remote_1987}. This point was attributed to the possibility of collecting the spontaneous Raman signal in low-loss higher order modes, in addition to the fundamental mode. This is a unique opportunity not possible with hollow capillaries due to the very high loss of higher-order modes \cite{marcatili_hollow_1964}. Further research on improving the detection sensitivity in HC-PCF for minute volumes of various samples is being carried on \cite{buric_improved_2009,yang_direct_2012,yang_high_2013,hanf_fast_2014}.

\paragraph{Spontaneous Raman scattering: liquid phase}

A liquid-core hollow waveguide may offer low-loss propagation if, by careful choice of the filling liquid and the waveguide material, the conditions for total internal reflection is satisfied. \textcite{ippen_low-power_1970} demonstrated the use of a liquid-core hollow capillary for SRS. For Raman spectroscopy, enhancement in the sensitivity by factors 100--1000 compared to free space are also reported in long ($\sim$20~m) index-guiding, liquid-filled hollow capillaries (fused silica, core diameter 75~\um{}). However, this method is limited to liquid samples with refractive indices higher than fused silica at the excitation and Raman signal wavelengths \cite{walrafen_intensification_1972}. The use of the Teflon fibers (refractive index $\sim$1.35) for enhancing the spontaneous Raman signal in aqueous samples (refractive index $\sim$1.3445 at 400~nm) and organic liquids, were also demonstrated although with much higher propagation losses \cite{walrafen_optische_1974,tian_raman_2007,dallas_light_2004}. 

\textcite{fini_microstructure_2004} suggested a novel design for a broadband-guiding liquid-core PCF in which guidance relied on the total internal reflection at the interface between the low-index liquid core and the hybrid air-silica cladding with a high air-filling fraction. In this design the presence of air enclosures in the cladding lowers its effective refractive index compared to the refractive index of the liquid. \textcite{nielsen_selective_2005} realized this experimentally. A different approach for optical sensing in liquids based on complete filling (core + cladding) of a HC-PCF was demonstrated by \textcite{cox_liquid-filled_2006}, using a broadband-guiding polymer HC-PCF. Further increase in the sensitivity was obtained by the development of new strategies for using HC-PCFs as surface-enhanced Raman scattering (SERS) probes \cite{yan_hollow_2006,zhang_liquid_2007,cox_surface_2007}, see also \textcite{yang_hollow-core_2011} and references therein.

\paragraph{CARS in HC-PCFs\label{sec:cars}}

Several studies have dealt with realizing nonlinear spectroscopic methods such as coherent anti-Stokes Raman scattering (CARS) in HC-PCF for an enhanced sensitivity \cite{fedotov_coherent_2004,konorov_phase-matched_2005,fedotov_raman-resonance-enhanced_2006,zheltikova_toward_2006}. HC-PCFs are ideally suited for gas-phase CARS spectroscopy because they offer a long interaction length, a high laser-matter spatial overlap, a low nonlinearity due to the negligible overlap with glass, and a tunable dispersion, e.g. by changing the gas pressure \textcite{travers_ultrafast_2011}. In CARS, specific molecular transitions are excited through interaction between a pump (P) and a broadband or tunable Stokes signal (S). The resulting coherence wave (see Section~\ref{sec:coherentraman}) of the molecular oscillation is probed by pump photons, resulting in the generation of a blue-shifted anti-Stokes (AS) signal. CARS is efficient if phase-matching is fulfilled, i.e., $\mathit{\Delta}\beta =2{\beta }_{\mathrm{P}}-{\beta }_{\mathrm{S}}-{\beta }_{\mathrm{AS}}\approx 0$ for each Raman transition. This is not a trivial task in the collinear fiber geometry due to the large Raman shifts which are typical of gases (up to 125 THz or 4166~cm${}^{-1}$ for hydrogen). \textcite{ziemienczuk_intermodal_2012} studied this problem and showed the possibility of phase-matching and intermodal Raman scattering using the slightly multi-mode nature of a \pgpcf{}. The body of work following this demonstration has clearly shown the importance of higher-order waveguide modes for phase-matching and efficient generation of higher-order Stokes and anti-Stokes components \cite{trabold_efficient_2013,trabold_amplification_2013,bauerschmidt_supercontinuum_2014} as well as the dynamics of SRS in HC-PCFs. In a neat demonstration \textcite{bauerschmidt_dramatic_2015} revisited parametric gain suppression is SRS \cite{Duncan:86} using an H${}_{2}$-filled \bhc{}, showing clear first Stokes generation in a higher-order mode while operating at the exact phase-matching point for the pump, Stokes and anti-Stokes in the fundamental mode.

Although it is possible to arrange phase-matching for a specific Raman transition in a single gas species, for example, by using higher order modes \cite{trabold_efficient_2013}, this is not suitable for CARS in multi-component analytes. Using pressure-tunable zero dispersion wavelength of a H${}_{2}$-filled \bhc{} \textcite{bauerschmidt_broadband-tunable_2015} demonstrated all-fundamental mode (LP${}_{01}$) collinear phase-matched frequency shifting of a broadband supercontinuum. This paved the way for achieving phase-matching over a broad wavelength range for multi-species CARS spectroscopy in \bhc{s} as recently demonstrated by \textcite{hupfer_multi-species_2016}.

\subsubsection{Coherent Raman effect\label{sec:coherentraman}}

Despite the incoherent nature of the spontaneous Raman scattering, it is possible to use the Raman effect in the stimulated regime in order to place the molecules in a coherent superposition of the molecular levels involved in the scattering process. This coherent Raman effect (CRE) is of paramount importance in nonlinear optics since the coherently prepared molecules oscillate in-phase at the molecular vibrational/rotational frequency, ${\mathrm{\Omega }}_{\mathrm{R}}$, macroscopically manifesting their synchronous motion as a fast varying refractive index, or coherence wave. Mathematically this is represented by the rapidly oscillating, off-diagonal elements of the density matrix $\rho_{if}=\rho_{fi}^*\propto e^{\mathrm{i}{\mathrm{\Omega }}_{\mathrm{R}}t}$ where $i\neq f$. The time-dependent refractive index, or molecular modulator, modulates an incident field at the incredibly high molecular oscillation frequencies, approaching $\sim$100's of terahertz in the case of diatomic gases, causing a cascaded generation of new frequencies. This enables generation and modulation of ultrashort pulses \cite{zhavoronkov_generation_2002,noack_generation_2005,chan_synthesis_2011}.

Due to the large energy spacing between their vibrational/rotational states, when interacting with an external field, simple diatomic gases behave to a good approximation as two-level systems. Assuming that the electric field, $E(t)$ is linearly polarized and parallel to the \textit{z} axis, the interaction of an ultrashort pulse described by $E(t)$ with a two-level medium beyond the slowly varying envelope approximation is given by a set of Bloch equations \cite{belli_vacuum-ultraviolet_2015},

\begin{align}
\label{eqn:bloch_coh}
\left (\frac{\partial}{\partial t}+\frac{1}{T_2}-i\Omega_{\mathrm{R}} \right )\rho_{if}&=\frac{i}{2\hbar}\left[ \left ( [\alpha_{zz}^{ii}-\alpha_{zz}^{f\!f}\right )\rho_{if}+\alpha_{zz}^{if} w \right ]E(t)^2,\\
\label{eqn:bloch_pop}
\frac{\partial w}{\partial t}+\frac{w+1}{T_1}&=\frac{i}{\hbar}\alpha_{zz}^{if}\left ( \rho_{if} - \rho_{if}^* \right ) E(t)^2,
\end{align}
where $w={\rho}_{f\!f}-{\rho}_{ii}$ represents the population difference between the initial and final states. Here $\alpha_{zz}^{mn}$, where subscript \textit{zz} represents the \textit{zz}--component of the polarizability tensor, are the elements of the $2\times2$ transition polarizability matrix with diagonal, ${\alpha}_{zz}^{ii}$ and ${\alpha}_{zz}^{f\!f}$, and nondiagonal elements ${\alpha}_{zz}^{if}=({\alpha}_{zz}^{fi})^*$, responsible for the Stark shift and the Raman scattering, respectively (see \textcite{long_raman_2002}). Note that based on our definition of $\boldsymbol{\mathrm{\alphaup }}$, the polarizability has the SI units of [F.m${}^{2}$] \cite{butcher_elements_1991}. Its value may be converted to cgs unit (${\alpha }_{\mathrm{cgs}}[\mathrm{c}{\mathrm{m}}^{\mathrm{3}}]$) by dividing the SI value by $4\pi {\varepsilon }_0\times {10}^{-6}$, and to atomic unit by dividing the SI value by $1.64878\times {10}^{-41}$. \textit{T}${}_{2}$ is the dephasing time of the Raman-induced nonlinear polarization and is inversely proportional to the Raman-gain bandwidth (FWHM) $\mathit{\Delta}{\nu }_{\mathrm{R}}={\left(\pi T_2\right)}^{-1}$. \textit{T}${}_{1}$ is the relaxation time of the population. 

For the case that the change in the population of the initial state in the course of the interaction could be neglected, i.e. $w \approx -1$, Eq.~\ref{eqn:bloch_coh} may be solved to obtain ${\rho }_{if}$,

\begin{equation}
\label{eqn:coh}
{\rho }_{if}\left(t\right)=\frac{{{\alpha }_{zz}^{if}}}{8{\pi }^2\hslash}{\iint^{+\infty }_{-\infty }{\frac{\tilde{E}\left(\omega \right){\tilde{E}}^*\left(\omega'\right)e^{\mathrm{i}\left(\omega'-\omega\right)t}}{{\mathrm{\Omega }}_{\mathrm{R}}-\left(\omega'-\omega\right)+i/T_2}\mathrm{d}\omega \mathrm{d}\omega'}},
\end{equation}
where $\tilde{E}\left(\omega \right)=\int^{+\infty }_{-\infty }{E\left(t\right)e^{i\omega t}\mathrm{d}t}$. The change in the refractive index is then given by $\delta n\left(t\right)=\frac{N_t}{{\varepsilon }_0n_0}{\alpha}_{zz}^{if}\mathfrak{Re}\left\{{\rho }_{if}\right\}$ where $n_0$ is the refractive index.

\paragraph{Methods of excitation of the coherent Raman effect}

Studies of CRE are mainly use two techniques for the excitation of the Raman coherence: \textit{i)} impulsive stimulated Raman scattering (ISRS) \cite{de_silvestri_femtosecond_1985,yan_impulsive_1985,nazarkin_generation_1999}, and \textit{ii)} two-color pumping (TCP) of a molecular ensemble \cite{imasaka_generation_1989,yoshikawa_new_1993,harris_broadband_1997,harris_subfemtosecond_1998}. The reason can be seen from the Eq.~\ref{eqn:coh}. In the TCP scheme the frequency difference between the two driving pulsed lasers, typically several nanoseconds in duration which is much longer than one cycle of the molecular oscillation, $T_{\mathrm{m}}=\frac{2\pi }{{\mathrm{\Omega }}_{\mathrm{R}}}$ typically lying in the femtosecond range, is brought to the close proximity of ${\mathrm{\Omega }}_{\mathrm{R}}$. This increases the denominator of the integrand in Eq.~\ref{eqn:coh} thus increasing the coherence. In TCP the two driving lasers could be independent or mutually coherent, i.e. coherent pumping \cite{chan_synthesis_2011,katsuragawa_efficient_2010}. It is well known that the TCP establishes a strong molecular modulator that can very efficiently phase modulate the incident electric field in order to collinearly generate Raman sidebands that are separated in frequency by the modulation frequency, $T^{-1}_{\mathrm{m}}$. In ISRS, however, a pulse with a duration, ${\tau }_{\mathrm{p}}$ shorter than $T_{\mathrm{m}}$, i.e. ${\tau }_{\mathrm{p}} \ll T_{\mathrm{m}}$, creates non-stationary rotational/vibrational wave packets in the molecular ensemble by exerting an impulse on the molecules. Because ${\tau }_{\mathrm{p}} \ll T_{\mathrm{m}}$ then in the frequency domain the bandwidth of the pulse is broad enough to already include pairs of frequencies that their difference is in resonance with the Raman frequency shift, ${\mathrm{\Omega }}_{\mathrm{R}}$. In contrast to the TCP, in the impulsive regime the driving pulse experiences a continuous increase in its bandwidth accompanied by a frequency downshift of its frequency centroid \cite{korn_observation_1998}.

\subsubsection{Early developments of coherent Raman effects in HC-PCF}

\begin{figure}
    \centering
    \includegraphics[width=0.98\linewidth]{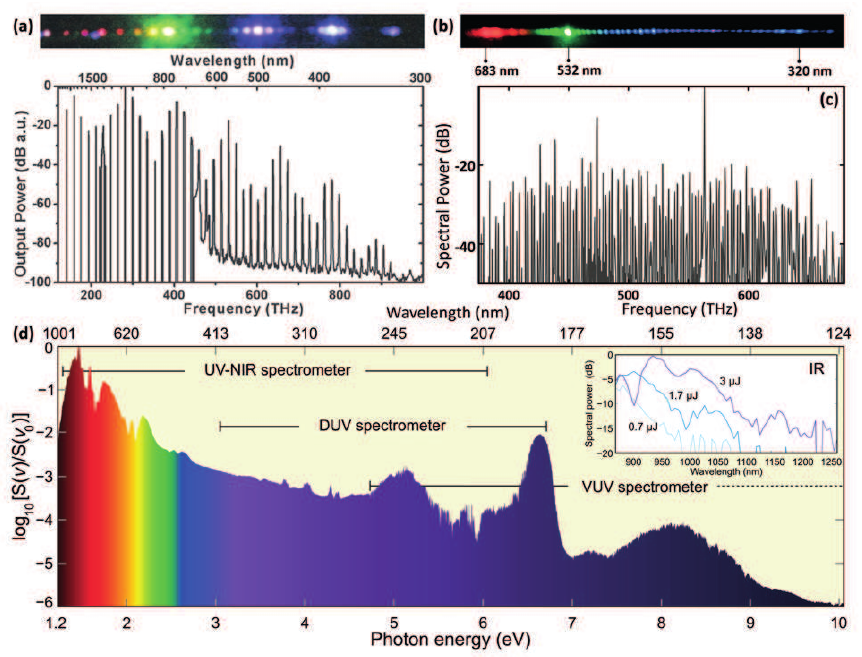}
    \caption{Several major results for Raman frequency comb generation and broadening in H${}_{2}$-filled HC-PCFs. (a) \textcite{couny_generation_2007}: image and spectrum of the generated and transmitted ro-vibrational Raman comb through a $\sim$1 m long hydrogen-filled ($\sim$20 bar) \bhc{} for a circularly polarized laser input (12 ns, 1064 nm). (b) \textcite{hosseini_two-octave-wide_2016}: Image of a Raman comb generated in a $\sim$3 m long \bhc{} filled with a mixture of H${}_{2}$ (5 bar) and D${}_{2}$ (8 bar) for a circularly polarized laser input (1 ns, 532 nm). (c) \textcite{hosseini_two-octave-wide_2016}: Spectrum of a Raman comb generated in a \bhc{} filled with a mixture of H${}_{2}$, D${}_{2}$, and Xe. More than 135 lines are generated in visible spectral region. (d) \textcite{belli_vacuum-ultraviolet_2015}: a vacuum UV till IR supercontinuum generated in a $\sim$15~cm long hydrogen-filled (5 bar) \bhc{} pumped with ultrashort pulses (50~fs, 800~nm).}
    \label{fig:ramresults}
\end{figure}

Although both techniques of ISRS and TCP are well developed, studies of CRE in gas-filled HC-PCFs have so far progressed with a slow pace with pump-probe experiments in the impulsive Raman regime emerging only recently \cite{abdolvand_impulsive_2015,belli_vacuum-ultraviolet_2015,belli2017control}. This can be perhaps attributed to the high values of the Raman gain offered by HC-PCFs which allow straightforward and easy generation of octave-spanning Raman combs, even when the process starts from quantum fluctuations \cite{couny_optical_2009,wang_compact_2010,tani_generation_2015,benoit_over-five_2015}. As a result, the focus of the early research on CRE in HC-PCFs, with a very few exceptions \cite{konorov_phase-matched_2005,ivanov_frequency-shifted_2006}, has been mainly on pumping a gas-filled HC-PCF with quasi-continuous pump sources (nanosecond pulses to continuous wave). In a beautiful demonstration \textcite{couny_generation_2007} used a hydrogen-filled \bhc{} for the generation and photonic guidance of multi-octave optical-frequency combs of (rotational-) vibrational Raman sidebands covering the spectral range from 325~nm to 2300~nm, using 12~ns pulses of 40~kW peak power, Fig.~\ref{fig:ramresults}(a). Remarkably the generation of this wide Raman comb from quantum fluctuations and its subsequent amplification was performed in a single pass along the fiber, combining the generation and amplification stages of the SRS. For spectroscopic applications, however, it is desirable to have a closer line spacing. This can be achieved by using a mixture of several Raman-active gases. Recently \textcite{hosseini_two-octave-wide_2016} have successfully demonstrated this using a mixture of H${}_{2}$, D${}_{2}$, and Xe, Fig.~\ref{fig:ramresults}(b). Mixing of Xe as a non-Raman-active gas allows for controlling the dispersion without changing the ratio of the overall Raman gain offered by H${}_{2}$ and D${}_{2}$.

Using the classical description of SRS, \textcite{nazarkin_optimizing_2009} highlighted the role of the parametric coupling and phase-locking between Stokes and anti-Stokes components in the efficient frequency up-conversion of the pump in \hc{s}. This explained some of the peculiarities of the 2002 demonstration of SRS in \bhc{} by \textcite{benabid_stimulated_2002}, namely the remarkably high conversion efficiency of 3\% to the first anti-Stokes in the presence of a large phase mismatch. By developing the quantum theory of Raman comb generation in gas-filled HC-PCF, the existence of a self- and mutual coherence between the Raman sidebands was predicted theoretically \cite{couny_generation_2007,wu_quantum_2010}. The presence of these phase correlations is despite the inherent phase fluctuations of the individual sidebands from pulse to pulse. The existence of these phase correlations was subsequently verified experimentally \cite{wang_quantum-fluctuation-initiated_2010,abdolvand_generation_2012}.

Although \pgpcf{}s were quite successful in reducing the threshold requirement for SRS, it was realized since the early stages that their limited transmission windows, in the range of $\sim$50~THz bandwidth, are restricting for increasing the spectral extend of multi-component Raman combs \cite{beaudou_matched_2010}. The development of \bhc{s} with exceedingly large transmission windows, covering the vacuum ultraviolet to infrared range, has offered an excellent solution to this issue (see below). However, the spectral filtering effect in \pgpcf{}s has itself been exploited to demonstrate unique capabilities of these fibers in enabling ultra-low power gas-phase nonlinear optics and studies of the long-distance dynamics of SRS. \textcite{benabid_ultrahigh_2004} used this to demonstrate pure rotational SRS using $\sim$1~ns pulses of only $\sim$1~nJ energy in a 35~m long H${}_{2}$-filled \pgpcf{}. The spectral filtering of \pgpcf{} prevented energy transfer to the normally dominant vibrational Stokes band of H${}_{2}$ as well as higher-order rotational Raman sidebands, all of which were lying outside the transmission window of the fiber. In yet another demonstration \textcite{couny_subwatt_2007} showed sub-watt threshold Raman laser action in a 30~m long \pgpcf{} filled with H${}_{2}$ at 10~bar.

\textcite{abdolvand_solitary_2009} used the spectral filtering effect in a \pgpcf{} to demonstrate the long-distance dynamics of the Stokes pulse in the backward transient stimulated Raman scattering (TSRS) in hydrogen. By TSRS we mean that the duration of either the pump or the Stokes is shorter than \textit{T}${}_{2}$ \cite{raymer_quantum_1985}. They demonstrated late-stage coherent amplification and shortening of the backward Stokes pulse well below \textit{T}${}_{2}$. They showed that before the pulse reaches its asymptotic solitary shape \cite{maier_backward_1969}, the amplification saturated due to the formation of a reshaped, stable pulse envelope, with the peak of the pulse envelop rolling with superluminal velocity over its (compact) support \cite{icsevgi_propagation_1969}. \textcite{nazarkin_direct_2010} used a similar H${}_{2}$-filled \pgpcf{} in order to study the long-distance dynamics of the Stokes pulse in the forward TSRS. In their 2010 paper \citeauthor{nazarkin_direct_2010} reported the first experimental observation of self-similar solutions of forward TSRS, also known as accordions (see \textcite{levi_self-similarity_1994} and the references therein), in a ``clean'' system of two frequencies, ``pump + Stokes'' coupled through the Raman nonlinearity. Accordions are the non-solitonic solutions of the sine-Gordon equation associated to TSRS. In the presence of memory they dominate the long-distance evolution of the system instead of solitons \cite{menyuk_long-distance_1993,menyuk_asymptotic_1989,menyuk_self-similarity_1992}. Two difficulties had severely hampered the previous attempts for observing accordions in free-space interaction geometry, namely designing an experiment of sufficient length to observe the phenomenon, while avoiding the parasitic interference from the higher-order Raman sidebands \cite{duncan_transient_1988}. By offering both a long interaction length and eliminating the higher order Stokes and anti-Stokes bands, \pgpcf{s} provided an elegant solution to these problems.

\subsubsection{Recent developments of coherent Raman effects in HC-PCF\label{sec:vuvraman}}

Coherent Raman scattering holds promise for the generation of ultrashort pulses in the visible and ultraviolet spectral region \cite{baker_femtosecond_2011}. Generation of a broad Raman comb via TCP of an ensemble of molecular hydrogen in a room-temperature gas cell at atmospheric pressure has enabled researchers to generate a train of ultrashort pulses of $\sim$1~fs duration with full control over their carrier envelope phase \cite{chan_synthesis_2011}. By driving the vibrational Raman transition in H${}_{2}$ with a pair of $\sim$3~ns pulses at 2.4~\um{} and its second harmonic (1.2~\um{})---commensurate with the fundamental vibrational frequency in H${}_{2}$ at ${\nu }_{\mathrm{R}}=125\ \mathrm{THz}$---researchers insured a well-defined phase relation between the driving fields, i.e. pump at 1.2~\um{} and Stokes at 2.4~\um{}. In contrast, the Raman comb in the work of \textcite{couny_generation_2007} originated from quantum zero-point fluctuations, bearing a large phase and energy fluctuations from pulse to pulse. However, propagation in the high-gain transient regime offered by the HC-PCF established a robust, mutual phase correlation between the comb lines. This opens up the possibility of attosecond pulse train generation with non-classical correlations between the amplitudes and phases of their individual spectral constituents. 

In contrast to the steady state, Raman gain in the transient regime is rather insensitive to the increase in the Raman gain bandwidth \cite{raymer_quantum_1993}. By exploiting this property and using a \bhc{} filled with a mixture of H${}_{2}$-D${}_{2}$-Xe \textcite{hosseini_two-octave-wide_2016} have been able to generate a spectral cluster of more than 135 ro-vibrational Raman sidebands covering the visible spectral region with an average spacing of only 2 THz, Fig.~\ref{fig:ramresults}(c). 

Although in recent years Raman comb generation via molecular modulation has been the subject of intense studies, by virtue of the Fourier transform theorem this method is incapable of constructing single isolated ultrashort pulses. On the other hand, impulsive excitation of the coherent molecular oscillations with femtosecond pulses has proven successful for the generation of isolated femtosecond pulses in the visible and ultraviolet spectral region \cite{zhavoronkov_generation_2002}. In this regard, fast molecular oscillations such as vibrations of light diatomic molecules, e.g. H${}_{2}$ or D${}_{2}$, are more favorable as they offer a larger modulation bandwidth (${\nu }_{\mathrm{R}}$ $\sim$100~THz, $T_m\sim 10$ fs) for the generation of shortest achievable pulses using this method. However, since impulsive excitation of a molecular oscillation generally requires a pump pulse with a duration, ${\tau }_{\mathrm{p}}$ shorter or comparable to $T_{\mathrm{m}}$, i.e. ${\tau }_{\mathrm{p}}/T_{\mathrm{m}}\le 1$, pulse durations required for efficient impulsive excitation of these fast vibrations are expected to be very short, ${\tau }_{\mathrm{p}}$ $\sim$10~fs. As an example for ortho-H${}_{2}$, the time scales of the molecular rotational and vibrational motions are $T^{\mathrm{rot}}_{\mathrm{m}}\sim 57$~fs, and $T^{\mathrm{vib}}_{\mathrm{m}}\sim 8$~fs, respectively. While pulse duration below 57~fs may directly be obtained from normal Ti-Sapphire amplifiers, generation and faithful control and delivery of high-energy pulses of less than 8~fs to a gas target is still challenging. This has significantly hampered studies of ISRS in light diatomic gases

HC-PCFs offer a neat solution to this problem. As shown in Section~\ref{sec:fsscomp}, the soliton propagation dynamics of ultrashort pulses in the anomalous dispersion regime of a gas-filled HC-PCF is accompanied by temporal self-compression. This offers a novel way to access fast molecular oscillations without a need for originally ultrashort pulses below $T_{\mathrm{m}}$. In an elegant demonstration, starting from 50~fs ultrashort pulses, \textcite{belli_vacuum-ultraviolet_2015} showed how using the soliton self-compression effect one can efficiently excite fast fundamental vibrational mode of hydrogen. Direct impulsive excitation of the fundamental vibrational mode in hydrogen would have required 8~fs pump pulses. By using self-compression, which is enhanced by the Raman contribution to the nonlinear refractive index, the 50~fs pump pulses used by \textcite{belli_vacuum-ultraviolet_2015} compressed to below 2~fs, thus generating a broad supercontinuum expanding from vacuum ultraviolet (VUV; $\sim$124~nm) to beyond 1200~nm in the infrared, Fig.~\ref{fig:ramresults}(d). This was the both the first demonstration of VUV guidance, and also the first VUV supercontinuum generated, in \bhc{.} The mechanism proceeds as follows. First the pump pulse emits a RDW in the VUV ($\sim$180~nm). Subsequently, the strong Raman coherence created via impulsive Raman scattering of the pump pulse modulates and broadens the RDW further into the VUV, down to 124~nm. The mechanism behind the VUV extension of the supercontinuum in the work of \textcite{belli_vacuum-ultraviolet_2015} is a unique demonstration of the interplay between VUV RDW generation and a pre-existing Raman coherence. More recently, \textcite{belli2017control} used the creation of a Raman coherence by a leading pulse to exert precise control on a probe pulse in a \bhc{}, even when the probe pulse contains similar energy to the pump. By timing the relative delay of the probe, either enhanced temporal probe compression, probe broadening, or steepening of the leading or trailing edge of the probe pulse was achieved. In addition, the frequency of deep-UV RDW emission was also modulated using the Raman coherence. Such coherent control of deep-UV pulses can be useful, for example, to advanced ultrafast spectroscopy in the UV.

With regard to soliton dynamics, the work of \textcite{belli_vacuum-ultraviolet_2015} is one manifestation of the interaction between the Raman response of gases and soliton dynamics in HC-PCF. In fact some of the earliest reports on the propagation of solitons in \pgpcf{} demonstrated the presence of a Raman-soliton self-frequency shift as described in Section~\ref{sec:solsinpbf}. In the context of \bhc{}, which can be expected to support much wider shifts, this was further explored by \textcite{saleh_tunable_2015}, who numerically predicted a tuning range of 1.3~\um{} to 1.7~\um{} in a hydrogen filled fiber. Further numerical results indicate that gas-filled \bhc{} offers a very rich environment for nonlinear soliton-Raman interactions, such as temporal analogues of condensed matter physics \cite{saleh_strong_2015,saleh_raman-induced_2015}, which have yet to be experimentally explored.

\subsection{Near and mid infrared supercontinuum generation\label{sec:mirscg}}
Many of the self-compression, RDW emission, plasma and Raman driven effects described above gave rise to extreme spectral broadening that can be classed as supercontinuum generation. These were predominantly concentrated in the visible, UV and the short-edge of the near-infrared spectral regions.

The extraordinary low-loss guidance fibers achieved in the MIR by \textcite{yu_spectral_2013,yu_negative_2016}, even with silica glass, has stimulated investigation of the possibilities of near and mid-infrarred (MIR) nonlinear dynamics and supercontinuum in gas-filled, silica-glass \bhc{}.

Numerical simulations based on \bhc{} filled with supercritical Xe and pumped at 3.7~\um{}  predicted supercontinuum generation spanning 1.85 to 5.20~\um{} \cite{Hasan:16}. Alternatively, numerical simulations of plasma dynamics in \bhc{} filled with 1.2~bar Xe and pumped at 3~\um{} predicted supercontinuum generation spanning 1 to 4~\um{} \cite{habib2017multiple}.

Experimentally, using simply an air-filled \bhc{}, \textcite{mousavi2017exploring} demonstrated a variety of nonlinear effects, including supercontinuum formation between 850 to 1600~nm, when pumping at 1030~nm. A more exotic supercontinuum mechanism, based on the dispersion caused by the resonances in a \bhc{} has also recently been reported \cite{sollapur2017soliton}, leading to broadening extending from 200~nm to 1.7~\um{}.

Experimental results extending into the MIR have been obtained by \textcite{Cassataro:17} who achieved supercontinuum extending from 270~nm to 3.1~\um{} in an Kr-filled \bhc{} pumped at 1.7~\um{}, and \textcite{kottig2017novel} who achieved a 4.7-octave supercontinuum spanning from 180~nm to 4.7~\um{} in a Ar-filled \bhc{} pumped with 27~fs pulses at 1030~nm.

Further nonlinear mechanisms and experiments in this important spectral region are to be expected.

\subsection{Modulational instability\label{sec:gas_mi}}
The use of modulational instability\footnote{For a description of this process in conventional optical fibers see the literature following the first proposal \cite{hasegawa1980tunable}, demonstration \cite{TaiObservation1986}, or early history in general \cite{ZAKHAROV2009540}.} (MI) to obtain a broad and smooth continuum in HC-PCF, albeit without temporal coherence, was first proposed by \textcite{travers_ultrafast_2011}. The essential idea was to extend the MI techniques used in solid-core fibers\footnote{ See \textcite{dudley_supercontinuum_2006} for a general review, and \textcite{travers_blue_2010} for a specific review of extending MI supercontinua towards the blue in solid-core PCF.} to the UV region using the weaker dispersion and UV transmission of gas-filled HC-PCF. This Kerr-driven MI is distinct from the plasma-driven MI predicted by \textcite{SalehMI2012} and described previously.

\textcite{tani_phz-wide_2013} used an 18~\um{} core diameter, 10~bar Xe filled \bhc{}, pumped with 500~fs, 5~$\muup$J pulses at 800~nm to obtain a supercontinuum extending from 320~nm to 1300~nm. The corresponding input soliton order was \textit{N}~=~210. Clear MI sidebands were observed in agreement with numerical simulations, which predicted that temporal soliton-like structures would emerge with a sub-cycle duration of $\sim$1~fs and peak power $\sim$20~MW. The extended blue-side of the continuum was formed by dispersive-wave emission from these structures, and the red-side by the resulting spectral recoil of the solitons.

\textcite{azhar_raman-free_2013} used an 18~\um{} core diameter \bhc{} filled with very high pressure Ar, up to 150~bar, to study the role and evolution of MI and dispersive-wave emission by tuning the zero dispersion wavelength and soliton order with pressure, while keeping the pump parameters fixed at 140~fs pulses at 800~nm with energies up to $\sim$450~nJ.

\subsubsection{Quantum Optics Applications.}
\textcite{finger_raman-free_2015} used the high-pressure filled \bhc{} system for twin-beam generation---a quantum optical light source of two beams, which may have a very uncertain number of photons, but where the number of photons in each beam is always exactly the same. Twin beams have applications in quantum metrology, imaging, and key distribution among others. Twin beam sources based on MI or FWM had been previously demonstrated in solid-core fibers, but always in the presence of Raman scattering, which in this case acts as a source of noise. Raman is absent in the high pressure Ar system, and in addition, the sideband locations can be tuned either by the pump energy or, more easily and widely, by the gas pressure. The source demonstrated by \citeauthor{finger_raman-free_2015} was the brightest twin-beam source yet demonstrated, and also contained the fewest spatiotemporal modes, a requirement of precise quantum optical metrology. Further work showed how the time-frequency mode structure of the twin-beams could be tuned using the gas pressure, pump pulse chirp and fiber length \cite{finger_twin_2017}.

\subsection{Spatial nonlinear effects and filamentation}
Almost all HC-PCFs designed or manufactured to date are multimode. Certainly the \bhc{} widely used for ultrafast nonlinear optics supports a plethora of modes. This means that spatial effects can occur through nonlinear interactions coupling the modes. Multimode effects have been included in numerical models of ultrafast pulse propagation in \bhc{} \cite{tani_multimode_2014}. At the extreme limit, nonlinear coupling between modes manifests as self-focusing\footnote{Analysis of the critical power for self-focusing in HCF, which is directly applicable to HC-PCF can be found by the papers of \textcite{fibich_critical_2000,milosevic_optical_2000,zheltikov_self-focusing_2013}.}.

In HC-PCF, \textcite{konorov_self-channeling_2004} observed spatial effects when using 4~$\muup$J, 30~fs pulses to pump a 14~\um{} core diameter \pgpcf{} fiber filled with air or Ar. These effects occurred when the peak powers reached between 0.5 to 0.9 time the critical power for self-focusing. The output spatial profile was observed to become smooth and symmetric and this was attributed to the formation of a nonlinear waveguide within the HC-PCF.

\textcite{azhar_nonlinear_2013} pumped \bhc{} filled with Xe at supercritical pressures, and hence with very high densities. As the energy was increased a threshold like change in transmission occurred at a certain input energy. The effect was ascribed to self-focusing in the input gas-cell, but the mechanism could not be fully explained for their parameters.

\textcite{tani_multimode_2014} modelled and experimentally demonstrated the emission of RDWs in higher order modes and modelled previous experimental results on intermodal THG and FWM. They also noted a form of filamentation-like balance occurring in the core of the \bhc{}, for parameters corresponding to their previous result on MI \cite{tani_phz-wide_2013}. When modelling the propagation without ionization, the beam was predicted to self-focus inside the fiber through efficient emission of energy into higher order modes through RDW emission; whereas including ionization terms prohibited this, and the beam was guided very close to the shape expected for linear propagation.

In their soliton self-compression experiments, \textcite{balciunas_strong-field_2015} noted that the output beam shrunk due to a small self-focusing effect at higher energies.

\subsection{Nonlinear wave mixing and harmonic generation}
\subsubsection{ Third harmonic generation}
Third harmonic generation was the first ultrafast nonlinear experiment to be performed in \bhc{}. \textcite{nold_pressure-controlled_2010} used 2~$\muup$J, 30~fs pulses at 800~nm, to generate pulses around 275~nm. To obtain phase-matching, the UV light needed to be emitted in a higher order mode. They targeted the HE${}_{13}$ mode with 5~bar Ar and demonstrated pressure tuning of the phase-matched THG emission. Overall the conversion efficiency was very low, at 10${}^{-5}$, and does not compete with RDW emission discussed in
Section~\ref{sec:dwave_kgm}. \textcite{tani_multimode_2014} used this experiment as a test for their multimode propagation model.

\subsubsection{ Second harmonic generation}
\textcite{menard_phase-matched_2015} demonstrated second harmonic generation in Xe filled \bhc{}. While Xe is macroscopically inversion symmetric and hence does not offer an intrinsic \textit{$\chi$}${}^{(2)}$ response, an effective \textit{$\chi$}${}^{(2)}$ can be obtained using the widely explored technique of applying a quasi-static electric field $E_{\mathrm{stat}}$, such that ${\chi }^{(2)}_{\mathrm{eff}}={E_{\mathrm{stat}}\chi }^{(3)}$. Similar to the THG work by  \textcite{nold_pressure-controlled_2010}, \textcite{menard_phase-matched_2015}  used intermodal phase-matching to increase the efficiency of the process. They obtained an efficiency of 10${}^{-7}$ when converting 30~$\muup$J, 2~ns pump pulses at 1064~nm to 532~nm in the HE${}_{12}$ mode. In a follow-up paper \textcite{menard_broadband_2016} improved the conversion efficiency, by three orders of magnitude, by using femtosecond pump pulses with higher peak power and improving the electrode design to increase $E_{\mathrm{stat}}$ inside the fiber core. Additionally, they demonstrated the use of quasi-phase matching, with patterned electrodes, to obtain conversion to the fundamental fiber mode.

While even the improved version still has very low conversion efficiency of 0.02\%, there are a number of interesting prospects for this technique. Firstly, it can be used to obtain signals at more exotic wavelengths, such as the VUV or MIR or THz range. Secondly, as the position of the SHG generation can be controlled by an external electric field, it is possible to vary the generation point along the fiber and hence noninvasively measure the modal attenuation at shifted wavelengths without cutting the fiber.

\subsubsection{Four wave mixing}
Four wave mixing (FWM) in HC-PBG was demonstrated and optimized in a series of papers by \textcite{konorov_enhanced_2003,konorov_phase-matched_2004,konorov_phase-matched_2005-1}. Due to the narrow guidance bands in HC-PBG this required very special fiber designs. \textcite{konorov_enhanced_2003} designed a fiber designed to transmit 1060~nm, its second harmonic at 530~nm, and its third harmonic at 353~nm. A 9~cm fiber was pumped with 30-50~$\muup$J, 30~ps pulses both at $1060$~nm and its second harmonic, and the UV signal generated using four-wave mixing in the configuration 3$\omegaup$ = 2$\omegaup$ + 2$\omegaup$ -- $\omegaup$. In a comparison to tight-focusing in free-space, the authors claimed an 800 times enhancement. A second fiber design enabled the same experiment to be performed starting from a fundamental wavelength of 1250~nm from a Cr:forsterite laser. In this case the fiber was optmized for phase-matching. Finally, by scaling to a larger core size \citeauthor{konorov_phase-matched_2005-1} scaled the pulse energy to the mJ range with ns pulse durations \cite{konorov_phase-matched_2005-1}.

Broadband guiding \arpcf{}, which can have guidance bands spanning the UV, visible, near-IR spectrum should be even better hosts for such FWM interactions. The first result in \arpcf{}, was by \textcite{azhar_nonlinear_2013}. During their experiments on filling \arpcf{} with supercritical Xe they observed intermodal FWM before the onset of self-focusing effects.

Ultrafast FWM for the efficient generation of deep-ultraviolet pulses, as developed in conventional HCF \cite{misoguti_generation_2001}, was recently demonstrated by \textcite{belli_ultrafast_2016}. They used a 26~\um{} diameter Ar-filled \bhc{} to obtain a >30\% conversion efficiency from 400~nm to and an energy of 490~nJ at 266~nm, when seeding at the idler wavelength of 800~nm. The authors noted a number of advantages compared to RDW emission, in particular that the UV pulse chirp could be controlled and that the bandwidth of the emitted band could be tuned independently of the central frequency. This is hard to achieve with RDW emission as the dynamics of high frequency emission are so tightly coupled with extreme pulse compression. The advantage compared to HCF is the ability to use much lower pump energies, in the $\mu$J region, and hence the possibility of scaling the repetition rate and average power of the deep-UV emission by making use of ultrafast fiber and thin-disk laser systems.

In principle this scheme could be extended to the VUV region. In a theoretical study, \textcite{im_microjoule_2015} considered chirped FWM to the VUV region. Predicting the generation of sub-10 fs VUV pulses with an energy of up to hundreds of $\muup$J by broad-band chirped idler pulses at 830 nm and MW pump pulses with narrow-band at 277~nm. Such a source would strongly compete with RDW emission.

\subsubsection{High harmonic generation\label{sec:hhg}}
Given that HC-PCF is capable of guiding intense ultrafast light pulses, such that they can partially ionize the filling gas, and in addition that the dispersion of the gas and waveguide system can be tuned, it is no surprise that it has been considered for the generation of high harmonics (HHG) in the extreme UV and soft-Xray spectral region.

\paragraph{Phase-matching}
\textcite{serebryannikov_phase-matching_2004} performed a detailed study on the possibilities of phase-matching HHG generation in \pgpcf{} fiber. Noting that the waveguide contribution to the dispersion is much stronger than in conventional HCF, various schemes, such as emitting the harmonics in high order modes, or using specific dispersion variation at the edges of a band-gap were proposed. In a further study, \textcite{serebryannikov_broadband_2008}, discussed using the nonlinear phase modulation resulting from soliton propagation and ionization in the fiber core, similar to the phase matching of RDW discussed in Sections \ref{sec:dwave_kgm} and \ref{sec:dwave_mir}. In this way it was predicted that a broad band of harmonics, from the 19${}^{th}$ to 129${}^{th}$, could be simultaneously phase-matched from the center of the band-gap, hence providing low-loss guidance of the pump field. Alternatively, \textcite{ren_quasi-phase-matched_2008} discussed using quasi-phase-matching with a counter-propagating laser, similar to that demonstrated in conventional HCF \cite{zhang_quasi-phase-matching_2007} could be used. So far neither of these techniques has been experimentally demonstrated.

Very recently the idea of achieving quasi-phase-matching by using interference between fiber modes was proposed and preliminary demonstrated \cite{Anderson:17}, as discussed below.

\paragraph{Experiment}
The first report on HHG inside a HC-PCF is by \textcite{heckl_high_2009}, who obtained the 13${}^{th}$ harmonic of 800~nm (at 60~nm) using 4.2~$\muup$J and the 7${}^{th}$ harmonic with as little as 200~nJ pump pulses. They used 1.5~cm of 15~\um{} core diameter \bhc{} filled with up to 30~mbar Xe. As noted above, recently \textcite{Anderson:17} demonstrated the 60-fold enhanced emission of high-harmonics at 30~eV. The authors claim that this enhancement was achieved by a form of quasi-phase-matching in which the interference between different fiber modes in a control pulse, and a delayed driving pulse, causes amplitude fluctuations with a period such that the harmonic emission is coherently enhanced.

An alternative combination of HC-PCF and HHG is to use a \bhc{} for soliton effect pulse compression or other soliton dynamics and use the output for direct HHG in a gas jet. This was the approach of \textcite{fan_integrated_2014} who used the self-compressed source described in Section~\ref{sec:fsscomp} with 30~$\muup$J in a single-cycle at 1.7~\um{} to generate an EUV continuum extending to beyond 48~eV. Recently \textcite{tani_wavelength-tunable_2017} showed that plasma blue-shifting solitons in a \bhc{} with an He pressure gradient (described in Section~\ref{sec:plasmasfbs}) can produce continuously blue-shifting HHG when the \hc{} output is placed directly in front of a Xe gas jet. The harmonic wavelengths could be continuously tuned over the range 25 to 60~nm.

Soliton self-compression can lead to HHG driving pulses of just 1~fs duration. Such a short driving field has not been widely explored as it has peviously been difficult to achieve. Prelimiary numerical simulations show that novel time-frequency phenomena can be expected in the EUV and XUV under such conditions \cite{ChuNear2016}.

\subsection{Gas based stimulated emission, amplification and lasing}
As fiber lasers based on doped solid-glass cores have been so successful, a number of groups have asked if it is possible to build a gas laser inside an HC-PCF. While apparently no discharge based gas-laser has yet been demonstrated (although see Section~\ref{sec:micro} for discharge emission), there have been a number of successful attempts at building optically pumped gas lasers. The first evidence of stimulated emission was provided by \textcite{jones_mid-infrared_2011}, using an HC-PCF filled with acetylene. Single pass conversion from 5~ns pulses at 1521.05~nm to two emission lines at 3123.2 nm and 3162.4 nm was achieved, with about 10${}^{-6}$ conversion efficiency from 1~$\muup$J pump pulses. This approach has recently been energy-scaled to achieve 1.4~$\muup$J at 3~\um{} for an absorbed pump energy of 8.2~$\muup$J \cite{Dadashzadeh:17}. \textcite{wang_efficient_2014} reported 3.1-3.2 \um{} mid-infrared emission from an acetylene-filled HC-PCF pumped with a diode laser. \textcite{nampoothiri_cw_2015} demonstrated continuous-wave lasing 1280--1340 nm region in a cavity surrounding an iodine filled HC-PCF, pumped at 532~nm. Most recently, \textcite{abu_hassan_cavity-based_2016} reported both continuous-wave and synchronously pumped operation of a ring-cavity mid-IR fiber laser based on acetylene filled HC-PCF and pumped with telecom diodes operating at 1.5~\um{}. Up to 9\% slope efficiency is achieved in the 3~\um{} region. Further enhancement has scaled the output power to 0.47~W and the sing-pass slope efficiency above 18\% \cite{Xu:17}.

\section{\label{4}Conclusions and Perspectives} 

In this article, we reviewed recent progress in silica-based hybrid PCFs fabricated through various post-processing methods, which allow the integration of novel and functional materials inside conventional solid- and HC-PCFs. Use of liquids, glasses, metals, semiconductors and gases have opened the door to a plethora of new applications in critical areas such as communications, sensing, bio-photonics, ultrafast science and quantum optics. 

Sections II.A. and B. overviewed the most important reports on hybrid PCFs for the development of tunable linear devices suitable for sensing, tunable filters, switching elements and other active and passive fiber components. We also presented a number of selected experimental studies focused on anti-resonant devices and plasmonics using highly nonlinear chalcogenide glasses and metals. Section II.C. highlighted the main nonlinear effects achieved in hybrid PCFs infiltrated with nonlinear liquids. Self-defocusing, spatial soliton formation, nonlocality, supercontinuum generation and nonlinear directional couplers are some of the main nonlinear PCF devices that have been developed over the past few years.

However, the main question is perhaps "what is next?" The answer to this question is directly linked to and dependent on two main aspects. The first one involves the development of novel functional nanomaterials that can be used to further enhance existing or reveal new optical properties when interacting with light. The second involves the development of new fabrication strategies that will enable of successfull post-integration of novel materials inside solid-core PCF as well as HC-PCF, which due to thermo-mechanical incompatibility are not suitable for direct fiber drawing. For example, the solution-processed approach of soft-glass deposition presented in Section II.A.2 could be combined with the pressure assisted-method for the integration of multi-materials with novel functionalities into a single cladding hole of the PCF. Such an approach would significantly extend the design and material flexibility of hybrid PCFs. As stated in the introduction, the results we presented in this review article were exclusively targeting silica glass. Considering different host materials, such as chalcogenide glasses, hybrid PCFs could find their way to mid-infrared (MIR) wavelength range  (known also as "molecular fingerprint" region), a very attractive but not yet fully explored area. The MIR region has been proved to be quite important due to the presence of a multitude of strong characteristic absorption lines of many important molecules, making this region crucial for applications in materials processing, chemical and biomolecular sensing as well as spectroscopy for non-destructive MIR optical coherence tomography (OCT) and pollution monitoring. With the rapid development of MIR sources such as MIR supercontinuum, tunable optical parametric amplifiers (TOPAs) and quantum cascade lasers (QCLs), it is expected that MIR monolithic fiber devices such as sensors, filters, fiber couplers, etc. will also play an important role in this area in the near future. Indeed by using appropriate MIR-compatible materials, the hybrid fiber devices presented in this article could straightforwardly be adopted in the MIR region.     

Section III highlights the most recent advances in the field of gas-filled HC-PCF for gas-based nonlinear optics. The main advantage of using gases inside HC-PCF is that both the dispersion and the nonlinearity can be tuned by simply changing the gas pressure while guiding high peak intensities without damaging the host glass material of the fiber. The strong confinement of light in the core of a HC-PCF and the strong light-gas interaction length has enabled investigation of previously inaccessible interaction regimes. Exciting new applications and results have emerged over the past few years, such as light generation in the extreme wavelength regions of vacuum ultraviolet, sub-cylce pulse compression, soliton-plasma interactions, high-harmonic generation to name a few. While there are many further directions to explore, here we outline three that are clear. (i) As for the solid and liquid-filled hybrid PCFs, the MIR region is relatively unexplored for gas-filled HC-PCF. With some of the extraordinary low-loss guidance fibers achieved in the MIR by \textcite{yu_spectral_2013,yu_negative_2016}, even with silica glass, there are certainly many possibilities. The recently demonstated MIR resonant dispersive-wave emission (Section~\ref{sec:dwave_mir}) hinted at this and there have been some initial work as discussed in Section~\ref{sec:mirscg}. It appears reasonable that many of the ultrafast dynamics explored to date in gas-filled HC-PCF can be shifted to the MIR, but also that new phenomena will occur, similarly to the new regimes currently being explored in MIR filamentation \cite{panagiotopoulos2015super}. (ii) Another, relatively unexplored direction, is scaling the pulse repetition rate. This can usually straightforwardly increase the average power of the system without significant alteration of the nonlinear dynamics if thermal management if effective, and a number of pulse compression experiments at high repetition rate (MHz range) have been explored \cite{kottig_high_2015,kottig2017generation}. However, given the long recombination times, plasma effects can significantly alter the dynamics as the time period between subsequent pulses is reduced. In this case it is reasonable to imagine building up plasma densities inside a HC-PCF using avalanche ionization and heating of the plasma over multiple shots. This could change the dispersion landscape of the system (i.e. a plasma filled fiber could break the fairly uniform scaling of linear dispersion and nonlinearity found in the noble gases), or be used for more exotic nonlinear experiments such as Raman-like interactions in plasmas. This area is quite unexplored and ripe for further investigation given the now ready availability of high repetition rate femtosecond laser sources with multi-$\mu$J pulse energies. (iii) Almost all of the nonlinear experiments reported in HC-PCF have been based on linearly polarized pump lasers and interactions. Utilizing the vector nature of the light fields in gas-filled HC-PCF to either enhance or alter the balance between nonlinear effects, or to create light sources with specific tailored characteristics is an area of research yet to receive much attention. In particular, tuning the ellipticity of the DUV RDW emission would be an important additional tool for spectroscopy (to e.g. spin-resolve photo-emitted electrons). 

In section III.I. we overviewed the use of HC-PCFs filled with Raman-active gases and liquids as Raman probes and generators. Efficient collection of the Raman signal, accumulation of the signal over the length of the fiber and operation in the short wavelength spectral regions of visible and UV without suffering from silica Raman background signal or photo-damage make a HC-PCF an excellent choice as a Raman probe. Extending the Raman spectroscopy in HC-PCFs to the nonlinear interaction regime, e.g. coherent anti-Stokes Raman scattering, is certainly of considerable interest. Extremely high Raman gains offered by HC-PCFs also allow straightforward generation of an octave-spanning Raman comb of narrowband spectral lines. Combination of different gases inside a single HC-PCF pumped with high energies may offer a new route to the generation of dense spectral clusters across the ultraviolet-visible wavelength range and could constitute another direction for novel nonlinear effects. Efficient molecular modulation with ultrashort pulses in HC-PCFs have shown for the first time the impulsive excitation of extremely fast molecular vibrations of light diatomic gases such as H${}_{2}$ and D${}_{2}$ and generation of extremely broad supercontinua. Developments in this direction are of interest for the generation and manipulation, via coherent Raman of ultrashort pulses in the visible and ultraviolet spectral regions.

In conclusion, novel photonic materials combined with PCFs as a versatile platform have empowered a wide arc of unique fiber-based linear and nonlinear devices and architectures. Hybrid PCFs will inevitably remain the "lion's share" of the research on PCFs and their applications in the new optical frontiers of UV and MIR over the next few years and thus a bright future is anticipated. We hope that this article serves as introductory point to the vast literature that has already accumulated in this field.

\begin{acknowledgments}
C.M. would like to thank Ayman F. Abouraddy, Kristian Nielsen and S{\o}ren Michael M{\o}rk Friis for fruitful discussions. J.C.T and A.A. thank Federico Belli for useful discussions. C.M. and O.B. acknowledge financial support from the Danish Research Council (FTP) (4184-00359B), Innovation Fund Denmark ShapeOCT (4107-00011A) and Carlsberg Foundation (CF14-0825). J.C.T. has received funding from the European Research Council (ERC) under the European Union’s Horizon 2020 research and innovation programme (grant agreement HISOL 679649). B.J.E. acknowledges the support of the Australian Research Council (ARC) through the Laureate Fellowship (FL120100029) and  Center of Excellence CUDOS, (CE110001018).

\end{acknowledgments}

\end{document}